\documentclass[pre,showpacs,showkeys,floatfix,nofootinbib,longbibliography,onecolumn]{revtex4-1}

\usepackage{graphicx}
\usepackage{amsmath}
\usepackage{amsfonts}
\usepackage{bm}
\usepackage{bbm}
\usepackage{amssymb}
\usepackage{mathrsfs}
\usepackage{dsfont}
\usepackage{subfigure}
\usepackage{xcolor}

\newlength{\irrl}
\newlength{\irrw}

\newcommand{\arc}{\operatorname{arctanh}}
\newcommand{\tr}{\operatorname{tr}}

\newcommand{\diver}{\operatorname{div}}
\newcommand{\divs}{\operatorname{div_s}}
\newcommand{\nablas}{\nabla_\mathrm{s}}
\newcommand{\n}{\bm{n}}
\newcommand{\normal}{{\bm{\nu}}}
\newcommand{\radial}{\bm{r}}
\newcommand{\radiale}{\radial_\mathrm{e}}

\newcommand{\ave}[1]{\left\langle{#1}\right\rangle}
\newcommand{\area}[1]{a({#1})}
\newcommand{\arean}{\area{\n}}
\newcommand{\areanu}{\area{\normal}}

\newcommand{\A}{\mathbf{A}}

\newcommand{\va}{\bm{a}}

\newcommand{\vu}{\bm{u}}
\newcommand{\vv}{\bm{v}}
\newcommand{\e}{\bm{e}}
\newcommand{\ep}{\bm{e}_\varphi}
\newcommand{\er}{\bm{e}_r}
\newcommand{\ex}{\bm{e}_x}
\newcommand{\ey}{\bm{e}_y}
\newcommand{\ez}{\bm{e}_z}

\newcommand{\Frame}{(\e_x,\e_y,\e_z)}
\newcommand{\tang}{\bm{t}}

\newcommand{\dip}{\bm{d}}
\newcommand{\m}{\bm{m}}
\newcommand{\x}{\bm{x}}
\newcommand{\xxi}{\bm{\xi}}
\newcommand{\y}{\bm{y}}

\newcommand{\sphere}{{\mathbb{S}^2}}
\newcommand{\ball}{\mathbb{B}^3}
\newcommand{\balle}{\ball_\ve}
\newcommand{\surface}{\mathscr{S}}
\newcommand{\evo}[2]{V_\mathrm{e}\{{#1},{#2}\}}
\newcommand{\evop}{\evo{\body_1}{\body_2}}
\newcommand{\evou}{V_\mathrm{e}}
\newcommand{\mdn}{\bm{m}\cdot\bm{n}}
\newcommand{\mdnu}{\bm{m}\cdot\bm{\nu}}
\newcommand{\mdm}{\bm{m}_1\cdot\bm{m}_2}

\newcommand{\body}{\mathscr{B}}
\newcommand{\bodyu}{\mathscr{B}_1}
\newcommand{\bodyt}{\mathscr{B}_2}
\newcommand{\bodysum}{\body_1+\body_2}
\newcommand{\bodypair}{[\body_1,\body_2]}
\newcommand{\Ku}{K^{(1)}}
\newcommand{\Kt}{K^{(2)}}
\newcommand{\ru}{\radial_1}
\newcommand{\rt}{\radial_2}
\newcommand{\cone}[1]{\mathscr{C}^{#1}}
\newcommand{\conea}{\cone{\alpha}}
\newcommand{\sphero}{\mathscr{S}^{\eta}}
\newcommand{\coneu}{\cone{\alpha}_1}
\newcommand{\conet}{\cone{\alpha}_2}
\newcommand{\disk}{\mathscr{D}}
\newcommand{\ridge}{\mathfrak{R}}
\newcommand{\ridgee}{\ridge_\varepsilon}

\newcommand{\boundary}{{\partial\body}}
\newcommand{\eucl}{\mathscr{E}}

\newcommand{\conv}{\mathscr{K}}
\newcommand{\convp}{\conv^+}
\newcommand{\Va}{\ave{\evou}}
\newcommand{\Vc}{V_\mathrm{c}}
\newcommand{\Vs}{V_\mathrm{s}}
\newcommand{\exc}[2]{\mathscr{B}_\mathrm{e}\{{#1},{#2}\}}
\newcommand{\excs}{\mathscr{B}_\mathrm{e}}
\newcommand{\excp}{\exc{\body_1}{\body_2}}
\newcommand{\replica}{{\body_2}}
\newcommand{\replicaa}{{{\body_2^\ast}}}
\newcommand{\iap}{\int_{\frac\pi2-\alpha}^\pi}
\newcommand{\iat}{\int_{\theta_1}^{\theta_2}}
\newcommand{\iad}{\int_0^\pi}
\newcommand{\ione}{\int_{-1}^1}
\newcommand{\denu}{1+(\eta^2-1)u^2}
\newcommand{\denx}{\eta^2+(1-\eta^2)\xi^2}
\newcommand{\vt}{\vartheta}
\newcommand{\vp}{\varphi}
\newcommand{\ve}{\varepsilon}
\newcommand{\Ja}{P_{n-2}^{(2,2)}}

\newcommand{\rhouu}{\rho_1^{(1)}}
\newcommand{\rhotu}{\rho_2^{(1)}}
\newcommand{\rhout}{\rho_1^{(2)}}
\newcommand{\rhott}{\rho_2^{(2)}}

\newcommand{\euu}{\e_1^{(1)}}
\newcommand{\etu}{\e_2^{(1)}}
\newcommand{\eut}{\e_1^{(2)}}
\newcommand{\ett}{\e_2^{(2)}}
\newcommand{\Angle}{\phi}
\newcommand{\dens}{\varrho}
\newcommand{\ndens}{\rho_0}
\newcommand{\vir}{\mathsf{B}}
\newcommand{\ecc}{\epsilon}

\newcommand{\frax}[2]{\textstyle\frac{#1}{#2}}

\newcommand{\argmin}{\arg\!\min}

\begin{document}

\title{Explicit excluded volume of cylindrically symmetric convex bodies}
\author{Marco \surname{Piastra}}
\email[e-mail: ]{marco.piastra@unipv.it} \affiliation{ Dipartimento di  Ingegneria
Industriale e dell'Informazione, Universit\`a di Pavia, via Ferrata 1, I-27100
Pavia, Italy}
\author{Epifanio G. \surname{Virga}}
\email[e-mail: ]{eg.virga@unipv.it}
\affiliation{ Dipartimento di Matematica, Universit\`a di Pavia, Via Ferrata 5, I-27100 Pavia, Italy}

\date{\today}

\begin{abstract}
We represent explicitly the excluded volume $\evop$ of two generic cylindrically symmetric, convex rigid bodies, $\bodyu$ and $\bodyt$, in terms of a family of shape functionals evaluated separately on $\bodyu$ and $\bodyt$. We show that $\evop$ fails systematically to feature a dipolar component, thus making illusory the assignment of any shape dipole to a tapered body in this class. The method proposed here is applied to cones and validated by a shape-reconstruction algorithm. It is further applied to spheroids (ellipsoids of revolution), for which it shows how some analytic estimates already regarded as classics should indeed be emended.
\end{abstract}

\pacs{61.30.-v; 61.30.Cz; 47.57.J-}

\keywords{Excluded volume; Second virial coefficient; Hard-body interactions; Brunn-Minkowski theory; Minkowski functionals; Colloids.}

\maketitle
\section{Introduction}\label{sec:introduction}
Onsager's celebrated paper \cite{onsager:effects} on the effect of shape on the interaction between hard particles has perhaps been the most influential contribution to colloidal sciences of the last century \cite{frenkel:perspective}. There, entropic forces alone were first recognized as capable of inducing a structural ordering transition with no involvement of whatever cohesion force may be present. The typical prototype of such an ordering transition remains indeed the isotropic-to-nematic transition predicted in \cite{onsager:effects} for an assembly of slender hard rods as their number density is increased beyond a critical value (falling within a narrow gap of phase coexistence). As paradoxical as it may appear at a superficial glance, such an ordering transition is duly accompanied by an increase in entropy, since the loss in orientational disorder attached to the rods' alignment is outbalanced by the gain in translational disorder made possible by the increase in the volume available for the particles' centers of mass \cite{frenkel:perspective,frenkel:entropy}. The conjugated counterpart of this volume is the \emph{excluded volume}.

The excluded volume of two rigid bodies is the volume in space that any one point in one body cannot access by the very presence of the other body. This definition is delusively simple as it conceals a formidable mathematical task which can seldom be accomplished in an exact analytic form.\footnote{We learn from \cite{palffy:distance_2D} that Viellard-Baron, who took an early interest in this problem \cite{vieillard-baron:phase}, ``was reportedly greatly disturbed by the difficulties he encountered.''} Of course, there are exceptions to this general statement, but they are very few. Noticeable among these are the excluded volume of circular cylinders \cite{onsager:effects}, sphero-cylinder \cite{vroege:phase}, sphero-platelets \cite{mulder:solution}, and sphero-zonotopes \cite{mulder:excluded}.\footnote{Isihara~\cite{isihara:theory} is often credited with having provided an explicit formula for the excluded volume of ellipsoids of revolution. In Sec.~\ref{sec:spheroids} below, we shall discuss this case in some detail.}

Despite its technical difficulties, the excluded volume remains a key ingredient of both Onsager's original theory and its most recent extensions. In all of these, the per-particle free energy $F$ of an assembly of hard bodies (appropriately made dimensionless) is a functional of the single-body local density $\dens$. A number of papers have interpreted Onsager's original theory in the light of the modern density functional theories; here we refer the reader to the most recent review on the subject \cite{mederos:hard-body}, which is mostly concerned with hard-body systems that exhibit liquid crystalline phases.\footnote{A general reference for simple liquids is still the classical book \cite{hansen:theory}, now enriched by an addition on complex fluids.} $F[\dens]$ differs from the free-energy functional for an ideal gas by the addition of an \emph{excess} free energy $F_\mathrm{ex}[\dens]$, which characterizes the interactions of anisometric particles. In general, $F_\mathrm{ex}[\dens]$ is not known explicitly, but it can always be expressed as a power series in the total number density $\ndens$, which is often called the \emph{virial} expansion. The first non-trivial term of such an expansion is $\ndens\vir_2[\dens]$, where the functional $\vir_2$ is the \emph{second} virial coefficient, which is nothing but the ensemble average of the excluded volume,
\begin{equation}\label{eq:second_virial_coefficient}
\vir_2[\dens]:=\frac12\int_{\Omega^2}\evou(\omega,\omega')\dens(\omega)\dens(\omega')d\omega d\omega'.
\end{equation}
In \eqref{eq:second_virial_coefficient}, $\Omega$ is the \emph{orientational manifold}, which describes all possible orientations of a particle in the system and $\evou(\omega,\omega')$ is the excluded volume of two particles with orientations $\omega$ and $\omega'$, respectively. Higher powers of $\ndens$ bear higher virial coefficients $\vir_n$, which however are even more difficult to compute than $\vir_2$.

Onsager \cite{onsager:effects} remarkably estimated that for rods sufficiently slender $\vir_2$ actually prevails over all other $\vir_n$'s. This makes Onsager's theory virtually exact, as was also subsequently confirmed directly by numerical computations \cite{frenkel:onsager,frenkel:onsager_erratum}. Nevertheless, even when the second virial coefficient $\vir_2[\dens]$ cannot be proved to be dominant, it remains a viable approximation to $F_\mathrm{ex}[\dens]$ in establishing, at least qualitatively, the variety of possible equilibrium phases in a hard-body system and the entropy-driven transitions between them. To this end, explicit formulas for the excluded volume of rigid bodies are to be especially treasured.

This is the motivation for our study. Our objective is to express $\evop$, the excluded volume for two rigid bodies, $\bodyu$ and $\bodyt$, in terms of \emph{shape} functionals depending solely on the individual bodies $\bodyu$ and $\bodyt$. We shall accomplish this task for bodies both convex and cylindrically symmetric, for which $\evop$ can be given with no loss in generality as the sum of a series of Legendre polynomials $P_n$,
\begin{equation}\label{eq:excluded_volume_series}
\evop=\sum_{n=0}^\infty B_nP_n({\mdm}),
\end{equation}
where $\m_1$ and $\m_2$ are unit vectors along the symmetry axes of $\bodyu$ and $\bodyt$, respectively.\footnote{Following Isihara~\cite{isihara:theory}, we denote by $B_n$ the Legendre coefficients of $\evou$, though often in more recent literature this symbol is used to designate the virial coeffients, here denoted as $\vir_n$.} The shape functionals involved in our explicit representation will be natural extensions of the classical functionals on which was largely based the celebrated Brunn-Minkowski theory of convex bodies.\footnote{Besides the original sources \cite{brunn:thesis,minkowski:volumen}, the general books \cite{bonnesen:theory,schneider:convex} are highly recommended. We also collected a number of relevant results phrased in the same mathematical language employed here in Appendix~A to our earlier study on this subject \cite{piastra:octupolar}. Finally, a different but equivalent approach is presented in \cite{singh:molecular}.} The major advantage of the method proposed here is the explicit computability of such extended Minkowski functionals, which makes our representation formula directly applicable to bodies $\bodyu$ and $\bodyt$ not necessarily congruent, possibly representing particles of different species.

The paper is organized as follows. In Sec.~\ref{sec:volume_averages}, we set the scene for our development by showing that the Legendre coefficients $B_n$ of the representation formula \eqref{eq:excluded_volume_series} can be expressed as appropriate anisotropic volume averages. Section~\ref{sec:no_dipole} is devoted to the coefficient $B_1$ of the first Legendre polynomial $P_1(\mdm)=\mdm$ in \eqref{eq:excluded_volume_series}. We attach a special meaning to this, as it represents the \emph{dipolar} contribution to $\evop$ which would possibly arise from tapered, cylindrically symmetric, convex bodies, if only one could unambiguously assign a \emph{shape dipole} to them. The somewhat surprising conclusion will be that $B_1$ \emph{vanishes} identically on this class of bodies, making the very notion of shape dipole void, despite its intuitive appeal. Section~\ref{sec:Extended_Minkowski_functionals} is concerned with the extended Minkowski functionals, in terms of which, once evaluated on the bodies $\bodyu$ and $\bodyt$, we can write in closed form all coefficients $B_n$ in \eqref{eq:excluded_volume_series}. An explicit application of our method is illustrated in Sec.~\ref{sec:cones}, where we evaluate the extended Minkowski functionals for a generic circular cone and validate our evaluations through a direct computation of the coefficients $B_n$ made possible by an independent shape-reconstruction algorithm, appropriately modified to tackle efficiently the cone's sharp ridge. Likewise, in Sec.~\ref{sec:spheroids}, we determine the extended Minkowski functionals for a spheroid, that is, an ellipsoid of revolution, either prolate or oblate. In Sec.~\ref{sec:conclusions}, we collect the main conclusions of our work, looking back afresh to some of them, also in the light of possible future developments that they may suggest.

We shall endeavor to make our presentation as free as possible from unwanted technical details that might obscure both the outcomes of our study and the strategy adopted to obtain them. To provide, however, the interested reader with enough information to appreciate the mathematical infrastructure underlining this paper, we collect in two closing appendices the details of both the mathematical theory and the shape-reconstruction algorithm.

\section{Anisotropic volume averages}\label{sec:volume_averages}
It was proved by Mulder~\cite{mulder:excluded} that the excluded volume of $\evop$ of two bodies, $\bodyu$ and $\bodyt$, be they convex or not, can be expressed as
\begin{equation}\label{eq:Mulder_equality}
\evop=V[\bodyu+\bodyt^\ast],
\end{equation}
where $V$ is the volume functional, $\bodyt^\ast$ is the central inverse (relative to a specified origin $o$) of the body $\bodyt$, and $+$ denotes the Minkowski addition (to the definition of which concurs the origin $o$).\footnote{We shall often call \eqref{eq:Mulder_equality} Mulder's identity. The reader is referred to the primer on the Brunn-Minkowski theory of convex bodies in Appendix~A of \cite{piastra:octupolar}. A short recapitulation of this theory is also given in Appendix~\ref{sec:essentials} below to make our paper self-contained.} Letting both $\bodyu$ and $\bodyt$ be cylindrically symmetric bodies with axes $\m_1$ and $\m_2$, respectively, since $\evop$ is an isotropic scalar-valued function, by a theorem of Cauchy,\footnote{See, for example, Sec.~113.1 of \cite{gurtin:mechanics}.} we can say that $\evop$ is a function (still denoted as) $\evou$ of the inner product $\mdm$. Setting $\mdm=\cos\vt$, the function $\evou(\cos\vt)$ can be expanded as the sum of a series of Legendre polynomials (see, for example, Secs.~18.2 and 18.3 of \cite{NIST:DLMF}):
\begin{equation}\label{eq:V_e_expansion}
\evou(\cos\vt)=\sum_{n=0}^\infty B_nP_n(\cos\vt),
\end{equation}
where
\begin{equation}\label{eq:B_n_definitions}
B_n:=\frac{2n+1}{2}\int_0^\pi\evou(\cos\vt)P_n(\cos\vt)\sin\vt d\vt
\end{equation}
are the \emph{Legendre coefficients} of $\evou$. We record for future use a few basic properties of the orthogonal polynomials $P_n$ (see, in particular, Secs.~18.6.1 of \cite{NIST:DLMF} and 8.917.1 of \cite{gradshteyn:table}):
\begin{equation}\label{eq:P_n_properties}
P_n(-x)=(-1)^nP_n(x),\quad P_n(1)=1,\quad|P_n(x)|\leqq1.
\end{equation}

There is another way of expressing the coefficients $B_n$, which we find illuminating. Consider the average
\begin{equation}\label{eq:V_e_anistropic_average}
\ave{P_n\evou}\bodypair:=\ave{P_n(\mdm)\evou(\mdm)}_\replica
\end{equation}
computed for fixed $\bodyu$ over all possible replicas of $\bodyt$ obtained by rotating arbitrarily $\bodyt$ in space. By the cylindrical symmetry of $\bodyt$, the average \eqref{eq:V_e_anistropic_average} also acquires the equivalent form
\begin{equation}\label{eq:V_e_anistropic_average_equivalent}
\ave{P_n\evou}\bodypair=\ave{P_n(\mdm)\evou(\mdm)}_{\m_2},
\end{equation}
where, for any function $f(\e)$ defined on the unit sphere $\sphere$,
\begin{equation}\label{eq:average_over_Sphere_definition}
\ave{f}_{\e}:=\frac{1}{4\pi}\int_{\sphere}f(\e)da(\e)
\end{equation}
and $da(\e)$ denotes the area element with unit normal $\e$. Representing $\m_2$ in polar spherical coordinates with polar axis $\m_1$ and combining \eqref{eq:V_e_anistropic_average_equivalent} and \eqref{eq:B_n_definitions}, we readily arrive at
\begin{equation}\label{eq:V_e_anistropic_average_and_B_n}
\ave{P_n\evou}\bodypair=\frac12\int_0^\pi P_n(\cos\vt)\evou(\cos\vt)\sin\vt d\vt=\frac{1}{2n+1}B_n.
\end{equation}
Since both functions $\evou$ and $P_n$ are symmetric under the exchange of $\m_1$ and $\m_2$, the average $\ave{P_n\evou}\bodypair$ is also symmetric under the exchange of bodies $\bodyu$ and $\bodyt$:
\begin{equation}\label{eq:V_e_anistropic_average_symmetry}
\ave{P_n\evou}\bodypair=\ave{P_n\evou}[\bodyt,\bodyu].
\end{equation}

Equation \eqref{eq:Mulder_equality} allows us to express the Legendre coefficients $B_n$ of the excluded volume of two cylindrically symmetric bodies in a way directly related to the anisotropic averages of the volume of a Minkowski sum. Combining \eqref{eq:V_e_anistropic_average_and_B_n}, \eqref{eq:V_e_anistropic_average}, and \eqref{eq:Mulder_equality}, we readily see that
\begin{equation}\label{eq:B_n_averages}
\begin{split}
B_n&=(2n+1)\ave{P_n(\mdm)V[\bodyu+\bodyt^\ast]}_{\replica}
=(2n+1)(-1)^n\ave{P_n(\m_1\cdot\m_2^\ast)V[\bodyu+\bodyt^\ast]}_{\replica}\\
&=(2n+1)(-1)^n\ave{P_n(\m_1\cdot\m_2^\ast)V[\bodyu+\bodyt^\ast]}_{\replicaa},
\end{split}
\end{equation}
where $\m_2^\ast=-\m_2$ is the symmetry axis of the central inverse $\bodyt^\ast$ of $\bodyt$ and use has been made of \eqref{eq:P_n_properties} and the fact that averaging over $\replica$ is just the same as averaging over $\replicaa$. Thus, to obtain all coefficients $B_n$ in \eqref{eq:V_e_expansion}, we need to learn how to compute the \emph{anisotropic volume averages}
\begin{equation}\label{eq:anisotropic_volume_average_definition}
\ave{P_nV}\bodypair:=\ave{P_nV[\bodysum]}_{\replica},
\end{equation}
as then \eqref{eq:B_n_averages} would simply reduce to
\begin{equation}\label{eq:B_n_averages_central}
B_n=(2n+1)(-1)^n\ave{P_nV}[\bodyu,\bodyt^\ast],
\end{equation}
which obeys the same symmetry relation as in \eqref{eq:V_e_anistropic_average_symmetry}. Equation \eqref{eq:B_n_averages_central} is the basic building block of our development.

Although \eqref{eq:B_n_averages_central} is as general as \eqref{eq:Mulder_equality} for cylindrically symmetric bodies, this paper will solely be concerned with the excluded volume of \emph{convex} cylindrically symmetric bodies. For $n=0$, the average in \eqref{eq:anisotropic_volume_average_definition} becomes isotropic as $P_0\equiv1$ and its expression has long been know for generic convex bodies:\footnote{A derivation of \eqref{eq:isotropic_volume_average} can be found in \cite{singh:molecular}. Moreover, Kihara~\cite{kihara:coefficients,kihara:isihara} credits Isihara~\cite{isihara:determination} and Isihara and Hayashida~\cite{isihara:theory_I,isihara:theory_II} for having proved \eqref{eq:isotropic_volume_average}, although he also seems aware that a proof had already been contained in the classical work of Minkowski~\cite{minkowski:volumen}.}
\begin{equation}\label{eq:isotropic_volume_average}
\ave{V}\bodypair=V[\bodyu]+V[\bodyt]+\frac{1}{4\pi}\left(M[\bodyu]S[\bodyt]+M[\bodyt]S[\bodyu]\right),
\end{equation}
where $M$ is the \emph{total mean curvature} functional in \eqref{eq:M_functional} and $S$ is the \emph{surface area} functional in \eqref{eq:S_functional}. Since both $M[\body]$ and $S[\body]$ are invaraint under central inversion of $\body$, it follows from \eqref{eq:B_n_averages_central} and \eqref{eq:isotropic_volume_average} that
\begin{equation}\label{eq:B_0}
B_0=\ave{V}\bodypair.
\end{equation}
Here our challenge is to extend the neat classical formula \eqref{eq:isotropic_volume_average} for the isotropic average of the volume of the Minkowski sum of convex bodies to the anisotropic averages needed in \eqref{eq:B_n_averages_central}. This will be achieved in the two following sections with the aid of appropriate extensions of the classical Minkowski functionals $M$ and $S$. We anticipate that they are invariant under central body inversion like the classical Minkowski functionals, so that, in complete analogy with \eqref{eq:isotropic_volume_average} and \eqref{eq:B_0}, we shall be able to express the excluded volume $\evop$ of cylindrically symmetric bodies $\bodyu$ and $\bodyt$ in terms of functionals evaluated separately on $\bodyu$ and $\bodyt$.

As recalled in Appendix~\ref{sec:mathematical_details}, there is no loss in generality in limiting attention to the class $\convp$ of convex bodies with smooth boundaries and strictly \emph{positive} principal curvatures, as $\convp$ is dense in the whole class $\conv$ of convex bodies (see Appendix~\ref{sec:essentials}). Thus, our strategy will be to compute first the anisotropic volume averages in $\convp$ and then extend them by continuity to the whole of $\conv$. In the following section, we shall first accomplish our task for $\ave{P_1V}\bodypair$; this will lead us to conclude that $B_1\equiv0$, a general result of some import. In Sec.~\ref{sec:Extended_Minkowski_functionals}, we shall compute $\ave{P_nV}\bodypair$ for all $n\geqq2$ and arrive at the expected general explicit formula for all $B_n$'s.

\section{No shape dipoles}\label{sec:no_dipole}
\begin{figure}[h]
\centering
   \includegraphics[width=0.14\linewidth]{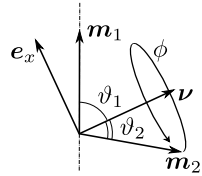}
\caption{Sketch representing the unit vectors $\normal$, $\m_1$, and $\m_2$. With $\normal$ and $\m_1$ fixed, $\m_2$ as represented by \eqref{eq:m_2_representation} describes a cone around $\normal$ in the first step of the averaging process described in the text.}
\label{fig:conical}
\end{figure}

Here our task is to compute $B_1$. To this end we remark that
\begin{subequations}\label{eq:B_1_preliminaries}
\begin{align}
\ave{P_n(\mdm)}_{\replica}&=\ave{P_n(\mdm)}_{\m_2}=0,\label{eq:B_1_preliminaries_one}\\
\ave{P_n(\mdm)V[\bodyt]}_{\replica}&=V[\bodyt]\ave{P_n(\mdm)}_{\replica}=0,\quad n\geqq1,\label{eq:B_1_preliminaries_two}
\end{align}
\end{subequations}
the former following from \eqref{eq:average_over_Sphere_definition} and the orthogonality of Legendre polynomials, and the latter also from the invariance of the volume functional under rotations. Then we represent $\m_2$ in a Cartesian frame $\Frame$ fixed in $\bodyu$. Letting $\ez=\normal$, where $\normal$ is the outer unit normal to $\bodyu$ at a selected point on $\partial\bodyu$, and choosing $\ey$ orthogonal to the plane $(\m_1,\normal)$, we have that
\begin{subequations}\label{eq:m_representations}
\begin{align}
\m_1&=\sin\vt_1\ex+\cos\vt_1\normal,\label{eq:m_1_representation}\\
\m_2&=\cos\Angle\sin\vt_2\ex+\sin\Angle\sin\vt_2\ey+\cos\vt_2\normal,\label{eq:m_2_representation}
\end{align}
\end{subequations}
the latter of which represents all possible orientations of $\m_2$, for given $\vt_1$ and $\vt_2$, the angles that $\m_1$ and $\m_2$ make with $\normal$ (see Fig.~\ref{fig:conical}).
An easy, but important consequence of \eqref{eq:m_representations} is that
\begin{equation}\label{eq:mdm}
\begin{split}
\mdm&=\sin\vt_1\sin\vt_2\cos\Angle+\cos\vt_1\cos\vt_2\\
&=\sin\vt_1\sin\vt_2\cos\Angle+(\m_1\cdot\normal)(\m_2\cdot\normal).
\end{split}
\end{equation}
Now, using also \eqref{eq:B_1_preliminaries}, we can derive from \eqref{eq:V_bodysum} the following expression\footnote{Unlike Mulder's identity \eqref{eq:Mulder_equality}, which is valid for general bodies, equation \eqref{eq:V_bodysum}, which is indeed one basic ingredient of our theory, has only been established for convex bodies.}
\begin{equation}\label{eq:P_1_V_bodypair}
\begin{split}
\ave{P_1V}\bodypair&=\frac13\left(\ave{\frac{\m_2\cdot\normal}{\Kt}}_\normal \int_\sphere(\normal\cdot\ru)(\normal\cdot\m_1)d\areanu+
\ave{(\normal\cdot\rt)(\normal\cdot\m_2)}_\normal\int_\sphere\frac{\m_1\cdot\normal}{\Ku}d\areanu\right)
\\
&+\frac16\ave{(\m_2\cdot\normal)\left(\rhout+\rhott\right)}_\normal \int_{\sphere}(\normal\cdot\ru)(\m_1\cdot\normal)\left(\rhouu+\rhotu\right)d\areanu\\
&+\frac16\ave{(\normal\cdot\rt)(\m_2\cdot\normal)\left(\rhout+\rhott\right)}_\normal \int_{\sphere}(\m_1\cdot\normal)\left(\rhouu+\rhotu\right)d\areanu,
\end{split}
\end{equation}
which results from computing the average over $\bodyt$ in two separate steps: first averaging over the angle $\Angle$ in \eqref{eq:mdm} which ranges in $[0,2\pi]$ and then averaging formally over $\normal$, meant as the outward unit normal to $\bodyt$, which ranges over $\sphere$. If the former average is taken over the process in which, with $\normal$ and $\m_1$ fixed, $\m_2$ is seen to describe a cone around $\normal$ (see Fig.~\ref{fig:conical}), the latter is nothing but the average over the independent process in which all different points of $\partial\bodyt$ come to be associated with one and the same fixed normal $\normal$. As in \eqref{eq:V_bodysum}, also in \eqref{eq:P_1_V_bodypair} $\rhouu$ and $\rhotu$ denote the principal radii of curvature of $\partial\bodyu$ and $\rhout$ and $\rhott$ denote the principal radii of curvature of $\partial\bodyt$; correspondingly, $\Ku=(\rhouu\rhotu)^{-1}$ and $\Kt=(\rhout\rhott)^{-1}$ are the Gaussian curvatures of $\partial\bodyu$ and $\partial\bodyt$ and $\ru$ and $\rt$ are the \emph{radial mappings} of $\bodyu$ and $\bodyt$ (see Appendix~\ref{sec:essentials} for more details).

Now, with the aid of the theory recalled in Appendix~\ref{sec:essentials}, we compute the new shape functionals featuring in \eqref{eq:P_1_V_bodypair}. It readily follows from \eqref{eq:surface_dilation_ratio} that for any body $\body\in\convp$
\begin{equation}\label{eq:B_1_shape_functional_one}
\int_\sphere\frac{\m\cdot\normal}{K}d\areanu=\int_\boundary\mdn\, d\arean=\int_\body\diver\m\, dv=0,
\end{equation}
where use has also been made of the classical divergence theorem (and the fact that $\m$ can be extended to the whole space as a uniform field). Likewise, \eqref{eq:surface_divergence_radial} and \eqref{eq:surface_dilation_ratio} imply that
\begin{equation}\label{eq:B_1_shape_functional_two}
\begin{split}
\int_\sphere(\m\cdot\normal)(\rho_1+\rho_2)d\areanu&=\int_\sphere(\m\cdot\normal)\frac1K\divs\n\, d\areanu\\
=\int_\boundary(\m\cdot\n)\divs\n\, d\arean&=\int_\boundary\divs\m\,d\arean=0,
\end{split}
\end{equation}
where use has also been made of the surface divergence theorem recalled in \eqref{eq:surface_divergence_theorem}. Combining \eqref{eq:B_1_shape_functional_one} and \eqref{eq:B_1_shape_functional_two}, we obtain from \eqref{eq:P_1_V_bodypair} that $\ave{P_1 V}\bodypair$ vanishes identically for all $\bodyu$ and $\bodyt$, and so, by \eqref{eq:B_n_averages_central},
\begin{equation}\label{eq:B_1_vanishes}
B_1=-3\ave{P_1V}[\bodyu,\bodyt^\ast]\equiv0.
\end{equation}

Equation \eqref{eq:B_1_vanishes} says that for cylindrically symmetric bodies, $\bodyu$ and $\bodyt$, the excluded volume $\evou$ in \eqref{eq:V_e_expansion} does not contain any dipolar contribution, no matter how tethered $\bodyu$ and $\bodyt$ can be, suggesting that \emph{no} shape dipole can be associated with them. It was already argued in \cite{piastra:octupolar} that a shape dipole cannot be unambiguously assigned to a body $\body$. Equation \eqref{eq:B_1_vanishes} shows that no matter how we endeavor to assign a shape dipole to $\body$ it plays no role in the hard-particle interactions governed by the excluded volume. Of course, polarity effects are also expected to be seen in these interactions. For example, it was proved in \cite{palffy:minimum} that the excluded volume of two congruent cylindrically symmetric convex bodies is minimized when the bodies are in the \emph{antiparallel} configuration, where $\m_2=-\m_1$. Such polar effects, however, cannot involve shape dipoles: as shown in \cite{piastra:octupolar}, they start being manifested through the shape \emph{octupole} that features in \eqref{eq:V_e_expansion} through the coefficient $B_3$. This and all higher order Legendre coefficients will be computed in the following section.

\section{Extended Minkowski functionals}\label{sec:Extended_Minkowski_functionals}
Computing the anisotropic volume averages $\ave{P_nV}\bodypair$ for $n\geqq2$ is technically more complicated than computing $\ave{P_1V}\bodypair$, although conceptually this task is not much different from that just accomplished in the preceding section. As shown in Appendix~\ref{sec:appendix_volume_averages}, this computation led quite naturally to the introduction of a number of shape functionals that extend the classical Minkowski functionals $M$ and $S$. They are defined for all $n\geqq2$ as follows:
\begin{subequations}\label{eq:M_and_S_functionals_definitions}
\begin{align}
M_n[\body]&:=\int_{\boundary}P_n(\mdn)Hd\arean,\label{eq:M_definition}\\
M_n'[\body]&:=\int_{\boundary}(\n\cdot\x) P_n(\mdn)Kd\arean, \label{eq:M_prime_definition}\\
M_n''[\body]&:=\int_{\boundary}[1-(\mdn)^2]\frax12(\sigma_1-\sigma_2)P_{n-2}^{(2,2)}(\mdn)d\arean, \label{eq:M_double_prime_definition}\\
S_n[\body]&:=\int_{\boundary}P_n(\mdn)d\arean, \label{eq:S_definition}\\
S_n'[\body]&:=\int_{\boundary}(\n\cdot\x) P_n(\mdn)Hd\arean, \label{eq:S_prime_definition}\\
S_n''[\body]&:=\int_{\boundary}(\n\cdot\x)[1-(\mdn)^2]\frax12(\sigma_1-\sigma_2)P_{n-2}^{(2,2)}(\mdn)d\arean. \label{eq:S_double_prime}
\end{align}
\end{subequations}
We shall often refer to them as the \emph{extended} Minkowski functionals.\footnote{More shortly, also as the extended $M$ and $S$ functionals.} They give $\ave{P_nV}\bodypair$ the following concise, explicit representation:
\begin{equation}\label{eq:anisotropic_volume_average_representation}
\begin{split}
\ave{P_nV}\bodypair&=
\frac{1}{12\pi}\left(M_n'[\body_1]S_n[\body_2]+M_n'[\body_2]S_n[\body_1]\right)
+\frac{1}{6\pi}\left(M_n[\body_1]S_n'[\body_2]+M_n[\body_2]S_n'[\body_1]\right)\\
&-\frac{1}{6\pi}\frac{(n-2)!(n+2)!}{(4n!)^2} \left(M_n''[\body_1]S_n''[\body_2]+M_n''[\body_2]S_n''[\body_1]\right).
\end{split}
\end{equation}

Strictly speaking, in Appendix~\ref{sec:appendix_volume_averages} we arrived at \eqref{eq:M_and_S_functionals_definitions} through the representation via radial mapping of the convex bodies in the special class $\convp$. However, the extended Minkowski functionals can also be extended by continuity to the whole of $\conv$. Moreover, as clearly shown by \eqref{eq:M_and_S_functionals_definitions}, their definition actually applies to any cylindrically symmetric body, be it convex or not. The extended $M$ and $S$ functionals are invariant under rotations. Their behavior under translations is further discussed in Appendix~\ref{sec:invariance_translations}.

Since the extended Minkowski functionals for a body $\body$ are invariant under central inversion of $\body$ (see Appendix~\ref{sec:essentials}), it follows from \eqref{eq:anisotropic_volume_average_representation} that $\ave{P_nV}[\bodyu,\bodyt^\ast]=\ave{P_nV}\bodypair$, and so equation \eqref{eq:B_n_averages_central} becomes \begin{equation}\label{eq:B_n_averages_final}
B_n=(2n+1)(-1)^n\ave{P_nV}\bodypair,
\end{equation}
which by \eqref{eq:anisotropic_volume_average_representation} expresses the Legendre coefficients of $\evou$ in \eqref{eq:V_e_expansion} in terms of shape functionals evaluated on the individual bodies $\bodyu$ and $\bodyt$. Formula \eqref{eq:B_n_averages_final} will be applied in the two following sections to special classes of bodies, namely, circular cones and ellipsoids of revolution.

As shown in Appendix~\ref{sec:reduction_formulae}, the functionals $M''_n$, $M'_n$, and $M_n$ are not independent of one another. If $M''_n$ is certainly related to $M_n$ through
\begin{equation}\label{eq:M_double_prime_M}
M''_n[\body]=\frac{4n}{n+2}M_n[\body]\quad\forall\ n\geqq2,
\end{equation}
for all cylindrically symmetric convex bodies $\body$, we expect the relation
\begin{equation}\label{eq:M_prime_M}
M'_n[\body]=-\frac{2}{(n-1)(n+2)}M_n[\body]
\end{equation}
to be valid at least for both classes of bodies studied in detail in this paper, having checked it by direct inspection for a large number of indices.\footnote{Of course, we are aware that this can by no means be considered as a proof of \eqref{eq:M_prime_M}, which remains for us a conjecture, though with a high likelihood of being true.} Whenever \eqref{eq:M_prime_M} applies, the anisotropic volume averages $\ave{P_nV}\bodypair$ in \eqref{eq:anisotropic_volume_average_representation} take on a much simpler form,
\begin{equation}\label{eq:eq:anisotropic_volume_average_representation_simplified}
\ave{P_nV}\bodypair=\frac{1}{6\pi}\left(M_n[\body_1]A_n[\body_2]+M_n[\body_2]A_n[\body_1]\right),
\end{equation}
where
\begin{equation}\label{eq:A_functional}
A_n[\body]:=S'_n[\body]-\frac{1}{(n-1)(n+2)}S_n[\body]-\frac{n+1}{4(n-1)}S''_n[\body],\quad\forall\ n\geqq2.
\end{equation}
In particular, for two congruent bodies, $\body_1\thicksim\body_2\thicksim\body$,\footnote{Meaning that $\body_1$ and $\body_2$ are images of $\body$ under the action of the full orthogonal group $O(3)$.} by \eqref{eq:B_n_averages_final}, $B_n$ can be given the following \emph{factorized} expression,
\begin{equation}\label{eq:B_n_multiplicative}
B_n=\frac{(2n+1)(-1)^n}{3\pi}M_n[\body]A_n[\body],
\end{equation}
which we shall assume to be valid in the following (and will be very convenient in our development below).\footnote{In the language of \cite{rosenfeld:desnity} and \cite{hansen-goos:fundamental}, once combined with \eqref{eq:excluded_volume_series}, \eqref{eq:B_n_multiplicative} would be called a \emph{convolution decomposition} (or simply a \emph{deconvolution}) of the excluded volume.}

\section{Circular cones}\label{sec:cones}
\begin{figure}[h]
\centering
\includegraphics[width=0.25\linewidth]{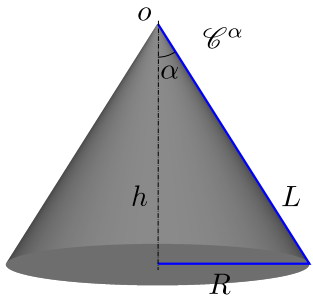}
\caption{(Color online) A circular cone with vertex in the origin $o$, semi-amplitude $\alpha$, radius $R$, height $h$, and slant height $L$.}
\label{fig:cone_text}
\end{figure}

We denote by $\conea$ a circular cone with semi-amplitude $\alpha\in[0,\frac\pi2]$, radius $R$, and height $h$, both related through \eqref{eq:cone_R_and_h} to the slant height $L$ (see Fig.~\ref{fig:cone_text}).
It is a simple matter to show that the classical Minkowski functionals for $\conea$ take the explicit forms (see also (A61) and (A62) of \cite{piastra:octupolar}),
\begin{subequations}\label{eq:cone_classical_Minkowski_functionals}
\begin{align}
M[\conea]&=\pi L\left[\cos\alpha+\left(\frac\pi2+\alpha\right)\sin\alpha\right],\label{eq:cone_M}
\\
S[\conea]&=\pi L^2\sin\alpha(1+\sin\alpha),\label{eq:cone_S}
\\
V[\conea]&=\frac13\pi L^3\cos\alpha\sin^2\alpha.\label{eq:cone_V}
\end{align}
\end{subequations}

\begin{figure}[h]
  \centering
  \subfigure[]{\includegraphics[width=.33\linewidth]{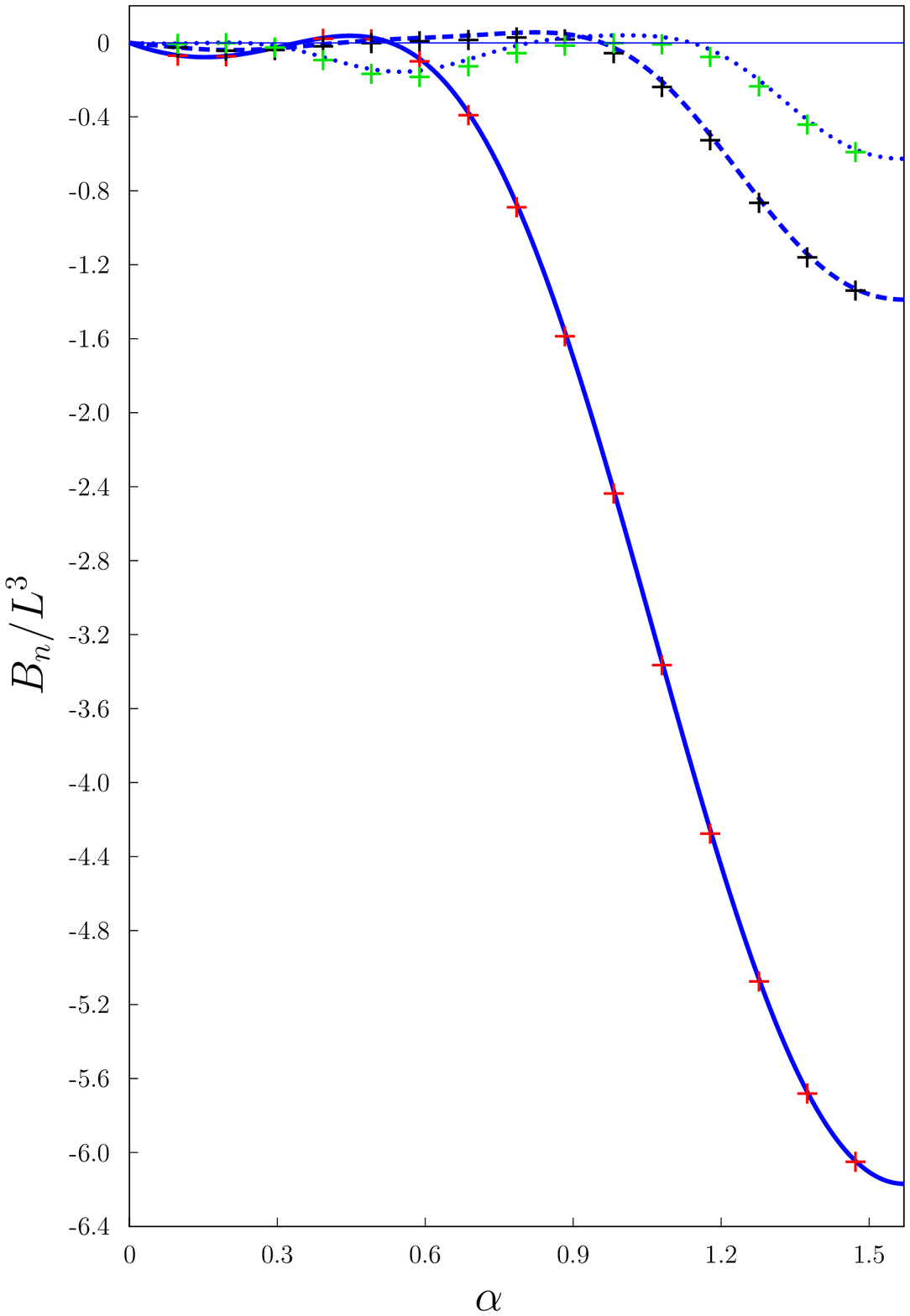}}
  \hspace{.02\linewidth}
  \subfigure[]{\includegraphics[width=.33\linewidth]{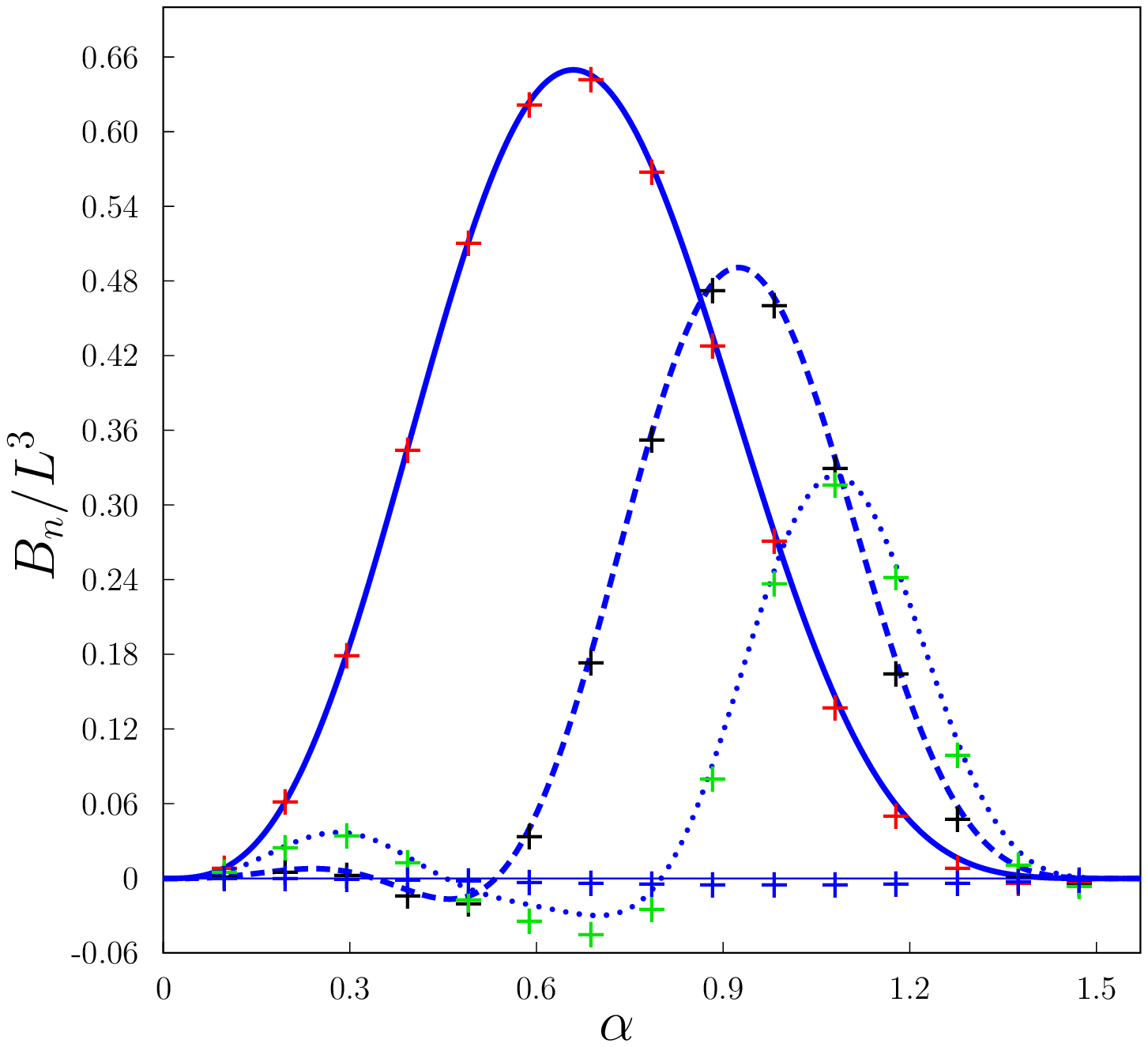}}
\caption{(Color online) (a) For two congruent circular cones, $\coneu$ and $\conet$, with slant height $L$ and semi-amplitude $\alpha$, the graphs of $B_n$ scaled to $L^3$ are plotted against $0\leqq\alpha\leqq\frac\pi2$ for $n=2$ (solid line), $n=4$ (dashed line), and $n=6$ (dotted line), according to \eqref{eq:cone_B_n}.
(b) For the same cones, $\coneu$ and $\conet$, the graphs of $B_n$ scaled to $L^3$ are plotted against $0\leqq\alpha\leqq\frac\pi2$ for $n=1$ (thin solid line), $n=3$ (solid line), $n=5$ (dashed line), and $n=7$ (dotted line).
In both panels, crosses represent the values computed numerically on the shape of the excluded body $\exc{\coneu}{\conet}$ reconstructed with the algorithm recalled in Appendix~\ref{sec:algorithm}.}
\label{fig:cone_B_n}
\end{figure}

\begin{figure}[h]
  \centering
  \subfigure[]{\includegraphics[width=.33\linewidth]{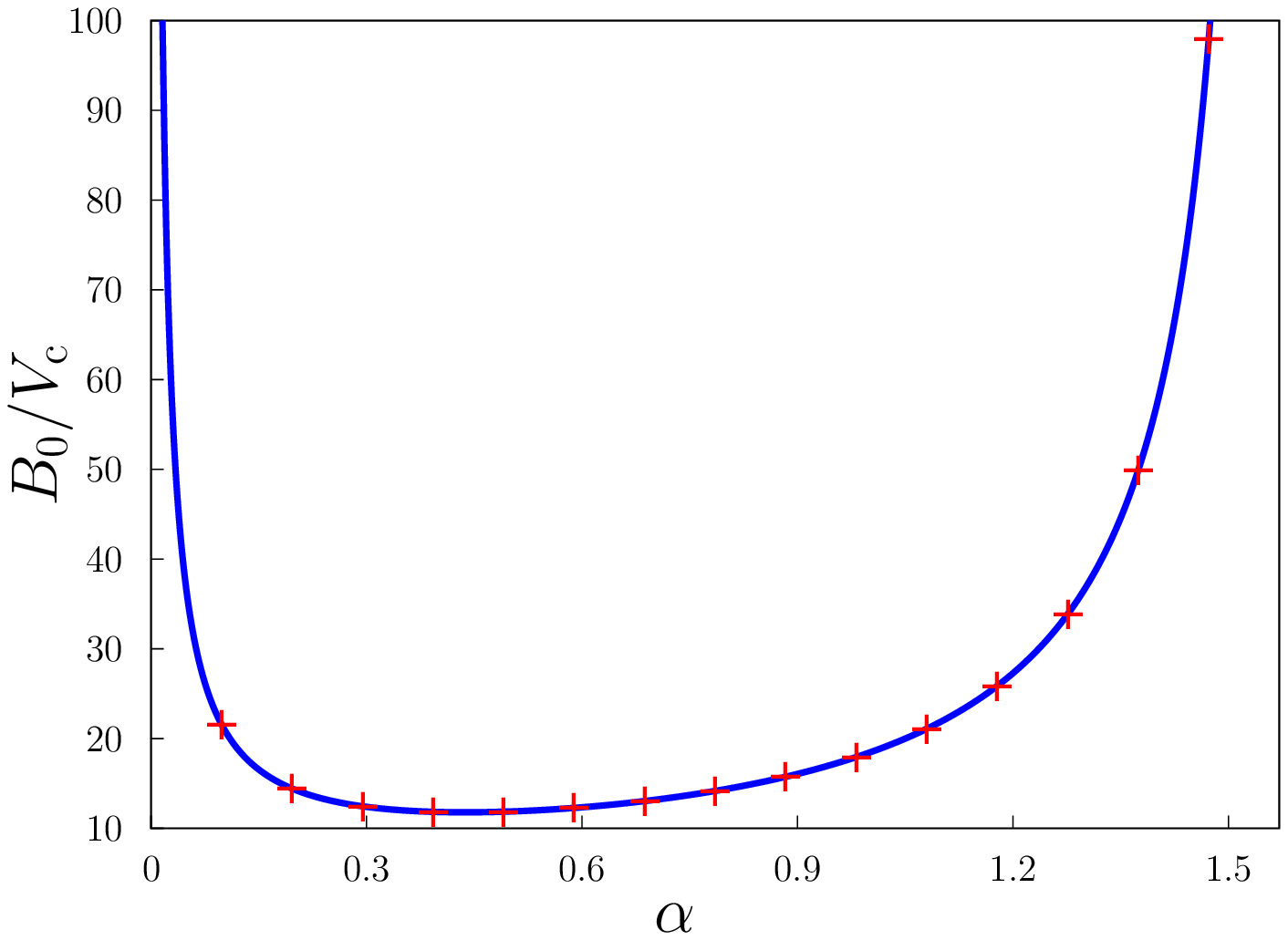}}
  \hspace{.02\linewidth}
  \subfigure[]{\includegraphics[width=.33\linewidth]{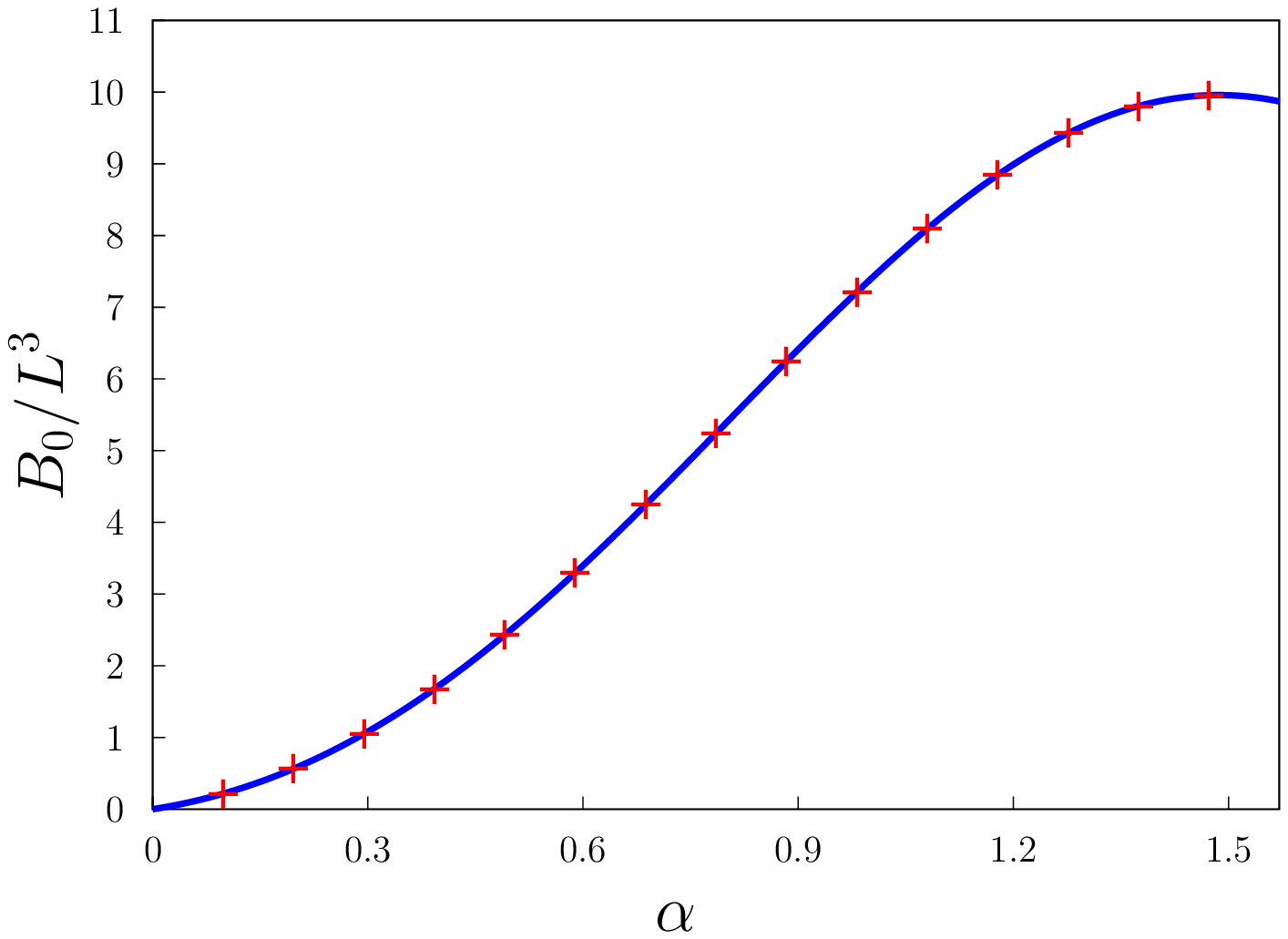}}
\caption{(Color online) (a) The isotropic average $B_0$ as in \eqref{eq:B_0} and \eqref{eq:isotropic_volume_average} normalized to the cone's volume $\Vc$ delivered by \eqref{eq:cone_V}; it attains its minimum at $\alpha\doteq0.14\pi$.
(b) $B_0$ normalized to $L^3$ like all other coefficients $B_n$'s shown in Fig.~\ref{fig:cone_B_n}; it attains its maximum at $\alpha\doteq0.47\pi$.
In both panels, crosses represent the volumes computed numerically to benchmark the shape-reconstruction algorithm described in Appendix~\ref{sec:algorithm}.}
\label{fig:B_Zero}
\end{figure}

As follows easily from \eqref{eq:cone_curvatures}, the Gaussian curvature $K$ vanishes identically on all smooth components of $\partial\conea$. Moreover, the contribution of the vertex $o$ to all the integrals in \eqref{eq:M_and_S_functionals_definitions} vanishes, as can be seen by replacing $o$ with a fitting spherical cap of radius $\ve$ (whose area surface scales like $\ve^2$) and then taking the limit as $\ve\to0^+$, in complete analogy to the method used in Appendix~\ref{sec:ridge_M_and_S} to compute the extended Minkowski functionals on a circular ridge $\ridge$. The formulae \eqref{eq:ridge_M_and_S_functionals} obtained there for a $\ridge$ can be directly applied here to the rim of the cone's base by simply setting $\theta_1=\frac\pi2-\alpha$ and $\theta_2=\pi$. Use of \eqref{eq:cone_R_and_h} finally leads us to
\begin{subequations}\label{eq:cone_M_S_functionals}
\begin{align}
M_n[\conea]&=\pi L\left(P_n(\sin\alpha)\cos\alpha +\sin\alpha\iap P_n(\cos\vt)d\vt\right), \label{eq:cone_M_n}
\\
M_n'[\conea]&=-2\pi L\iap\cos(\vt+\alpha)P_n(\cos\vt)\sin\vt d\vt, \label{eq:cone_M_n_p}
\\
M_n''[\conea]&=-\pi L\left(\Ja(\sin\alpha)\cos^3\alpha-\sin\alpha\iap\Ja(\cos\vt)\sin^2\vt d\vt\right), \label{eq:cone_M_n_p_p}
\\
S_n[\conea]&=\pi L^2\sin\alpha\left[P_n(\sin\alpha)+(-1)^n\sin\alpha\right], \label{eq:cone_S_n}
\\
S_n'[\conea]&=-\pi L^2\sin\alpha\iap\cos(\vt+\alpha)P_n(\cos\vt)d\vt, \label{eq:cone_S_n_p}
\\
S_n''[\conea]&=-\pi L^2\sin\alpha\iap\cos(\vt+\alpha)\Ja(\cos\vt)\sin^2\vt d\vt, \label{eq:cone_S_n_p_p}
\end{align}
\end{subequations}
for all $n\geqq2$. Inserting \eqref{eq:cone_M_S_functionals} in \eqref{eq:B_n_averages_final}, we obtain explicit, analytic formulae for the Legendre coefficients $B_n$ of the excluded volume of two congruent circular cones, $\coneu$ and $\conet$, which for completeness are recorded in \eqref{eq:cone_B_n} for the first seven indices $n\geqq1$. They are plotted in Fig.~\ref{fig:cone_B_n}
as functions of $\alpha$. Inserting \eqref{eq:cone_classical_Minkowski_functionals} in \eqref{eq:isotropic_volume_average}, we also obtain the isotropic average $B_0$ in \eqref{eq:B_0}, which is plotted in Fig.~\ref{fig:B_Zero} with two possible normalizations, relative to the volume $\Vc$ of each cone delivered by \eqref{eq:cone_V} in Fig.~\ref{fig:B_Zero}(a), and relative to $L^3$ in Fig.~\ref{fig:B_Zero}(b).
The even-indexed coefficients $B_n$'s are mostly negative, indicating by \eqref{eq:P_n_properties} a tendency for the corresponding terms in the sum \eqref{eq:V_e_expansion} to minimize $\evou$ for either $\vt=0$ or $\vt=\pi$, irrespectively. On the contrary, the odd-indexed coefficients are mostly positive (apart from $B_3$ which is never negative), indicating a tendency for the corresponding terms in \eqref{eq:V_e_expansion} to minimize $\evou$ for $\vt=\pi$, that is, when the cones $\coneu$ and $\conet$ are in the \emph{antiparallel} configuration, with $\m_2=-\m_1$. This suggests that the excluded volume of two congruent circular cones is minimized in the antiparallel configuration, as shown by direct computation in \cite{piastra:octupolar} in accord with the general minimum property established more recently in \cite{palffy:minimum}.

The crosses superimposed to the graphs in Fig.~\ref{fig:cone_B_n} represent the values of $B_n$ extracted numerically from the volume of the excluded body $\exc{\coneu}{\conet}$, the region in space that cone $\conet$ cannot access by the presence of cone $\coneu$. Determining $\exc{\coneu}{\conet}$ is indeed necessary for a direct determination of $\evo{\coneu}{\conet}$, as the general proper geometric definition of the excluded volume of bodies $\bodyu$ and $\bodyt$ is precisely the volume of the excluded body $\excp$, $\evop:=V[\excp]$ (see also \cite{piastra:octupolar}). Here $\exc{\coneu}{\conet}$ was obtained from the shape-reconstruction algorithm outlined in Appendix~\ref{sec:algorithm}. Our strategy was completely different from that adopted so far in this paper. For a given $\alpha$, we reconstructed $\exc{\coneu}{\conet}$ for a number of values of the angle $\vt$ made by the cones' axes $\m_1$ and $\m_2$; we computed numerically the excluded volume $\evou$ as a function of $\vt$ by applying \eqref{eq:V_functional} to a triangulation of $\partial\exc{\coneu}{\conet}$ and we extracted from this function the coefficients $B_n$ through \eqref{eq:B_n_definitions}. To what extent the two methods agree, thus granting support to each other, is left to the reader to judge from Fig.~\ref{fig:cone_B_n}. Quantitative details about both the shape-reconstruction algorithm employed here (including its adaptation to the specific case of cones, which with their sharp edge and pointed vertex required special attention) and the way the coefficients $B_n$ were computed can be found in Appendix~\ref{sec:algorithm} below.

Figure~\ref{fig:excluded_volume_cone} shows three graphs representing the excluded volume $\evou$ of
\begin{figure}[h]
\centering
\includegraphics[width=0.4\linewidth]{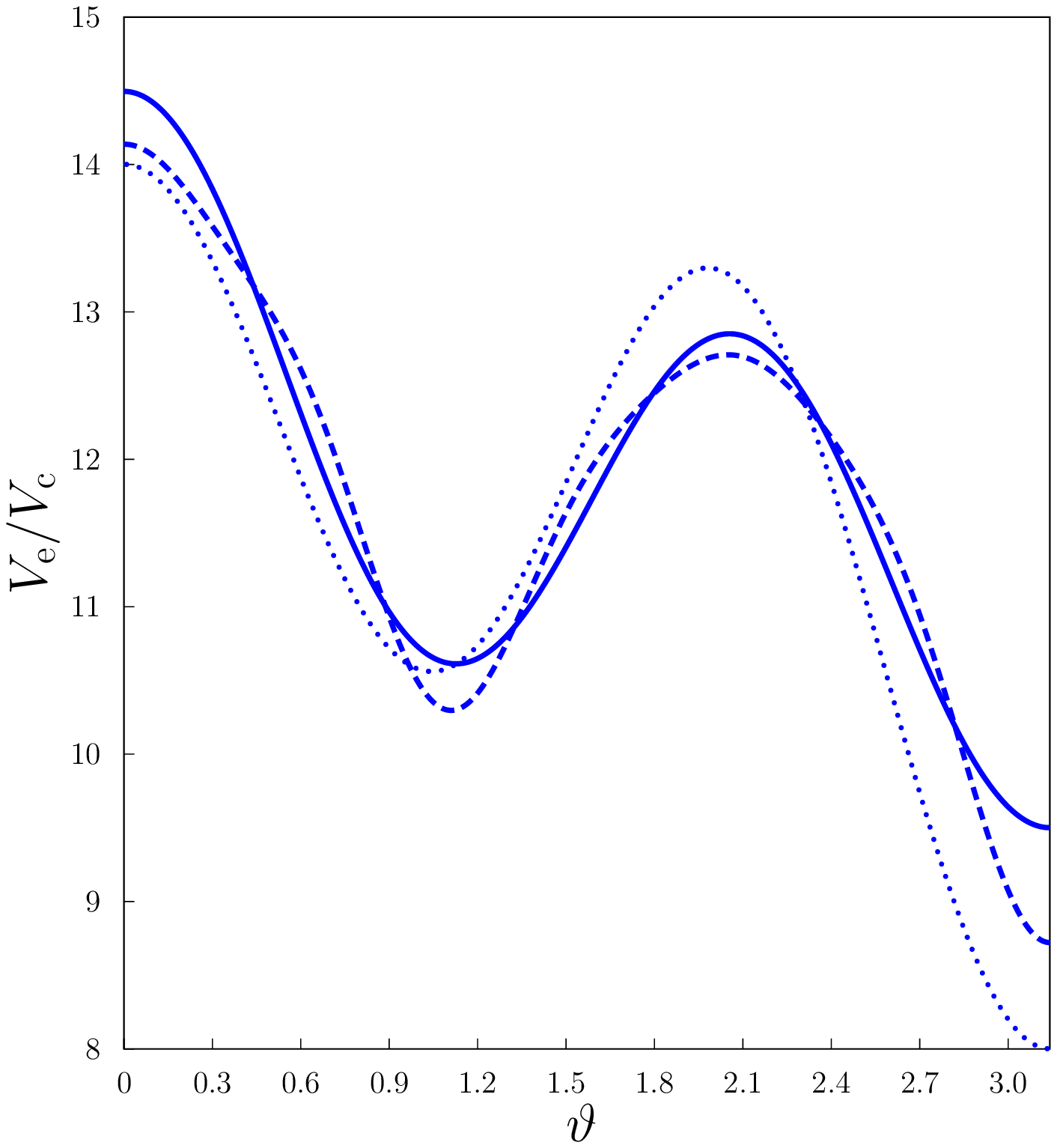}
\caption{Excluded volume $\evou$ of two congruent circular cones, $\coneu$ and $\conet$, with slant height $L$ and semi-amplitude $\alpha_0\doteq0.14\,\pi$ corresponding to the minimum value of the scaled average $\Va/\Vc$, where $\Vc$ is the volume of each cone. Two graphs, plotted against the angle $0\leqq\vt\leqq\pi$ made by the cones' axes, are delivered by \eqref{eq:V_e_expansion} truncated at $n=3$ (solid line) and $n=9$ (dashed line). The third graph (dotted line) represents the \emph{octupolar approximation} proposed in \cite{piastra:octupolar}, which interpolates the excluded volumes of parallel ($\vt=0$) and antiparallel ($\vt=\pi$) configurations.}
\label{fig:excluded_volume_cone}
\end{figure}
$\coneu$ and $\conet$ scaled to their common volume $\Vc$ (given by \eqref{eq:cone_V}) as a function of the angle $\vt$ between their axes. The semi-amplitude $\alpha$ of both cones is taken here to be $\alpha_0\doteq0.14\,\pi$, for which, as shown in Fig.~\ref{fig:B_Zero}, the isotropic average $\Va$ scaled to $\Vc$ takes on its minimum value. The graphs in Fig.~\ref{fig:excluded_volume_cone} correspond to the function in \eqref{eq:B_n_averages} truncated at $n=3$ and $n=9$; they are both contrasted against the \emph{octupolar approximation}, which in \cite{piastra:octupolar} was shown to be rather accurate. While, by construction, the latter takes on the exact values of $\evou$ at both $\vt=0$ (parallel cones) and $\vt=\pi$ (antiparallel cones), which are $14\Vc$ and $8\Vc$, respectively, both truncated expansions do not. Actually, as expected,\footnote{Since the expansion in \eqref{eq:V_e_expansion} is an approximation in the $L^2$-norm, and not pointwise.} the convergence of the series in \eqref{eq:V_e_expansion} at these points is rather slow: for example, a computation with $61$ terms was required to obtain
\begin{equation}\label{eq:V_e_computation}
\frac\evou\Vc\doteq14.01\quad\text{and}\quad\frac\evou\Vc\doteq8.153,
\end{equation}
at $\vt=0$ and $\vt=\pi$, respectively. Thus, if for cones the explicit octupolar approximation of the excluded volume could still be a good choice, for other cylindrically symmetric convex bodies, the general method proposed in this paper might be even a better choice.

\section{Spheroids}\label{sec:spheroids}
\begin{figure}[h]
  \centering
  \subfigure[]{\includegraphics[width=.25\linewidth]{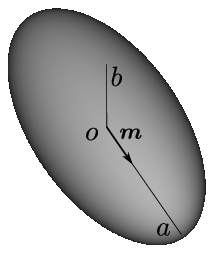}}
  \hspace{.1\linewidth}
  \subfigure[]{\includegraphics[width=.25\linewidth]{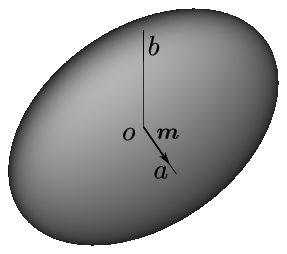}}
\caption{A spheroid is an ellipsoid of revolution. The symmetry axis is here denoted by $\m$;  $a$ and $b$ are the ellipsoid's semi-axes, in the direction of  $\m$ and in the direction orthogonal to $\m$, respectively. This spheroid is said to be prolate (a) if the aspect ratio $\eta:=b/a$ is less than unity; it is said to be oblate (b) if $\eta>1$.}
\label{fig:spheroid}
\end{figure}

Spheroids are cylindrically symmetric ellipsoids (see Fig.~\ref{fig:spheroid}).
Letting $a$ be the semi-axis of the spheroid along the symmetry axis $\m$ and $b$ the semi-axis orthogonal to $\m$, we set
\begin{equation}\label{eq:eta_definition}
\eta:=\frac{b}{a}
\end{equation}
and call it the \emph{aspect ratio} of the body. A spheroid with aspect ratio $\eta$ will denoted $\sphero$ for short; it is \emph{prolate} along the symmetry axis for $0<\eta<1$ and \emph{oblate} for $\eta>1$. Clearly, for $\eta=1$, $\sphero$ reduces to a sphere of radius $a$. Making use of the explicit representation of $\sphero$ described in Appendix~\ref{sec:appendix_spheroids}, we may write the classical Minkowski functionals as
\begin{subequations}\label{eq:spheroids_calssical_M_functionals}
\begin{align}
M[\sphero]=&\pi a\left(2+\ione\frac{\eta^2}{\denu}\right)du,\\
S[\sphero]=&2\pi a^2\eta\ione\sqrt{\denu}du,\\
V[\sphero]=&\frac{4\pi}{3}a^3\eta^2=:\Vs,
\end{align}
\end{subequations}
where $\Vs$ has been introduced as a shorthand for the spheroid's volume. It is often useful to describe how far $\sphero$ is from a sphere by defining its \emph{eccentricity} $\ecc$ as
\begin{equation}\label{eq:eccentricity}
\ecc:=
\begin{cases}
\sqrt{1-\eta^2}&\text{for}\quad 0\leqq\eta\leqq1,\\
\sqrt{1-\frac{1}{\eta^2}}&\text{for}\quad \eta\geqq1.
\end{cases}
\end{equation}
A relevant property of $\ecc$ is that the transformation $\eta\mapsto1/\eta$, which represents the \emph{reciprocal inversion} of $\sphero$ relative to its center, changes a prolate spheroid into an oblate spheroid with the same eccentricity. Though neither of the functionals \eqref{eq:spheroids_calssical_M_functionals} is invariant under reciprocal inversion of $\sphero$, all the ratios
\begin{equation}\label{eq:f_n}
f_n:=\frac{B_n}{\Vs}
\end{equation}
are expected to be so, as such a property should indeed be enjoyed by the ratio of the excluded volume $\evo{\sphero_1}{\sphero_2}$ of two congruent spheroids, $\sphero_1$ and $\sphero_2$, to their common volume.\footnote{Tjipto-Margo and Evans~\cite{tjipto-margo:onsager} credit  Ho{\l}yst and Poniewierski~\cite{holyst:study} for having proved analytically such an invariance property for uniaxial ellipsoids, but we were unable to retrace a convincing analytic proof in \cite{holyst:study}. Similarly, the extension of this property to biaxial ellipsoids was established numerically in \cite{tjipto-margo:onsager} by a Monte Carlo method. Contrariwise, the explicit analytic formula obtained by Mulder~\cite{mulder:solution,mulder:isotropic} for the excluded volume of spheroplatelets allows one to prove that its ratio to the individual spheroplatelet's volume is invariant under reciprocal transformation of the three unequal lengths that characterize these bodies. In any event, as shown in \cite{rigby:hard}, even for spheroids, this property does \emph{not} apply to higher-order virial coefficients.} As a consequence, all $f_n$'s should be functions of $\ecc$ alone. The expression for $f_0$ was already obtained by Isihara~\cite{isihara:determination},
\begin{equation}\label{eq:f_0}
f_0=2+\frac32\left(1+(1-\ecc^2)\frac{\arc\ecc}{\ecc} \right)\left(1+\frac{\arcsin\ecc}{\ecc\sqrt{1-\ecc^2}}\right),
\end{equation}
which is also known as the Isihara-Ogston-Winzor formula \cite{ogston:treatment,ambrosetti:percolative}.

The representation for $B_n$ in \eqref{eq:B_n_multiplicative} can appropriately be used to obtain all even-indexed functions $f_n$.\footnote{Clearly, all odd-indexed $f_n$ vanish identically since spheroids are symmetric under central inversion.} To this end, we first record the form taken on a spheroid $\sphero$ by the extended Minkowski functional (see Appendix~\ref{sec:spheroids_B_n} for more details):
\begin{subequations}\label{eq:M_and_S_spheroids}
\begin{align}
M_n[\sphero]=&\pi a\ione P_n(\xi)\frac{\eta^2[1+\denx]}{[\denx]^{\frac32}}d\xi,\label{eq:M_n_spheroids}\\
M'_n[\sphero]=&2\pi a\ione P_n(\xi)\sqrt{\denx}d\xi,\label{eq:M'_n_spheroids}\\
M''_n[\sphero]=&\pi a\ione\Ja(\xi)\frac{\eta^2(\eta^2-1)(1-\xi^2)^2}{[\denx]^{\frac32}}d\xi,\\
S_n[\sphero]=&2\pi a^2\ione P_n(\xi)\frac{\eta^4}{[\denx]^2}d\xi,\\
S'_n[\sphero]=&\pi a^2\ione P_n(\xi)\frac{\eta^2}{\denx}d\xi,\\
S''_n[\sphero]=&\pi a^2\ione \Ja(\xi)\frac{\eta^2(\eta^2-1)(1-\xi^2)^2}{\denx}d\xi.
\end{align}
\end{subequations}
For $n=2$, we obtained
\begin{equation}\label{eq:f_2}
f_2=\frac{15}{32}\frac{1}{\ecc^4}\left(\ecc^2-3+(\ecc^2+3)(1-\ecc^2)\frac{\arc\ecc}{\ecc}\right) \left(3-2\ecc^2+\frac{4\ecc^2-3}{\ecc\sqrt{1-\ecc^2}}\arcsin\ecc\right).
\end{equation}
It is worth noting that this formula coincides with that found by Isihara~\cite{isihara:theory} for oblate spheroids ($\eta\geqq1$).\footnote{See equations (48)--(50) of \cite{isihara:theory}.} For prolate spheroids, Isihara~\cite{isihara:theory} records a result which does not comply with the requirement of $f_2$ being invariant under the transformation $\eta\mapsto1/\eta$. For this reason, we deem it to be incorrect. This should not indeed surprise us, as Isihara's method delivers $B_n$ in the form of two separate power series in $\ecc$, one for the prolate case and the other for the oblate case,\footnote{See equations (29) and (47) of \cite{isihara:theory}.} which need then be resummed.\footnote{A similar discrepancy for $f_4$ is pointed out in Appendix~\ref{sec:spheroids_B_n} below.}

\begin{figure}[h]
\centering
\includegraphics[width=0.4\linewidth]{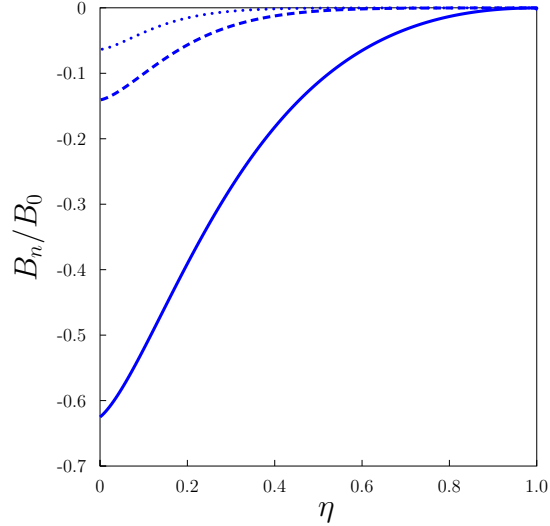}
\caption{The plots of $B_2$ (solid line), $B_4$ (dashed line), and $B_6$ (dotted line) normalized to $B_0$ for $0\leqq\eta\leqq1$. In the limit as $\eta\to0$ (needle-shaped spheroids), the ratios shown here tend to $B_2/B_0=-5/8\doteq-0.63$, $B_4/B_0=-9/64\doteq-0.14$, and $B_6/B_0=-65/1024\doteq-0.06$.}
\label{fig:B2oB0}
\end{figure}

Explicit formulae for both $f_4$ and $f_6$ are reproduced in Appendix~\ref{sec:spheroids_B_n}; here we shall be contented with showing in Fig.~\ref{fig:B2oB0} $B_6$, $B_4$, and $B_2$ normalized to $B_0$ as functions of $\eta$ for prolate spheroids (as for oblate spheroids these ratios also remain unchanged under the transformation $\eta\mapsto1/\eta$).
The graphs plotted in Fig.~\ref{fig:B2oB0} may help deciding how many terms to retain in \eqref{eq:V_e_expansion} for any given value of $\eta$.

\begin{figure}[h]
\centering
  \subfigure[]{\includegraphics[width=0.33\linewidth]{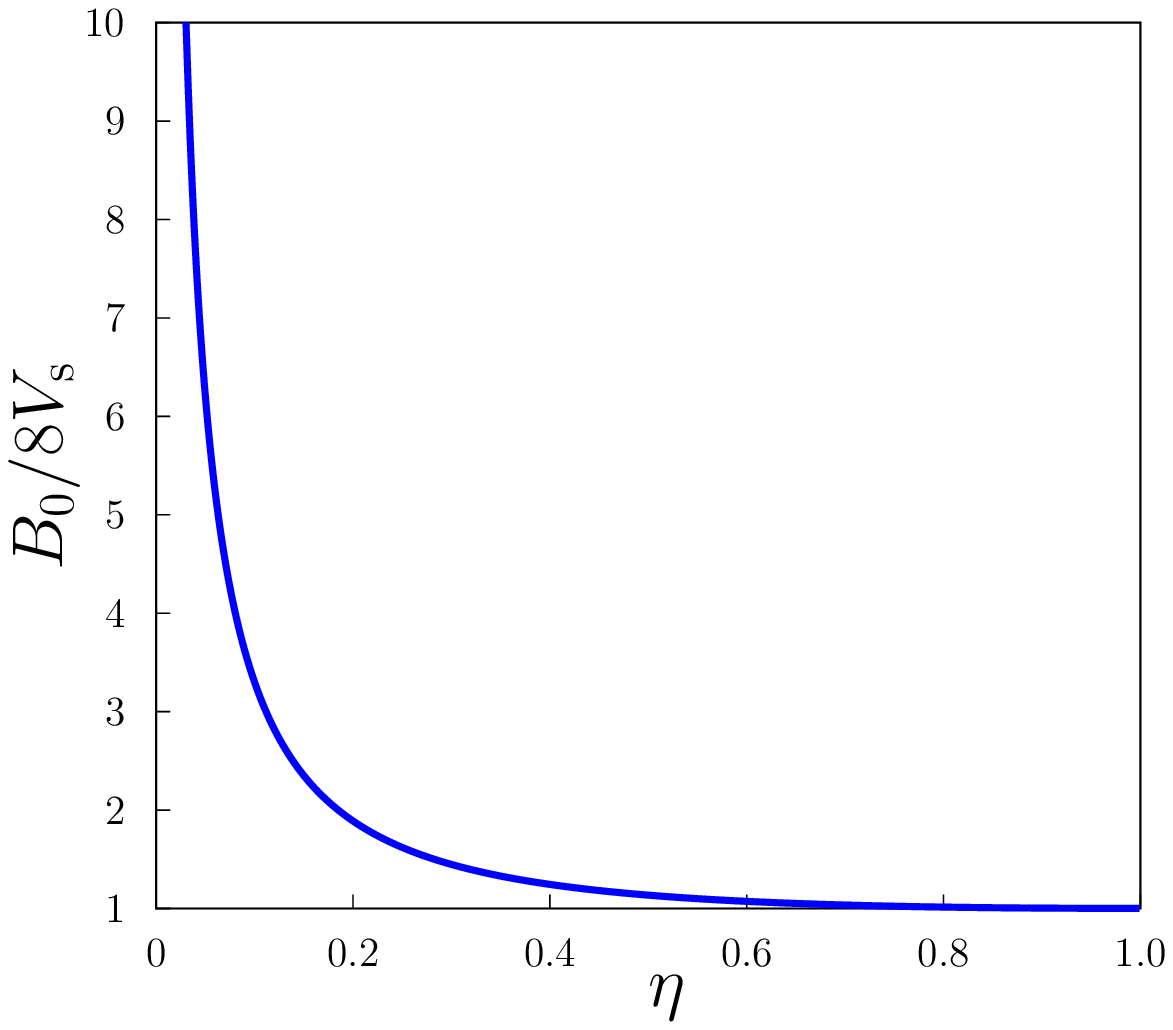}}
  \hspace{.02\linewidth}
  \subfigure[]{\includegraphics[width=0.30\linewidth]{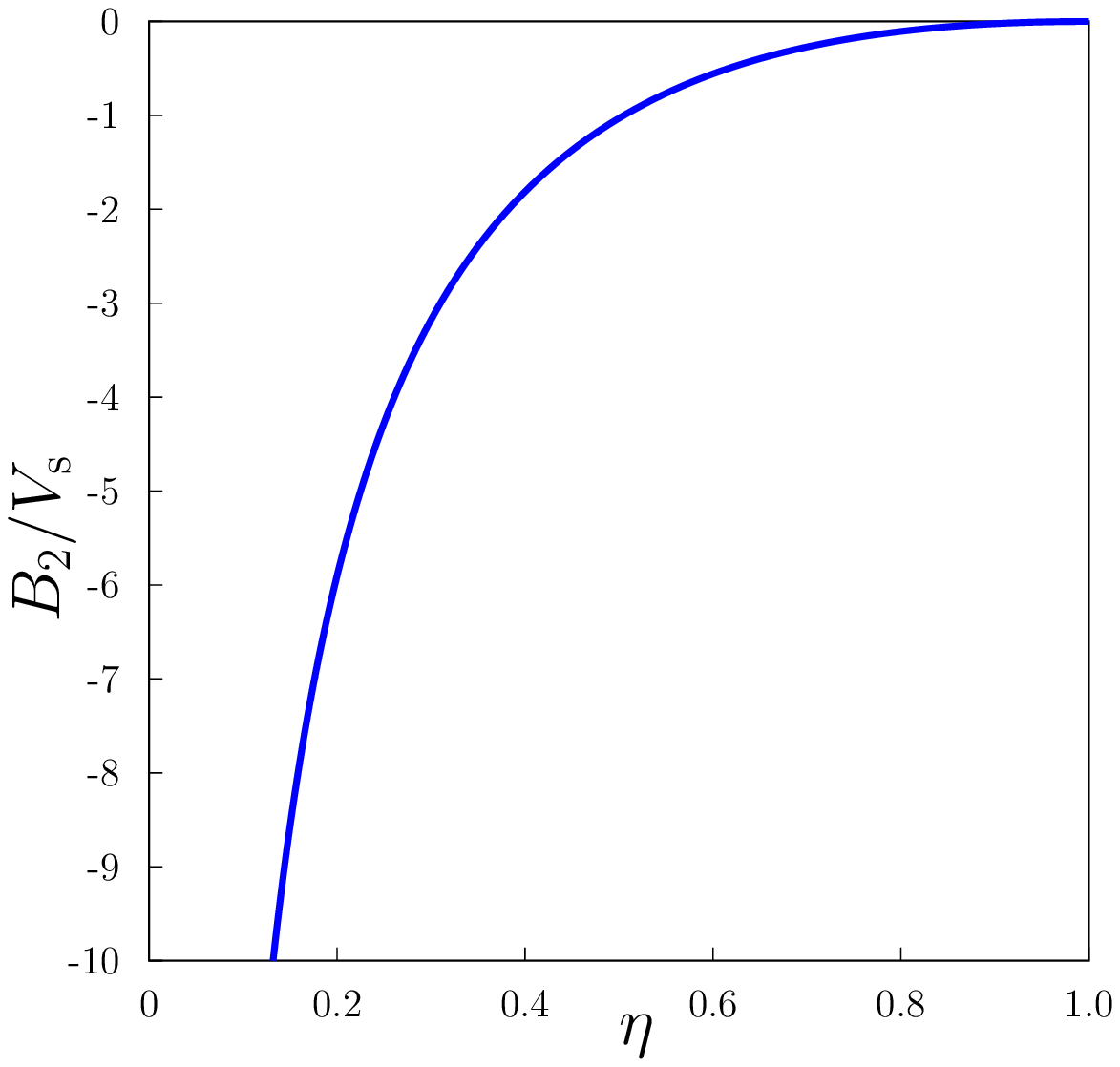}}
\caption{(Color online) (a) The plot of $B_0$ (normalized to $8\Vs$) for a prolate spheroid. It behaves like $3/32\eta$ as $\eta\to0$. (b) The plot of $B_2$ (normalized to $\Vs$) for a prolate spheroid. It behaves like $-15\pi/32\eta$ as $\eta\to0$. Both plots are easily extended to oblate spheroids by preserving their values under the transformation $\eta\mapsto1/\eta$.}
\label{fig:B0B2oV0}
\end{figure}

For completeness, we show in Fig.~\ref{fig:B0B2oV0} the graphs of the coefficients $B_0$ and $B_2$, the former of which is normalized to $8\Vs$, the minimum excluded volume of two congruent spheroids of volume $\Vs$ (attained when they are in the parallel configuration).

Over the past few decades, hard ellipsoids have been the object of many studies revisiting the classical Onsager theory of hard cylindrical rods. In all these studies, the excluded volume of ellipsoids plays by necessity a central role (occasionally, along with some higher-order virial coefficients which have also been computed).\footnote{We refer the reader to \cite{masters:virial} for a review. Other relevant information can be gathered from the works \cite{wertheim:third,wertheim:fluids_3,singh:geometry,rigby:virial,rigby:hard,baus:finite,colot:desnity}.} More recently, a formula was also obtained in \cite{ambrosetti:percolative} for the excluded volume of two congruent oblate spheroids, elaborating on the original method of Isihara~\cite{isihara:theory}. That formula\footnote{See equation (B12) of \cite{ambrosetti:percolative}.} is not directly comparable with ours, as it expresses the excluded volume as a power series of trigonometric functions of the angle between the bodies' symmetry axes, which unlike Legendre polynomials is not a system of orthogonal functions.

\section{Conclusions}\label{sec:conclusions}
The major objective of this paper was to express explicitly the excluded volume $\evop$ of two arbitrary cylindrically symmetric, convex bodies $\bodyu$ and $\bodyt$ (with symmetry axes $\m_1$ and $\m_2$), in terms of shape functionals to be evaluated separately for $\bodyu$ and $\bodyt$. We accomplished this task by relating the coefficients $B_n$ that represent $\evop$ in the basis of Legendre polynomials $P_n(\mdm)$ to certain anisotropic volume averages which, in complete analogy with the classical Minkowski formula for the isotropic average of the excluded volume, were expressed in terms of shape functionals that extend Minkowski's. As demonstrated by the examples of cones and spheroids, which we worked out in full details, the extended Minkowski functionals can be evaluated exactly. A large number of them might be required to obtain $\evop$  at a high degree of accuracy, but the proposed method provides them exactly in any desired number.

As witnessed by the case of cones, one motivation of our study was to explore the role of shape polarity in the excluded volume of tapered bodies. It has already been shown that when such  congruent bodies $\bodyu$ and $\bodyt$ are convex and cylindrically symmetric, $\evop$ attains its minimum in the \emph{antiparallel} configuration \cite{palffy:minimum}. Therefore, one could think of assigning a \emph{shape dipole} $\dip$ to these bodies by extracting from $\evop$ the dipolar component, $B_1\mdm$, and rewriting it formally as $\dip_1\cdot\dip_2$.\footnote{Actually, for selected $\m_1$ and $\m_2$ on the symmetry axes of the congruent bodies $\bodyu$ and $\bodyt$, one could either orient the vectors $\dip_1$ and $\dip_2$ along $\m_1$ and $\m_2$, respectively, or in the opposite directions, provided their orientations are reverted in both bodies.} Instead, we proved that $B_1\equiv0$, thus making elusive the definition of any shape dipole for a tapered, cylindrically symmetric, convex body. Clearly, the antipolar property revealed by the minimum of $\evop$ remains valid, but it can in general be read off from the coefficient $B_3$, and so properly speaking it is an \emph{octupolar} effect.

Cones indeed interested us because they are tapered, but they are not the easiest cylindrically symmetric, convex bodies for which one would compute the excluded volume. Perhaps, ellipsoids of revolution might come first in anyone's list. For this reason, we also applied our method to ellipsoids of revolution. Other methods have already been devised to compute the excluded volume of these bodies, such as the overlap criteria used in computer simulations \cite{vieillard-baron:phase,perram:statistical}, or the approximations stipulated in the Gaussian overlap model originally introduced in \cite{berne:gaussian},\footnote{An Onsager theory for hard-ellipsoids based on this approximation can be found in \cite{lee:onsager}, a paper well aware of the possible inaccuracies stemming from the hard-body modification of the simple Gaussian overlap model \cite{bhethanabotla:comparison}. See also \cite{singh:structure} for a recent review of the Gaussian overlap model for hard-ellipsoids.} but with the admirable exception of the classical factorized formulae of Isihara~\cite{isihara:theory} for the first Legendre coefficients $B_n$ and the closed form expression for the distance of closest approach for two ellipses in two space dimensions \cite{palffy:distance_2D},\footnote{Unfortunately, the extension to ellipsoids in three space dimensions of the method that was successful in two dimensions can only be performed numerically \cite{palffy:distance_3D}.} no explicit analytic representation was known for the excluded volume of ellipsoids of revolution. We hope that we have provided one, rooting on geometric grounds the multiplicative structure of Isihara's formulae and emending some of them.

Several other applications could be foreseen for our representation formula. In tune again with Onsager's paper \cite{onsager:effects}, we mention just one: the role of shape in steric interactions of filamentous viruses. This was indeed the original motivation of Onsager's work, which intended to provide a theoretical explanation for the liquid crystalline behavior of tobacco mosaic viruses, which were the first to be isolated and purified \cite{bawden:liquid}. An up-to-date review of the recent applications of Onsager's theory to viruses of various elongated shapes can be found in \cite{dogic:ordered}. We trust that our representation formula for the excluded volume could help making the role of viruses' shape more explicit.

\begin{acknowledgements}
One of us (EGV) is grateful to Peter Palffy-Muhoray for having raised the question about which would be the most appropriate definition of shape dipole for a cylindrically symmetric rigid body, which prompted the study presented here.
We are indebted to an anonymous Referee for a number of learned and constructive critiques of an earlier version of our paper, answering which has noticeably improved our work.
\end{acknowledgements}

\appendix
\section{Mathematical details}\label{sec:mathematical_details}
In this appendix we record for completeness the mathematical details needed to make our development rigorous, but which would have hampered our presentation if dispersed in the main body of the paper. We start by recalling the essentials of convex body geometry; they are extracted from the wider treatment presented in Appendix~A of \cite{piastra:octupolar}, to which the interested reader is referred for a better appreciation of the formalism adopted in this paper.

\subsection{Essentials of convex body geometry}\label{sec:essentials}
A convex body $\body$ in the three-dimensional space $\eucl$ is represented here through the \emph{radial} mapping $\normal\mapsto\radial(\normal)$, which associates to each unit vector $\normal$ in the unit sphere $\sphere$ of $\eucl$ the point on the boundary $\boundary$ of $\body$ where the outward unit normal is precisely $\normal$. Such a representation requires $\sphere$ to be mapped univocally onto $\boundary$, which is the case whenever $\body$ belongs to the class $\convp$ of convex bodies with smooth boundaries and strictly positive curvatures. Such an assumption is not a true limitation to our development, as $\convp$ is indeed dense in the whole class $\conv$ of convex bodies with respect to the Hausdorff metric. Thus, the values attained in $\conv\setminus\convp$ by a continuous functional defined in $\convp$ can be computed as limits on appropriate approximating sequences of bodies in $\convp$. This property is for example exploited in Sec.~\ref{sec:ridge_M_and_S} below to compute the contribution of a sharp ridge to the extended Minkowski functionals introduced in Sec.~\ref{sec:Extended_Minkowski_functionals}.\footnote{The very same property makes it possible to arrive at the expressions for the extended $M$ and $S$ functionals of a cone $\conea$ listed in Sec.~\ref{sec:cones}.}

\begin{figure}[h]
\centering
\includegraphics[width=0.25\linewidth]{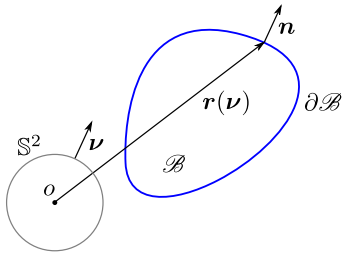}
\caption{(Color online) Sketch that describes how the radial mapping $\radial$ assigns to a
unit vector $\normal$ of $\sphere$ the translation that brings $o$ into  the
point on $\partial\body$ where $\normal$ is the unit outward normal to
$\partial\body$. The existence of such a mapping is guaranteed by the assumption
that $\body$  belongs to $\convp$.}
\label{fig:radial_mapping}
\end{figure}

\begin{figure}[h]
\centering
   \includegraphics[width=0.16\linewidth]{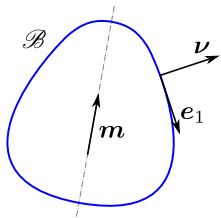}
\caption{(Color online) Cross section of a cylindrically symmetric body $\body$ through a plane containing its axis of symmetry $\m$. Both the outer unit normal $\n$ and the principal direction of curvature $\e_1$ along the local meridian are on this plane.}
\label{fig:cylindrically_symmetric_body}
\end{figure}

Figure~\ref{fig:radial_mapping} illustrates our representation of $\body$ through its radial mapping $\radial$.
It also shows that the unit outward normal $\n$ to $\boundary$, which by construction at the point $\radial(\normal)$ coincides with $\normal$, can also be regarded as a field on $\boundary$. Its surface gradient $\nablas\n$ is the \emph{curvature} tensor and can be represented as
\begin{equation}\label{eq:curvature_tensor}
\nablas\n=\sigma_1\e_1\otimes\e_1+\sigma_2\e_2\otimes\e_2,
\end{equation}
where the \emph{positive} scalars $\sigma_1$ and $\sigma_2$ are the \emph{principal curvatures} of $\boundary$, and the orthogonal unit vectors $\e_1$ and $\e_2$, both tangent to $\boundary$, designates the \emph{principal directions} of curvature. In this paper, fully devoted to cylindrically symmetric bodies, we have conventionally taken $\e_1$ along the local meridian, so that $\e_1$, $\n$, and the symmetry axis $\m$ of $\body$ are everywhere in one and the same plane (possibly varying with the point selected on $\boundary$). Figure~\ref{fig:cylindrically_symmetric_body} shows the geometric situation envisaged here.

The \emph{mean} curvature $H$ and the \emph{Gaussian} curvature $K$ are defined in terms of the principal curvatures as
\begin{equation}\label{eq:H_and_K}
H:=\frax12\left(\sigma_1+\sigma_2\right)\quad\text{and}\quad K:=\sigma_1\sigma_2.
\end{equation}
The former can also be expressed as
\begin{equation}\label{eq:H_div_s_n}
H=\tr\nablas\n=\frax12\divs\n,
\end{equation}
where $\tr$ is the trace operator and $\divs$ denotes the surface divergence. Similarly, letting $\A^\ast$ denote the \emph{adjugate} of a second-rank tensor $\A$,\footnote{$\A^\ast$ is characterized by requiring that $\A^\ast(\vu\times\vv)=\A\vu\times\A\vv$, for all vectors $\vu$ and $\vv$, where $\times$ denotes the cross product of vectors (see also Sec.~2.11 of \cite{gurtin:mechanics}).} we also have that
\begin{equation}\label{eq:curvature_tensor_adjugate}
(\nablas\n)^\ast=K\n\otimes\n=K\normal\otimes\normal.
\end{equation}

The surface gradient $\nablas\radial$ of the radial mapping $\radial$ over $\sphere$ has an expression similar to \eqref{eq:curvature_tensor},
\begin{equation}\label{eq:radial_mapping_surface_gradient}
\nablas\radial=\rho_1\e_1\otimes\e_1+\rho_2\e_2\otimes\e_2,
\end{equation}
where
\begin{equation}\label{eq:curvature_radii}
\rho_1:=\frac{1}{\sigma_1}\quad\text{and}\quad\rho_2:=\frac{1}{\sigma_2}
\end{equation}
are the principal \emph{radii} of curvature of $\boundary$. In complete analogy with \eqref{eq:curvature_tensor}, we have that
\begin{equation}\label{eq:gradient_radial_adjugate}
(\nablas\radial)^\ast=\frac1K\normal\otimes\normal,
\end{equation}
whence it follows that the surface dilation ratio induced by the mapping $\radial$ that sends $\sphere$ onto $\boundary$ is given by\footnote{See also Sec.~5.2 of \cite{gurtin:mechanics}.}
\begin{equation}\label{eq:surface_dilation_ratio}
\frac{d\arean}{d\areanu}=|(\nablas\radial)^\ast\normal|=\frac1K.
\end{equation}
Putting together \eqref{eq:radial_mapping_surface_gradient}, \eqref{eq:curvature_radii}, \eqref{eq:H_div_s_n}, and \eqref{eq:H_and_K}, we can also write
\begin{equation}\label{eq:surface_divergence_radial}
\divs\radial=\rho_1+\rho_2=\frac1K\divs\n.
\end{equation}
In the following, we shall also denote by $\x$ the position vector on $\boundary$. Formally, the fields $\normal$ and $\n$ are related through $\x$ by the relations
\begin{equation}\label{eq:nu_n_relations}
\normal=\n(\x)\quad\text{and}\quad\x=\radial(\normal).
\end{equation}

A theorem that we have often used in this paper is the surface-divergence theorem.\footnote{See also Sec.~5.2.3 of \cite{sonnet:dissipative}.} It says that
\begin{equation}\label{eq:surface_divergence_theorem}
\int_\surface\divs\vu d\arean=\int_\surface(\divs\n)\vu\cdot\n d\arean=2\int_\surface H\vu\cdot\n d\arean,
\end{equation}
for any continuously differentiable field $\vu$ defined on a \emph{closed} smooth surface $\surface$ with unit outer normal $\n$ and mean curvature $H$.

Three continuous functionals defined on the whole class $\conv$ of convex bodies were introduced by Minkowski. They are the \emph{total mean curvature} $M$, the \emph{surface area} $S$, and the \emph{volume} $V$. For a body $\body\in\convp$, they are defined and represented as follows:\footnote{See Appendix~A of \cite{piastra:octupolar}, for more details}
\begin{subequations}\label{eq:fundamental_functionals}
\begin{equation}\label{eq:M_functional}
M[\body]:=\int_{\boundary}Hd\arean=\int_{\sphere}\radial\cdot\normal d\areanu,
\end{equation}
\begin{equation}\label{eq:S_functional}
S[\body]:=\int_{\boundary}d\arean=\int_{\sphere}\frac1K d\areanu= \int_{\sphere}\normal\cdot(\nablas\radial)^\ast\normal d\areanu,
\end{equation}
\begin{equation}\label{eq:V_functional}
V[\body]:=\frac13\int_{\boundary}\n\cdot\x\, d\arean= \frac13\int_{\sphere}\radial\cdot\normal\frac1K d\areanu=\frac13\int_{\sphere}(\normal\cdot\radial)\normal\cdot(\nablas\radial)^\ast\normal d\areanu.
\end{equation}
\end{subequations}

As shown in greater details in \cite{piastra:octupolar}, one of the advantages of representing a body $\body$ in $\convp$ through its radial mapping $\radial$ is that the Minkowski sum $\bodysum$ of two bodies, $\body_1$ and $\body_2$, represented by the radial mappings $\radial_1$ and $\radial_2$, respectively, is represented by the radial mapping $\radial_{12}=\radial_1+\radial_2$. Correspondingly, the fundamental functionals in \eqref{eq:fundamental_functionals} evaluated on the Minkoski sum of two bodies, $\body_1$
and $\body_2$, of $\convp$ are delivered by\footnote{See (A25), (A43), and (A49) of \cite{piastra:octupolar}.}
\begin{subequations}\label{eq:fundamental_functionals_bodysum}
\begin{equation}\label{eq:M_bodysum}
M[\bodysum]=M[\body_1]+M[\body_2],
\end{equation}
\begin{equation}\label{eq:S_bodysum}
S[\bodysum]=S[\body_1]+S[\body_2]
+\int_{\sphere} \left[\sin^2\Angle\left(\rhouu\rhout+\rhotu\rhott \right)
+\cos^2\Angle\left(\rhouu\rhott+\rhotu\rhout\right)\right]da(\normal),
\end{equation}
\end{subequations}
\begin{equation}\label{eq:V_bodysum}
\begin{split}
V[\bodysum]&=V[\body_1]+V[\body_2]+\frac13\int_{\sphere}\left(\normal\cdot\radial_1\frac{1}{K^{(2)}} +\normal\cdot\radial_2\frac{1}{K^{(1)}}\right)da(\normal)\\
&+\frac13\int_{\sphere}(\normal\cdot\radial_1+\normal\cdot\radial_2)
\left[\sin^2\Angle\left(\rhouu\rhout+\rhotu\rhott \right)
+\cos^2\Angle\left(\rhouu\rhott+\rhotu\rhout\right)\right]da(\normal),
\end{split}
\end{equation}
where $\rhouu$ and $\rhotu$ are the principal radii of curvature of $\boundary_1$, $\rhout$ and $\rhott$ are those of $\boundary_2$, $\Angle\in[0,2\pi]$ is the angle of the rotation about $\normal$ that brings the pair of principal curvature directions $(\euu,\etu)$ of $\body_1$ into the pair of principal curvature directions  $(\eut,\ett)$ of body $\body_2$, and $K^{(1)}=(\rhouu\rhotu)^{-1}$, $K^{(2)}=(\rhout\rhott)^{-1}$ are the Gaussian curvatures of $\boundary_1$ and $\boundary_2$, respectively.

We finally remark that for a body $\body\in\convp$ represented by the radial mapping $\radial(\normal)$, the central inverse $\body^\ast$ (relative to the same origin $o$) is represented by the radial mapping $\radial^\ast$ defined by
\begin{equation}\label{eq:radial_inverse}
\radial^\ast(\normal):=-\radial(-\normal).
\end{equation}
As a result, if $\radial_1$ and $\radial_2$ are the radial mappings representing the bodies $\body_1$ and $\body_2$ in $\convp$, the body $\body_1+\body_2^\ast$, whose volume, by Mulder's identity \eqref{eq:Mulder_equality}, is the excluded volume $\evo{\body_1}{\body_2}$ of the pair $(\body_1,\body_2)$, is represented by the radial mapping
\begin{equation}\label{eq:radial_excluded_body}
\radiale(\normal):=\radial_1(\normal)-\radial_2(-\normal).
\end{equation}
It is not difficult to show with aid of \eqref{eq:radial_inverse} that the shape functionals defined in \eqref{eq:M_and_S_functionals_definitions} for $\body\in\convp$ are invariant under the transformation $\body\mapsto\body^\ast$.

Another consequence of Mulder's identity is that \eqref{eq:V_bodysum} bears a close resemblance to Wertheim's representation for Mayer's function \cite{wertheim:fluids_1}.\footnote{See, in particular, (36) of \cite{wertheim:fluids_1}, which in an incomplete form was also referred to as the \emph{convolution decomposition} of Mayer's function by Rosenfeld~\cite{rosenfeld:desnity} and later re-established in \cite{hansen-goos:fundamental} in its complete form, equivalent to Wertheim's original equation.} An important difference, however, between our method and Wertheim's is that the expansion in \eqref{eq:excluded_volume_series} with coefficients $B_n$ as in \eqref{eq:B_n_multiplicative} does not result from an expansion of the integrand in the last integral of \eqref{eq:V_bodysum}, thus avoiding the ambiguities acknowledged in \cite{wertheim:fluids_1}.\footnote{Compare, for example, (64) and (68) of \cite{wertheim:fluids_1}.}
\subsection{Anisotropic volume averages}\label{sec:appendix_volume_averages}
The anisotropic volume averages $\ave{P_nV}\bodypair$ are defined in \eqref{eq:anisotropic_volume_average_definition}. The first average $\ave{P_1V}\bodypair$ has been computed in Sec.~\ref{sec:no_dipole}; here we compute all others. The method employed will be the same as in Sec.~\ref{sec:no_dipole}, but to make it effective we need to replace \eqref{eq:mdm} with the more general \emph{addition formula} (see Sec.~18.18.9 of \cite{NIST:DLMF}),
\begin{equation}\label{eq:addition_formula}
\begin{split}
P_n(\mdm)&=P_n(\sin\vt_1\sin\vt_2\cos\Angle+\cos\vt_1\cos\vt_2)=P_n(\cos\vt_1)P_n(\cos\vt_2)\\
&+2\sum_{k=1}^n\frac{(n-k)!(n+k)!}{2^{2k}(n!)^2}(\sin\vt_1)^k(\sin\vt_2)^k P_{n-k}^{(k,k)}(\cos\vt_1)P_{n-k}^{(k,k)}(\cos\vt_2)\cos k\Angle,
\end{split}
\end{equation}
where $P_n^{(\alpha,\beta)}$ is the Jacobi polynomial of degree $n$ and indices $(\alpha,\beta)$. Jacobi polynomials are defined in the interval $[-1,1]$ and are orthogonal relative to the weight function $w(x)=(1-x)^\alpha(1+x)^\beta$. They enjoy the symmetry property $P_n^{(\alpha,\beta)}(-x)=(-1)^nP_n^{(\alpha,\beta)}(x)$ and can be represented as finite sums (see Sec.~18.5.8 of \cite{NIST:DLMF}),
\begin{equation}\label{eq:Jacobi_polynomials}
P_n^{(\alpha,\beta)}(x)=\frac{1}{2^n}\sum_{k=0}^n\binom{n+\alpha}{k}\binom{n+\beta}{n-k}(x-1)^{n-k}(x+1)^k.
\end{equation}
The first three Jacobi polynomials that interest us are
\begin{equation}\label{eq:Jacobi_functionals_three}
P_0^{(2,2)}(x)=1,\quad P_1^{(2,2)}(x)=3x,\quad P_2^{(2,2)}(x)=7x^2-1.
\end{equation}

With the aid of \eqref{eq:m_representations} and \eqref{eq:addition_formula}, we establish the identity,
\begin{equation}\label{eq:average_identity}
\begin{split}
&\frac{1}{2\pi}\int_0^{2\pi}P_n(\mdm)\left[\sin^2\Angle\left(\rhouu\rhout+\rhotu\rhott\right) +\cos^2\Angle\left(\rhouu\rhott+\rhotu\rhout\right) \right]d\Angle\\
&=P_n(\m_1\cdot\normal)P_n(\m_2\cdot\normal)\left(\rhouu+\rhotu \right)\left(\rhout+\rhott \right)\\
&-\frac{(n-2)!(n+2)!}{(4n!)^2}[1-(\m_1\cdot\normal)^2][1-(\m_2\cdot\normal)^2] \Ja(\m_1\cdot\normal)\Ja(\m_2\cdot\normal)\left(\rhouu-\rhotu\right)\left(\rhout-\rhott\right),
\end{split}
\end{equation}
where, as stipulated above, the principal directions of curvatures $\e_1^{(1)}$ and $\e_1^{(2)}$ for bodies $\bodyu$ and $\bodyt$, respectively, to which the principal radii of curvature $\rhouu$ and $\rhout$ are correspondingly associated, lie orderly on the planes $(\m_1,\n)$ and $(\m_2,\n)$. Use of \eqref{eq:B_1_preliminaries}, \eqref{eq:addition_formula}, and \eqref{eq:average_identity} in \eqref{eq:V_bodysum} leads us to
\begin{equation}\label{eq:P_n_bodypair}
\begin{split}
\ave{P_nV}\bodypair&= \frac13\left(\ave{\frac{P_n(\m_2\cdot\normal)}{\Kt}}_\normal \int_\sphere(\normal\cdot\ru) P_n(\m_1\cdot\normal)d\areanu +\ave{(\normal\cdot\rt) P_n(\m_2\cdot\normal)}_\normal \int_\sphere\frac{P_n(\m_1\cdot\normal)}{\Ku}\right)\\
&+\frac16\ave{P_n(\m_2\cdot\normal)\left(\rhout+\rhott\right)}_\normal \int_\sphere(\normal\cdot\ru) P_n(\m_1\cdot\normal \left(\rhouu+\rhotu\right)d\areanu\\
&+\frac16\ave{(\normal\cdot\rt) P_n(\m_2\cdot\normal \left(\rhout+\rhott\right)}_\normal \int_\sphere P_n(\m_1\cdot\normal)\left(\rhouu+\rhotu\right)d\areanu\\
&-\frac16\frac{(n-2)!(n+2)!}{(4n!)^2}\ave{[1-(\m_2\cdot\normal)^2]\Ja(\m_2\cdot\normal) \left(\rhout-\rhott\right)}_\normal\\
&\times \int_\sphere[1-(\m_1\cdot\normal)^2](\normal\cdot\ru)\Ja(\m_1\cdot\normal)\left(\rhouu-\rhotu\right)d\areanu\\
&-\frac16\frac{(n-2)!(n+2)!}{(4n!)^2}\ave{[1-(\m_2\cdot\normal)^2](\normal\cdot\rt)\Ja(\m_2\cdot\normal) \left(\rhout-\rhott\right)}_\normal\\
&\times\int_\sphere [1-(\m_1\cdot\normal)^2]\Ja(\m_1\cdot\normal) \left(\rhouu-\rhotu\right) d\areanu.
\end{split}
\end{equation}

To accomplish our task we need now compute all the integrals featuring in \eqref{eq:P_n_bodypair}. To make this easier, it is expedient to realize that they result from parameterizing some general shape functionals through the radial mappings $\ru$ and $\rt$ of bodies $\bodyu$ and $\bodyt$. For a cylindrically symmetric body $\body\in\convp$, by \eqref{eq:surface_dilation_ratio}, we see that
\begin{subequations}\label{eq:new_shape_functionals}
\begin{equation}\label{eq:new_S_functional}
\int_\sphere\frac{P_n(\m\cdot\normal)}{K}d\areanu=\int_\boundary P_n(\mdn)d\arean.
\end{equation}
Similarly, also by use of \eqref{eq:radial_mapping_surface_gradient}, \eqref{eq:curvature_radii}, and \eqref{eq:H_and_K}, we easily arrive at
\begin{align}
\int_\sphere(\normal\cdot\radial)P_n(\m\cdot\normal)d\areanu&= \int_\boundary(\normal\cdot\x) P_n(\mdn)Kd\arean,\label{eq:new_M_prime_functional}\\
\int_\sphere P_n(\mdnu)(\rho_1+\rho_2)d\areanu&= 2\int_\boundary P_n(\mdn)Hd\arean,\label{eq:new_M_functional}\\
\int_\sphere(\normal\cdot\radial)P_n(\mdnu)(\rho_1+\rho_2)d\areanu&= 2\int_\boundary(\n\cdot\x)P_n(\mdn)Hd\arean,\label{eq:new_S_prime_functional}\\
\int_\sphere[1-(\mdnu)^2](\normal\cdot\radial)\Ja(\mdnu)(\rho_1-\rho_2)d\areanu&= -2\int_\boundary[1-(\mdn)^2](\n\cdot\x)\Ja(\mdn)\frax12(\sigma_1-\sigma_2)d\arean, \label{eq:new_S_double_prime}\\
\int_\sphere[1-(\mdnu)^2]\Ja(\mdnu)(\rho_1-\rho_2)d\areanu&= -2\int_\boundary[1-(\mdn)^2]\Ja(\mdn)\frax12(\sigma_1-\sigma_2)d\arean \label{eq:new_M_double_prime_functional}
\end{align}
\end{subequations}
In formulae \eqref{eq:new_shape_functionals} we readily recognize the shape functionals defined in \eqref{eq:M_and_S_functionals_definitions}. With the aid of these definitions, we give \eqref{eq:P_n_bodypair} the form \eqref{eq:anisotropic_volume_average_representation} used in the main text.

\subsection{Generating curve}\label{sec:generating_curve}
\begin{figure}[h]
\centering
   \includegraphics[width=0.12\linewidth]{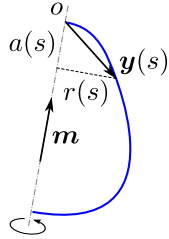}
\caption{(Color online) The plane curve $\y$, which generates $\boundary$ by a $2\pi$-rotation about $\m$, is parameterized in the generic scalar $s$ as $\y(s)=\x(s,0)$, where $\x(s,\vp)$ is given by \eqref{eq:generating_curve}. The origin $o$ is taken to coincide with the uppermost pole.}
\label{fig:generating_curve}
\end{figure}

Here we represent the boundary $\boundary$ of a cylindrically symmetric convex body $\body$ as generated by the $2\pi$-rotation of a plane curve, $\y(s)=r(s)\ex-a(s)\ez$, parameterized in the generic scalar $s$ (see Fig.~\ref{fig:generating_curve}).
Identifying $\m$ with the unit vector $\ez$ of a Cartesian frame $\Frame$, we can then represent  $\boundary$ as the surface
\begin{equation}\label{eq:generating_curve}
\x(s,\vp)=r(s)\er-a(s)\ez,
\end{equation}
where
\begin{subequations}
\begin{equation}\label{eq:e_r}
\er=\cos\vp\,\ex+\sin\vp\,\ey
\end{equation}
is the radial unit vector and
\begin{equation}\label{eq:e_varphi}
\ep=-\sin\vp\,\ex+\cos\vp\,\ey
\end{equation}
\end{subequations}
is the associate orthogonal unit vector in the plane $(\ex,\ey)$.

By letting $s$ and $\vp$ depend on a parameter $t$, we obtain a trajectory $t\mapsto\xxi(t):=\x(s(t),\vp(t))$ on $\boundary$. It follows from \eqref{eq:generating_curve} that
\begin{equation}\label{eq:dotted_surface_trajectory}
\dot\xxi=\dot{s}\sqrt{r'^2+a'^2}\tang+\dot\vp r\ep,
\end{equation}
where a prime $'$ denotes differentiation with respect to $s$, a superimposed dot denotes differentiation with respect to $t$, and
\begin{equation}\label{eq:generating_curve_tangent_vector}
\tang=\frac{r'\er-a'\ez}{\sqrt{r'^2+a'^2}}
\end{equation}
is the unit tangent vector to $\x(\cdot,\varphi)$, for given $\varphi$. From \eqref{eq:generating_curve_tangent_vector} and \eqref{eq:dotted_surface_trajectory}, we easily arrive at both the unit outward normal to $\boundary$,
\begin{equation}\label{eq:generating_curve_outward_normal}
\n=\frac{a'\er+r'\ez}{\sqrt{r'^2+a'^2}}
\end{equation}
and the surface area element
\begin{equation}\label{eq:generating_curve_area_element}
d\arean=r\sqrt{r'^2+a'^2}dsd\vp.
\end{equation}
By further differentiating $\n$ along the trajectory $\xxi(t)$, we obtain that
\begin{equation}\label{eq:generating_curve_dotted_n}
\dot\n=\frac{\dot{s}(a''r'-a'r'')}{r'^2+a'^2}\tang+\frac{\dot\vp a'}{\sqrt{r'^2+a'^2}}\ep.
\end{equation}
Since $\dot\n=(\nablas\n)\dot\xxi$ and, by \eqref{eq:dotted_surface_trajectory},
\begin{equation}\label{eq:generating_curve_dotted_s_varphi}
\dot{s}=\frac{\dot\xxi\cdot\tang}{\sqrt{r'^2+a'^2}}\quad\text{and}\quad\dot\vp=\frac{\dot\xxi\cdot\ep}{r},
\end{equation}
for $\dot\xxi$ is arbitrary, we conclude that
\begin{equation}\label{eq:generating_curve_curvature_tensor}
\nablas\n=\frac{a''r'-a'r''}{(r'^2+a'^2)^{3/2}}\tang\otimes\tang +\frac{a'}{r\sqrt{r'^2+a'^2}}\ep\otimes\ep,
\end{equation}
whence we read off at once the principal curvatures of $\boundary$.

\subsubsection{Cones}\label{sec:appendix_cones}
\begin{figure}[h]
\centering
\includegraphics[width=0.15\linewidth]{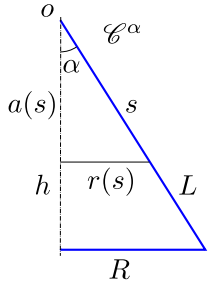}
\caption{(Color online) The generating curve of a circular cone $\conea$ with semi-amplitude $\alpha$, radius $R$ and height $h$, the two latter related to the slant height $L$ as in \eqref{eq:cone_R_and_h}. The parameter $s$ here represents the arc-length along the slant side of the cone.}
\label{fig:cone}
\end{figure}

Figure~\ref{fig:cone} depicts the generating curve for a circular cone $\conea$ with vertex in the origin $o$, semi-amplitude $\alpha$, radius $R$ and height $h$, which are related to the slant height $L$ through the equations
\begin{equation}\label{eq:cone_R_and_h}
R=L\sin\alpha,\quad h=L\cos\alpha.
\end{equation}
The functions $r(s)$ and $a(s)$ featuring in \eqref{eq:generating_curve} are correspondingly given by
\begin{equation}\label{eq:cone_r_and_a}
r(s)=s\sin\alpha\quad\text{and}\quad a(s)=s\cos\alpha,
\end{equation}
where now $s$ has been chosen as the arc-length along the slant height of the cone; it follows from \eqref{eq:generating_curve_curvature_tensor} that
\begin{equation}\label{eq:cone_curvatures}
\sigma_1=0\quad\text{and}\quad\sigma_2=\frac{\cot\alpha}{s}.
\end{equation}

\subsubsection{Spheroids}\label{sec:appendix_spheroids}
\begin{figure}[h]
\centering
  \includegraphics[width=0.1\linewidth]{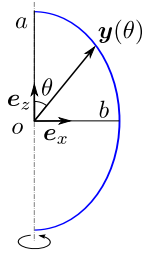}
\caption{(Color online) The half-ellipse with semi-axes $a$ and $b$ whose $2\pi$-rotation about the symmetry axis $\m=\ex$ generates a spheroid with aspect ratio $\eta=b/a$. The parameter $\theta$ featuring in \eqref{eq:a_r_spheroid} designates the angle between $\y(\theta)$ and $\ez$. The origin $o$ has been chosen in the center of the spheroid.}
\label{fig:generating_curve_spheroid}
\end{figure}

The generating curve for a spheroid $\sphero$ is illustrated in Fig.~\ref{fig:generating_curve_spheroid}; it is a half-ellipse with semi-axes $a$ and $b$, along $\ez$ and $\ex$ respectively, and centered in the origin $o$.
Letting the parameter $s$ be the angle $\theta$ ranging in $[0,\pi]$ and depicted in Fig.~\ref{fig:generating_curve_spheroid}, the functions $a(s)$ and $r(s)$ in \eqref{eq:generating_curve} are now written as\footnote{To avoid typographical clutter, we are guilty of using the same symbol for both the function $a(\theta)$ and the scaling semi-axis of the generating half-ellipse.}
\begin{equation}\label{eq:a_r_spheroid}
a(\theta)=-a\cos\theta\quad\text{and}\quad r(\theta)=b\sin\theta.
\end{equation}
By use of \eqref{eq:generating_curve_outward_normal} and \eqref{eq:generating_curve_curvature_tensor}, we readily arrive at
\begin{subequations}\label{eq:generating_curve_spheroid}
\begin{align}
\m\cdot\n=&\frac{\eta\cos\theta}{\sqrt{1+(\eta^2-1)\cos^2\theta}},\\
\n\cdot\x=&\frac{a\eta}{\sqrt{1+(\eta^2-1)\cos^2\theta}},\\
\sigma_1=&\frac{\eta}{a}\frac{1}{[1+(\eta^2-1)\cos^2\theta]^{\frac32}},\\
\sigma_2=&\frac{1}{a\eta}\frac{1}{\sqrt{1+(\eta^2-1)\cos^2\theta}},
\end{align}
\end{subequations}
where $\eta:=b/a$ is the spheroid's aspect ratio.

\subsection{Extended $M$ and $S$ functionals of a circular ridge}\label{sec:ridge_M_and_S}
\begin{figure}[h]
\centering
  \subfigure[]{\includegraphics[width=0.20\linewidth]{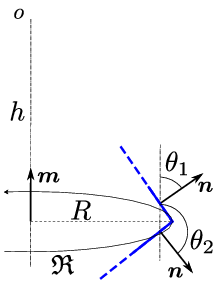}}
  \hspace{.02\linewidth}
  \subfigure[]{\includegraphics[width=0.20\linewidth]{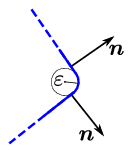}}
\caption{(Color online) (a) Circular ridge $\ridge$ of radius $R$ in the plane orthogonal to the symmetry axis $\m$ at the distance $h$ from the origin $o$. The unit outward normal $\n$ makes the angle $\theta_1$ with $\m$ on one side and angle $\theta_2$ on the other side. (b) The sharp corner of $\ridge$ is rounded off in a toroidal surface with meridian radius $\varepsilon$.}
\label{fig:ridge}
\end{figure}

Here we apply the formalism presented in Sec.~\ref{sec:generating_curve} to compute the extended $M$ and $S$ functionals defined in Sec.~\ref{sec:Extended_Minkowski_functionals} for a circular ridge $\ridge$ of radius $R$, where neither $H$ nor $K$ are defined. To this end, we replace $\ridge$ with a toroidal approximation  $\ridgee$ with equatorial radius $R$ and meridian radius $\varepsilon$, whose outer unit normal $\n$ spans the sector in which the angle $\theta$ that it makes with the symmetry axis $\m$ ranges in the interval $[\theta_1,\theta_2]$. To afford a greater generality (and in view of our application to cones in Sec.~\ref{sec:cones} above), we choose the origin $o$ on the symmetry axis at the generic distance $h$ from the ridge's plane (see Fig.~\ref{fig:ridge}).
Our strategy will be to compute the extended $M$ and $S$ functionals on $\ridgee$ and then take the limit as $\varepsilon\to0^+$. The functions $r(s)$ and $a(s)$ introduced in Sec.~\ref{sec:generating_curve} which here describe $\ridgee$ are
\begin{equation}\label{eq:ridge_r_and_a}
r(s)=R+\ve\sin\theta(s),\quad a(s)=h-\ve\cos\theta(s),
\end{equation}
where $\theta$ and $s$ are related through $s-s_1=\ve(\theta(s)-\theta_1)$, with $s_1$ an arbitrary constant. It easily follows from \eqref{eq:generating_curve_curvature_tensor} and \eqref{eq:ridge_r_and_a} that
\begin{equation}\label{eq:ridge_curvatures}
\sigma_1=\frac1\ve,\quad\sigma_2=\frac{\sin\theta}{R+\ve\sin\theta}.
\end{equation}
Moreover, \eqref{eq:generating_curve_area_element} yields
\begin{equation}\label{eq:ridge_area_element}
d\arean=(R+\ve\sin\theta)\ve d\theta d\vp.
\end{equation}

Using \eqref{eq:ridge_curvatures} and \eqref{eq:ridge_area_element} in the definitions of the extended $M$ and $S$ functionals in \eqref{eq:M_and_S_functionals_definitions}, and then taking the limit as $\ve\to0^+$, we arrive at the following expressions:
\begin{subequations}\label{eq:ridge_M_and_S_functionals}
\begin{align}
M_n[\ridge]&=\pi R\iat P_n(\cos\theta)d\theta,\label{eq:ridge_M}\\
M'_n[\ridge]&=2\pi\iat(R\sin\theta-h\cos\theta)\sin\theta P_n(\cos\theta)d\theta,\label{eq:ridge_M_prime}\\
M''_n[\ridge]&=\pi R\iat\sin^2\theta\Ja(\cos\theta)d\theta,\label{eq:ridge_M_double_prime}\\
S_n[\ridge]&=0\label{eq:ridge_S},\\
S'_n[\ridge]&=\pi R\iat(R\sin\theta-h\cos\theta)P_n(\cos\theta)d\theta,\label{eq:ridge_S_prime}\\
S''_n[\ridge]&=\pi R\iat(R\sin\theta-h\cos\theta)\sin^2\theta\Ja(\cos\theta)d\theta. \label{eq:ridge_S_double_prime}
\end{align}
\end{subequations}

\subsubsection{Extended $M$ and $S$ functionals for a disk}\label{sec:ridge_M_and_S}
\begin{figure}[h]
\centering
  \subfigure[]{\includegraphics[width=0.25\linewidth]{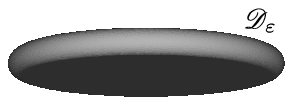}}
  \hspace{.02\linewidth}
  \subfigure[]{\includegraphics[width=0.25\linewidth]{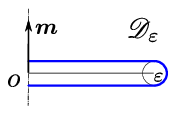}}
\caption{(Color online) (a) Spherodisk $\disk_\ve$ defined as the Minkowski sum of the disk $\disk$ and a ball $\balle$ of radius $\ve$ and same center $o$ as $\disk$. (b) The generating curve of $\disk_\ve$. The symmetry axis $\m$ is orthogonal to $\disk$.}
\label{fig:disk}
\end{figure}

Formulae \eqref{eq:ridge_M_and_S_functionals} are instrumental to obtaining the explicit expressions for the extended $M$ and $S$ functionals of a disk $\disk$ of radius $R$. As before, we start by replacing $\disk$ with an approximating rounded body, the \emph{spherodisk} $\disk_\ve$ defined as the Minkowski sum of $\disk$ and a ball $\balle$ of radius $\ve$ and center coincident with the center of $\disk$. Figure~\ref{fig:disk} illustrates both $\disk_\ve$ and the generating curve of its boundary.
The extended $M$ and $S$ functionals for $\disk$ will be obtained by taking the limit as $\ve\to0^+$ in those computed for $\disk_\ve$. $\partial\disk_\ve$ consists of two flat parallel disks, for which both principal curvatures vanish, and the toroidal approximation $\ridgee$ of the circular rim $\ridge$ of $\disk$, for which the angles $\theta_1$ and $\theta_2$ in Fig.~\ref{fig:ridge} are $\theta_1=0$ and $\theta_2=\pi$, respectively. Apart from the limit as $\ve\to0^+$ of $S_n[\disk_\ve]$, which is immediate to compute, for all other functionals this limit  follows directly from \eqref{eq:ridge_M_and_S_functionals} by setting $h=0$ and choosing $\theta_1$ and $\theta_2$ as above. We thus arrive at
\begin{subequations}\label{eq:disk_M_S_functionals}
\begin{align}
M_n[\disk]&=\pi R\iad P_n(\cos\theta)d\theta,\label{eq:disk_M_n}\\
M'_n[\disk]&=2\pi R\iad P_n(\cos\theta)\sin^2\theta d\theta,\label{eq:disk_M_prime}\\
M''_n[\disk]&=\pi R\iad\Ja(\cos\theta)\sin^2\theta d\theta,\label{eq:disk_M_double_prime}\\
S_n[\disk]&=\pi R^2\left(P_n(1)+P_n(-1)\right),\label{eq:disk_S}\\
S'_n[\disk]&=\pi R^2\iad P_n(\cos\theta)\sin\theta d\theta,\label{eq:disk_S_prime}\\
S''_n[\disk]&=\pi R^2\iad\Ja(\cos\theta)\sin^3\theta d\theta.\label{eq:disk_S_double_prime}
\end{align}
\end{subequations}
In particular, it follows from \eqref{eq:disk_M_S_functionals} that all extended $M$ and $S$ functionals with an odd index $n$ vanish for a disk.

\subsection{Invariance under translations}\label{sec:invariance_translations}
The anisotropic volume averages for which we found in \eqref{eq:anisotropic_volume_average_representation} an explicit representation in terms of the extended Minkowski functionals are clearly invariant under the full Euclidean group comprising both translations and rotations. On the other hand, as clearly shown by equations \eqref{eq:M_and_S_functionals_definitions}, while all extended $M$ and $S$ functionals are invariant under rotations, those that also appear to be invariant under translations are only $M_n$, $M''_n$, and $S_n$. $M'_n[\body]$, $S'_n[\body]$, and $S''_n[\body]$ are expressed as integrals over the boundary $\boundary$ of the body $\body$ of fields that depend explicitly on the origin $o$ through the position vector $\x$. Here we shall show that, despite all appearances, $M'_n$ is indeed invariant under translations, whereas both $S'_n$ and $S''_n$ are not. For the latter two, we shall also give explicit formulae that describe how they are affected by a translation. Of course, the combination of these functionals in \eqref{eq:anisotropic_volume_average_representation} must be translation-invariant. We shall exploit this fact in Sec.~\ref{sec:reduction_formulae} below to show that functionals $M''_n$ and $M_n$ are not independent, a conclusion which would be hard to reach by direct comparison of their definitions.

Translating a body $\body$ by the vector $\va$ is formally equivalent to taking the Minkowski sum $\body+\va$ of $\body$ and the point in space identified by $\va$. Moreover, since all extended $M$ and $S$ functionals are invariant under rotations, computed on $\body+\va$ for any given $\body$, they are isotropic functions of $\va$. It readily follows from \eqref{eq:M_prime_definition} that
\begin{equation}\label{eq:M_prime_pre_translation}
M'_n[\body+\va]=M'_n[\body]+\va\cdot\int_{\boundary}P_n(\mdn)K\n d\arean.
\end{equation}
The integral on the right side of \eqref{eq:M_prime_pre_translation} is an isotropic vector-valued function of $\m$; as such, by the Cauchy theorem on isotropic vector-valued functions, it must be proportional to $\m$. Thus, \eqref{eq:M_prime_pre_translation} becomes
\begin{equation}\label{eq:M_prime_translation}
M'_n[\body+\va]=M'_n[\body]+\va\cdot\m\int_{\boundary}(\mdn)P_n(\mdn)Kd\arean.
\end{equation}
For $\body\in\convp$, by use of \eqref{eq:surface_dilation_ratio}, we see that
\begin{equation}\label{eq:M_post_translation}
\int_{\boundary}(\mdn)P_n(\mdn)Kd\arean= \int_{\sphere}P_1(\mdnu)P_n(\mdnu)d\areanu=2\pi\int_{-1}^1P_1(x)P_n(x)dx=0\quad\forall\ n\geqq2,
\end{equation}
where the last equality follows from the orthogonality of Legendre polynomials. Since we have already proved in Sec.~\ref{sec:no_dipole} that $M_1[\body]$ vanishes identically for all $\body\in\convp$, by \eqref{eq:M_post_translation} we conclude that all functionals $M_n$ are invariant under translations.

This is not the case for both $S'_n$ and $S''_n$. Reasoning precisely as above and making use of the recurrence relations\footnote{See, for example, Sec.~18.9.1 of \cite{NIST:DLMF}.}
\begin{subequations}\label{eq:recurrence_relations}
\begin{align}
xP_n(x)&=\frac{n+1}{2n+1}P_{n+1}(x)+\frac{n}{2n+1}P_{n-1}(x),\label{eq:recurrence_relation_Legendre}\\
x\Ja(x)&=\frac{(n-1)(n+3)}{(n+1)(2n+1)}P_{n-1}^{(2,2)}(x) +\frac{n}{2n+1}P_{n-3}^{(2,2)}(x), \label{eq:recurrence_relation_Jacobi}
\end{align}
\end{subequations}
the latter valid for $n\geqq2$ and with the postulation that $P_{-1}^{(2,2)}\equiv0$, we arrive at
\begin{subequations}\label{eq:translation_S_prime_S_double_prime}
\begin{align}
S'_n[\body+\va]&=S'_n[\body]+\va\cdot\m\left(\frac{n+1}{2n+1}M_{n+1}[\body] +\frac{n}{2n+1}M_{n-1}[\body]\right),\label{eq:translation_S_prime}\\
S''_n[\body+\va]&=S''_n[\body]+\va\cdot\m\left(\frac{(n-1)(n+3)}{(n+1)(2n+1)}M''_{n+1}[\body] +\frac{n}{2n+1}M''_{n-1}[\body]\right),\label{eq:translation_S_double_prime}
\end{align}
\end{subequations}
the latter valid for $n\geqq2$ and with the postulation that $M''_1[\body]\equiv0$.

\subsection{Reduction formulae}\label{sec:reduction_formulae}
Here we take advantage of the general formulae \eqref{eq:translation_S_prime_S_double_prime} just established and of the specific expressions for the extended $M$ and $S$ functionals obtained in \eqref{eq:disk_M_S_functionals} to show that each functional $M''_n$ reduces to $M_n$ and to substantiate our conjecture that so should equally do each $M'_n$.

\subsubsection{$M''_n$ reduced to $M_n$}\label{sec:M_double_prime_reduction}
By requiring that the anisotropic volume averages, as expressed by \eqref{eq:anisotropic_volume_average_representation}, be invariant under translations for all bodies $\body_1$ and $\body_2$, a laborious but easy computation relying on \eqref{eq:translation_S_prime_S_double_prime} and the translation-invariance of $M'_n$ shows that
\begin{equation}\label{eq:a_double_prime_definition}
M''_n[\body]=a''_nM_n[\body],
\end{equation}
where the coefficients $a''_n$ must obey the recurrence equation
\begin{equation}\label{eq:a_double_prime_recurrence}
\frac{(n+3)(n+2)}{16(n+1)n}a''_{n+1}a''_n=1,
\end{equation}
whose explicit solution is
\begin{equation}\label{eq:a_double_prime_solution}
a''_n=\frac{4n}{n+2}.
\end{equation}
Combining \eqref{eq:a_double_prime_solution} with \eqref{eq:a_double_prime_definition}, we arrive immediately at \eqref{eq:M_double_prime_M}.

\subsubsection{$M'_n$ reduced to $M_n$}\label{sec:M_prime_reduction}
Inspired by \eqref{eq:a_double_prime_solution}, we computed the ratio $a'_n$ of $M'_n[\disk]$ to $M_n[\disk]$ for a disk $\disk$; interpolating with the aid of \eqref{eq:disk_M_n} and \eqref{eq:disk_M_prime} the values of $a'_n$ obtained for a number of indices $n$, we concluded that
\begin{equation}\label{eq:M_prime_cojectured}
a'_n=-\frac{2}{(n-1)(n+1)}\quad\forall\ n\geqq2,
\end{equation}
whence \eqref{eq:M_prime_M} follows at once.
Although we could not establish \eqref{eq:M_prime_M} on a firmer basis, we checked by use of \eqref{eq:cone_M_n} and \eqref{eq:cone_M_n_p} and of \eqref{eq:M_n_spheroids} and \eqref{eq:M'_n_spheroids} that it is valid for a large number of indices $n$ when $\body$ is taken to be either a cone $\conea$ or a spheroid $\sphero$, for all values of the semi-amplitude $\alpha$ and of the aspect ratio $\eta$. We are aware that \eqref{eq:M_double_prime_M} and \eqref{eq:M_prime_M} have a completely different standing, as the former has been proved rigorously, whereas the latter is only conjectured. Most of our development in the main body of the paper relies neither on \eqref{eq:M_double_prime_M} nor on \eqref{eq:M_prime_M}. What does depend on \eqref{eq:M_prime_M} is only the possibility of giving compact factorized formulas for the coefficients $B_n$ as those listed in \eqref{eq:cone_B_n} and \eqref{eq:spheroid_B_n} for cones and spheroids, respectively, both of which are expected to obey \eqref{eq:M_prime_M}.

\subsection{Legendre coefficients for the excluded volume of cones}\label{sec:cones_B_n}
Letting $\body_1$ and $\body_2$ be two congruent circular cones, $\coneu$ and $\conet$, with semi-amplitude $\alpha$, with the aid of \eqref{eq:isotropic_volume_average}, \eqref{eq:B_n_multiplicative}, and \eqref{eq:cone_M_S_functionals} we arrived at the following explicit formulae for the first eight Legendre coefficients $B_n$ plotted in Figs.~\ref{fig:cone_B_n} and \ref{fig:B_Zero}(b) as functions of $\alpha$:
\begin{subequations}\label{eq:cone_B_n}
\begin{align}
B_0&=\frax23\pi L^3\sin^2\!\alpha\cos\alpha+\frax12\pi L^3\sin\alpha\left[\left(\frax{\pi}{2}+\alpha\right)\sin\alpha+\cos\alpha\right](1+\sin\alpha),\label{eq:cone_B_0}
\\
B_1&=0,\label{eq:cone_B_1}
\\
B_2&=\frax{5}{64}\pi L^3\sin\alpha(2\alpha\sin\alpha+ \pi\sin\alpha  + 2\cos\alpha - 6\cos^{3}\!\alpha)
( 3\cos^2\!\alpha - 2 - 2\sin\alpha),
\\
B_3&=\frax{35}{12}\pi L^3\sin^3\!\alpha\cos^5\!\alpha,\label{eq:cone_B_3}
\\
B_4&=-\frax{3}{2048}\pi L^3 \sin\alpha(3\pi\sin\alpha  + 6\alpha\sin\alpha
 + 6\cos\alpha - 130\cos^3\!\alpha +
140\cos^5\!\alpha)\nonumber\\
&\times(35\cos^4\!\alpha - 40\cos^2\!\alpha + 8 + 8\sin\alpha),\label{eq:cone_B_4}
\\
B_5&=\frax{77}{960}\pi L^3\sin^3\!\alpha\cos^5\!\alpha( 27\cos^2\!\alpha - 20)
(9\cos^2\!\alpha - 8),\label{eq:cone_B_5}
\\
B_6&=
\frax{13}{65536} \pi L^3 \sin\alpha(5\pi\sin\alpha  + 10\alpha\sin\alpha  + 10\cos\alpha - 686\cos^3\!\alpha + 1876\cos^5\!\alpha - 1232\cos^7\!\alpha)\nonumber\\
&\times(231\cos^6\!\alpha - 378 \cos^4\!\alpha + 168\cos^2\!\alpha - 16  - 16\sin\alpha),\label{eq:cone_B_6}
\\
B_7&=\frax{3}{1792}\pi L^3 \sin^3\!\alpha\cos^5\!\alpha(280 - 924\cos^2\!\alpha + 715\cos^4\alpha)
(143\cos^4\!\alpha- 198\cos^2\!\alpha + 72),\label{eq:cone_B_7}
\end{align}
\end{subequations}
where $L$ is the cone's slant height. They are recorded here both for completeness and as an illustration of the method proposed in this paper.

\subsection{Legendre coefficients for the excluded volume of spheroids}\label{sec:spheroids_B_n}
To obtain the coefficients $B_n$ that express the excluded volume of congruent spheroids as a series of Legendre polynomials, we computed the extended Minkowski functionals in \eqref{eq:M_and_S_functionals_definitions} for the generating curve described by \eqref{eq:a_r_spheroid}. Use of \eqref{eq:generating_curve_spheroid} in \eqref{eq:M_definition} gave
\begin{equation}\label{eq:M_n_pre_spheroid}
M_n[\sphero]=\pi a\ione P_n\left(\frac{\eta u}{\sqrt{\denu}}\right)\left(\frac{\eta^2}{\denu}+1\right)du,
\end{equation}
where we have set $u:=\cos\theta$. The change of variables
\begin{equation}\label{eq:_xi_definition}
\xi:=\frac{\eta u}{\sqrt{\denu}}
\end{equation}
then led us from \eqref{eq:M_n_pre_spheroid} to \eqref{eq:M_n_spheroids}. The other formulae in \eqref{eq:M_and_S_spheroids} were obtained in precisely the same way.

\begin{figure}[h]
\centering
\includegraphics[width=0.4\linewidth]{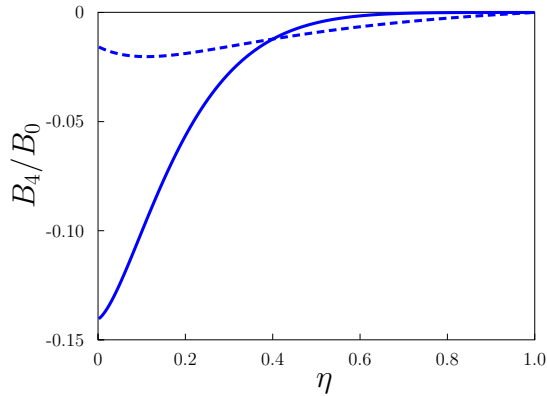}
\caption{(Color online) The ratio $B_4/B_0=f_4/f_0$ is plotted against $\eta$ both according to the expressions in \eqref{eq:f_4} and \eqref{eq:f_0} (solid line) and to formula (60) of \cite{isihara:theory} (dashed line).}
\label{fig:B4IoB0}
\end{figure}

Along with the expression for $f_2$ recorded in \eqref{eq:f_2}, we also obtained
\begin{subequations}\label{eq:spheroid_B_n}
\begin{equation}\label{eq:f_4}
\begin{split}
f_4=\frac{B_4}{\Vs}=&-\frac{9}{1024}\frac{1}{\ecc^8} \left(3\ecc^4-100\ecc^2+105 + 3(1-\ecc^2)(\ecc^4+10\ecc^2-35)\frac{\arc\ecc}{\ecc}\right)\\
&\times\left(8\ecc^4-110\ecc^2+105-(72\ecc^4-180\ecc^2+105)\frac{\arcsin\ecc}{\ecc\sqrt{1-\ecc^2}} \right)
\end{split}
\end{equation}
and
\begin{equation}\label{eq:f_6}
\begin{split}
f_6=\frac{B_6}{\Vs}=&-\frac{39}{32768}\frac{1}{\ecc^{12}}\left(5\ecc^6-581\ecc^4+1715\ecc^2-1155 -5\ecc^2(\ecc^6+20\ecc^4-210\ecc^2+420)\frac{\arc\ecc}{\ecc} \right) \\
&\times\left(16\ecc^6-616\ecc^4+1750\ecc^2-1165 -(320\ecc^5-1680\ecc^4+2520\ecc^2-1155)\frac{\arcsin\ecc}{\ecc\sqrt{1-\ecc^2}} \right).
\end{split}
\end{equation}
\end{subequations}
In Fig.~\ref{fig:B4IoB0}, using \eqref{eq:f_4} and \eqref{eq:f_0} we plotted the ratio $B_4/B_0=f_4/f_0$ against $\eta$.
It is there also contrasted against the function obtained for this ratio by Isihara~\cite{isihara:theory} (only for the prolate case). The two graphs fail to coincide, even dramatically so, away from $\eta=1$. In particular, we estimate that
\begin{equation}\label{eq:limit_Isihara}
\lim_{\eta\to0}\frac{B_4}{B_4^\mathrm{(I)}}=9,
\end{equation}
where $B_4^\mathrm{(I)}$ is $B_4$ as delivered by equation (60) of \cite{isihara:theory}.

\section{Shape-reconstruction method}
\label{sec:algorithm}
In this appendix we describe the method adopted for reconstructing the boundary  of the excluded body $\exc{\coneu}{\conet}$ for two congruent circular cones $\coneu$ and $\conet$ of semi-amplitude $\alpha$ and slant height $L$, hereafter simply denoted $\excs$ for short. More precisely, the method reconstructs a triangular surface mesh that, depending on a fundamental parameter to be described, approximates $\partial\excs$ at any degree of precision. From the surface mesh, the approximate value of the excluded volume $V[\excs]$ can be computed immediately.

The method adopted for this task is a pipeline of two algorithmic components:
\begin{enumerate}
  \item an online vector quantization algorithm that includes a generator of random point samples from $\partial\excs$ and which produces a configuration of reference vectors $\bm{W}$;
  \item a surface reconstruction algorithm that produces from $\bm{W}$  the triangulated surface mesh that represents an approximation to $\partial\excs$;
\end{enumerate}

The method described is similar to that in \cite{piastra:octupolar}. In particular, the random generator of point samples from $\partial\excs$ is essentialy the same. In that context, however, all target surfaces $\partial\excs$ were generated from \emph{sphero-cones} and could be assumed to be smooth, so that the reconstruction process could be embedded into step 1 above via the SOAM algorithm \cite{piastra:soam}. By contrast, in the case of cones considered here, the presence of ridges and cusps in $\partial\excs$ forces adopting a different strategy. In the rest of this appendix, the main aspects of this new strategy are discussed in detail.

Upon comparing the method described here with others computing the densest packing of particles of arbitrary shape \cite{deGraaf2011prl}, we heed in passing that our method determines directly the surface bounding the excluded region, with arbitrary degree of precision and in one run per pose, whereas those other methods typically require repeated Monte Carlo simulations \cite{deGraaf2012method}.

\subsection{Sampling the surface boundary}

Random point samples from $\partial\excs$ can be generated with a procedure based on equation \eqref{eq:radial_excluded_body}, reproduced here for convenience:
\begin{equation*}
\radiale(\normal) = \radial_1(\normal)-\radial_2(-\normal).
\end{equation*}
Here $\radiale$ reaches a point on $\partial\excs$ and $\bm{r}_1$ and $\bm{r}_2$,  in this specific case,  designate points on $\partial\coneu$ and $\partial\conet$, respectively. Random points on $\partial\excs$ can be obtained either by a generating a random vector reaching a point on $\partial\coneu$ and then finding a vector to a point on $\partial\conet$ that has opposite normal $-\bm{\nu}$ or by reverting this very procedure: the sum of the vectors thus obtained will belong to $\partial\excs$.

The main difficulty in implementing such a random generator is to guarantee positive sampling probability almost everywhere on $\partial\excs$, that is, apart from subsets of zero area measure. On all smooth components of a circular cone, in fact, the Gaussian curvature $K$ vanishes and this means that in general a normal vector $\bm{\nu}$ does not identify uniquely one point on the cone's surface.  Furthermore, the Minkowski sum of two straight lines on the boundary of each cone can result in a surface patch with positive area measure on $\partial\excs$, despite the fact that each line has zero area measure and thus no chances  of being sampled, unless specific provisions are introduced. Appendix B in \cite{piastra:octupolar} describes how these problems can be circumvented in actual computations.

\begin{figure}[h]
\centering
  \subfigure[]{\includegraphics[width=0.16\linewidth]{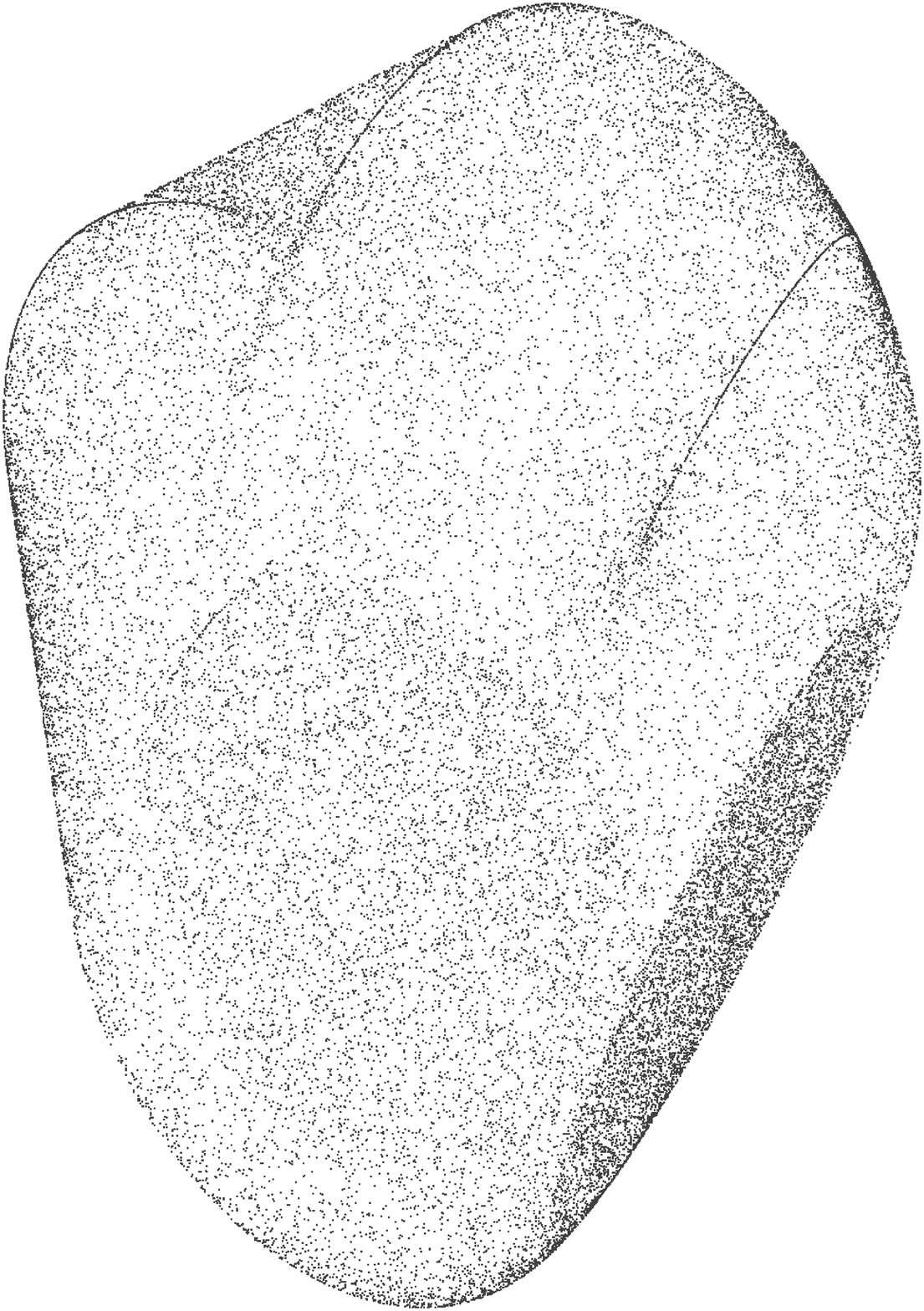}}
  \hspace{.02\linewidth}
  \subfigure[]{\includegraphics[width=0.16\linewidth]{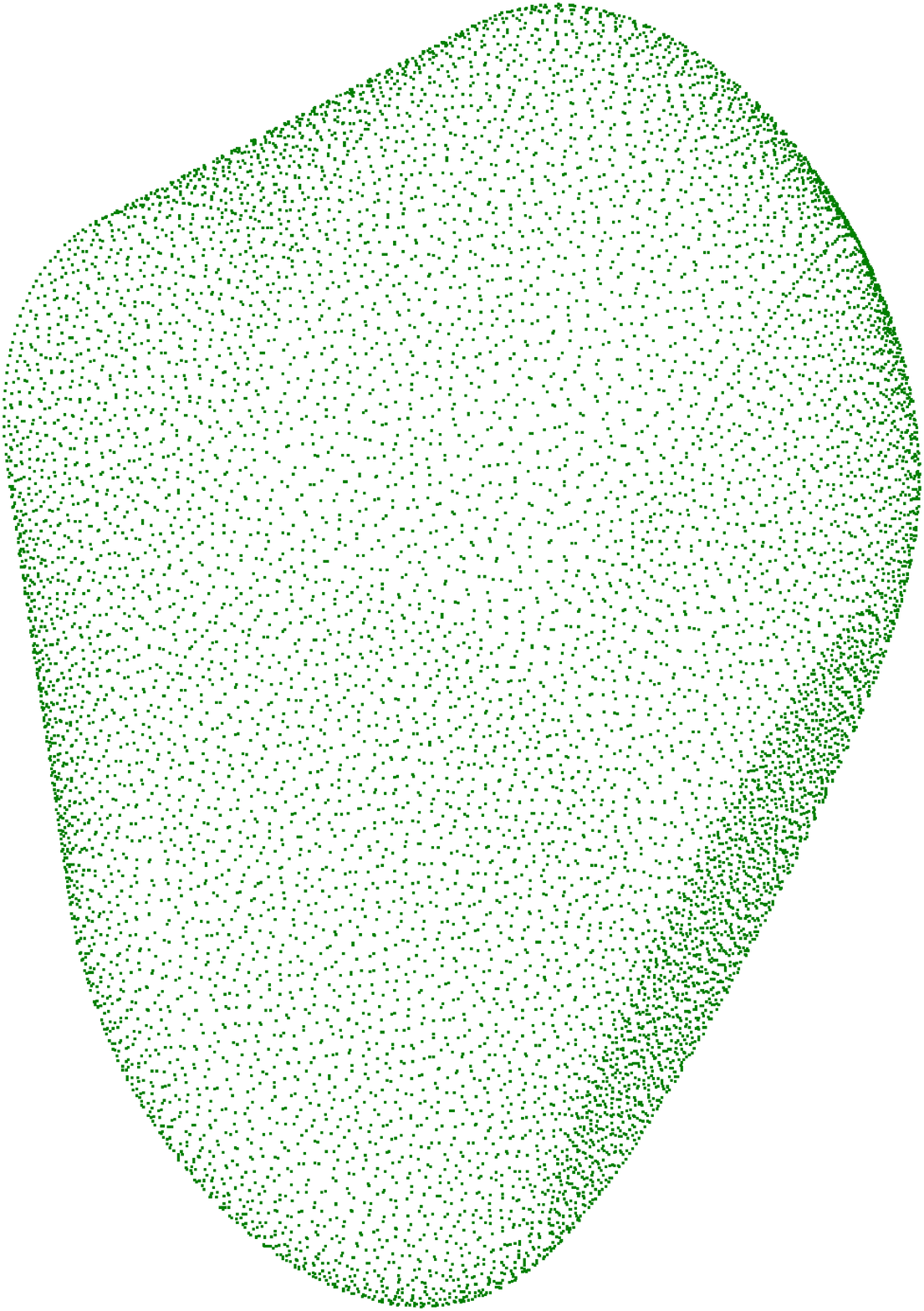}}
  \hspace{.02\linewidth}
  \subfigure[]{\includegraphics[width=0.16\linewidth]{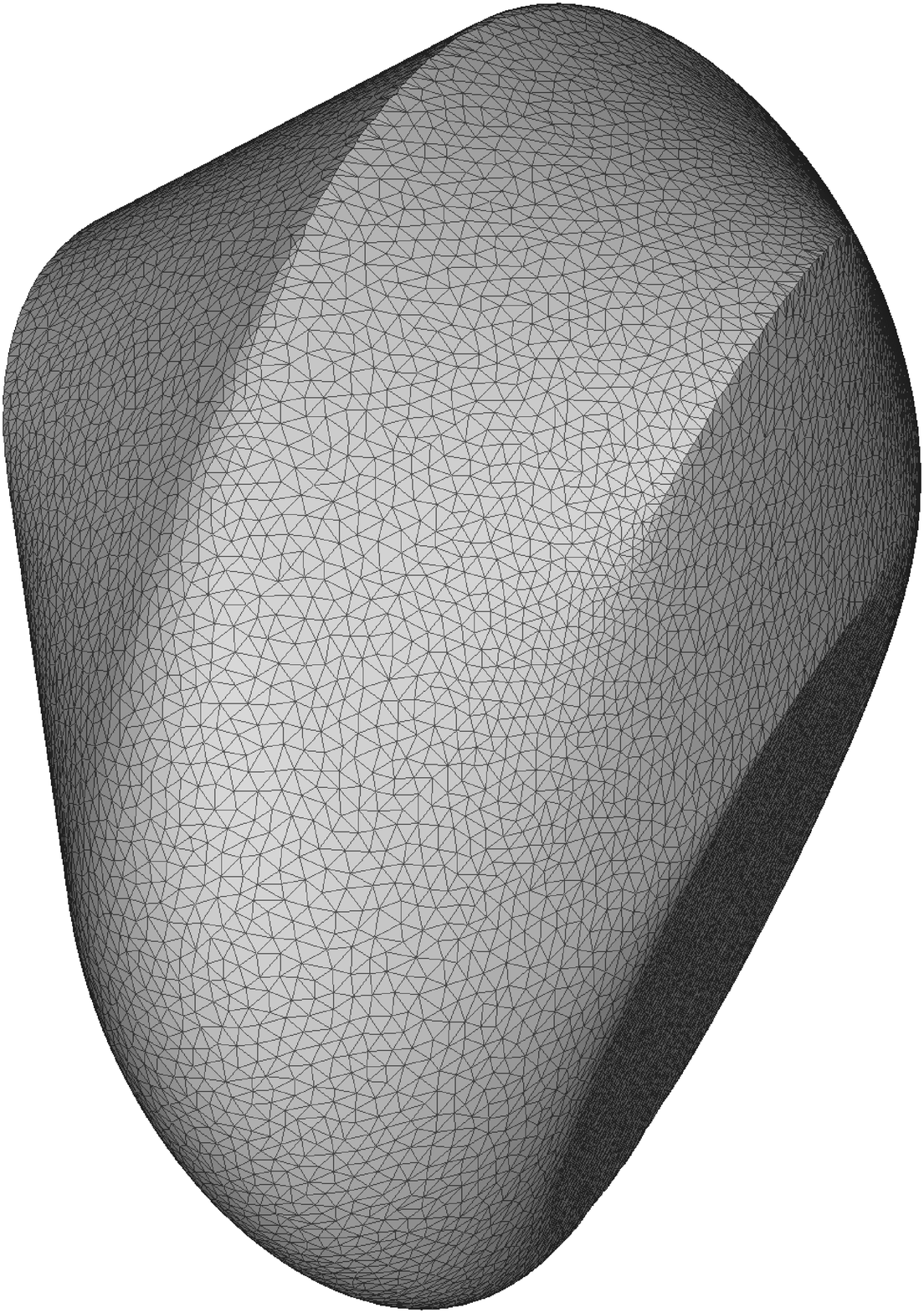}}
  \hspace{.02\linewidth}
  \subfigure[]{\includegraphics[width=0.16\linewidth]{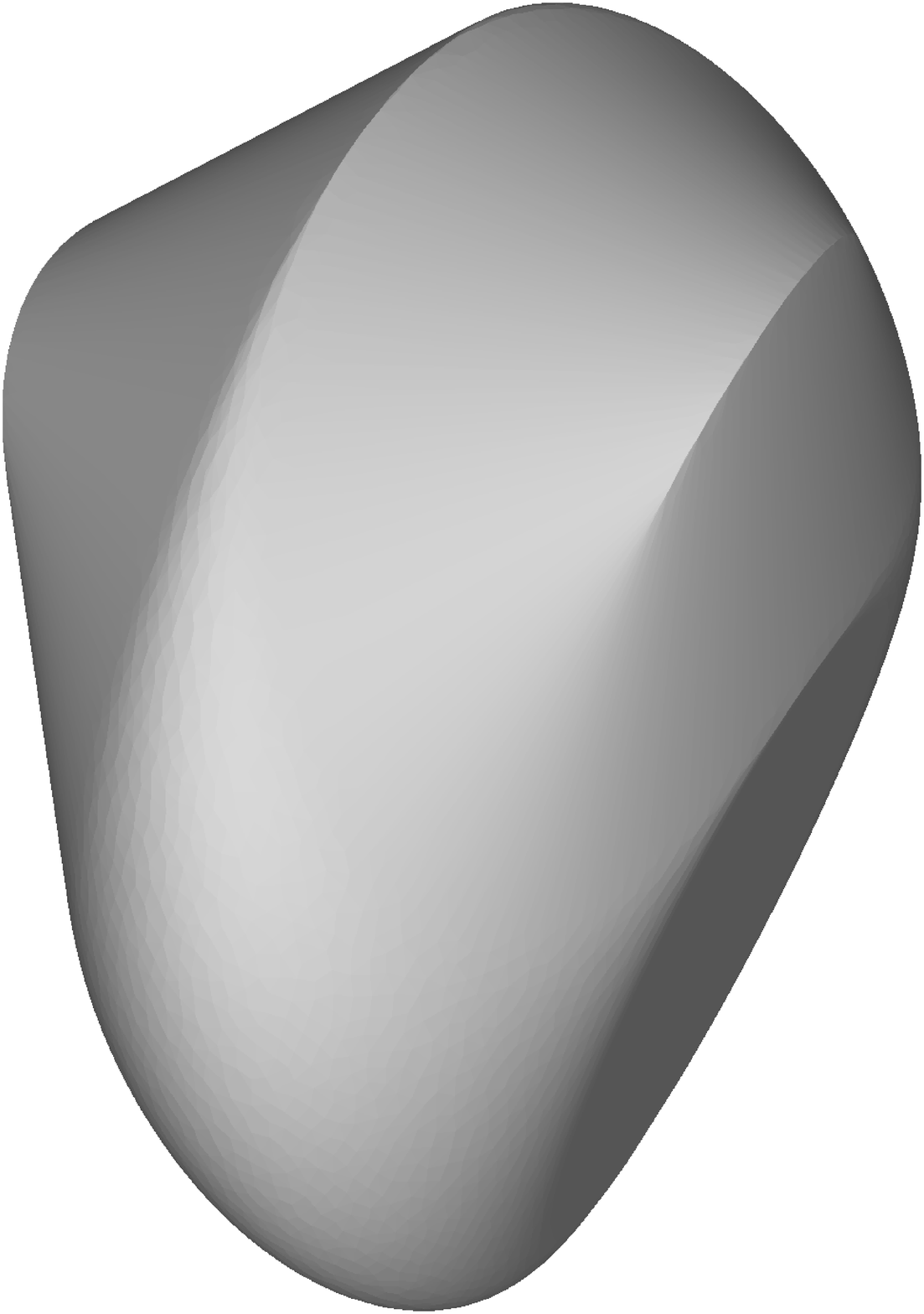}}
\caption{(Color online) (a) A set of $30\,\mathrm{K}$ random points generated from $\partial\excs$ with the method adopted here: the sampling of the surface is clearly non-uniform; (b) the final configuration of 10K reference vectors $\bm{W}$ produced by the adaptive NG algorithm is more uniformly distributed; (c) and (d) from $\bm{W}$, the ball-pivoting algorithm reconstructs the surface boundary $\partial\excs$ with no human intervention.}
\label{fig:method-overview}
\end{figure}

Although the requirement of positive sampling probability almost everywhere can be enforced in practice, no known method guarantees \emph{uniform} sampling probability over $\partial\excs$.\footnote{Known methods for uniform sampling presuppose knowledge of the surface's analytic description plus further specific conditions \cite{Arndt2008509}.} As shown in Fig.~\ref{fig:method-overview}(a), the overall sampling obtained with the chosen random point generation method is indeed non-uniform.

\subsection{Vector quantization: adaptive neural gas}
Many well-known algorithms for surface reconstruction work considerably better when the input point cloud is as close as possible to a uniform sample of the target surface and are often hampered when this is not the case.\footnote{More precisely, the relevant requisite in this respect is that the point sampling should be at least \emph{locally uniform} \cite{dey:CurveandSurfaceReconstruction}.} Apart from greater time complexity, these difficulties can lead in practice to the need for accurate verification of results and possibly to manual post-processing, to correct imperfections.

The intended purpose of a vector quantization algorithm in this context is to obtain both an improvement in the uniformity of sampling and a quantitative reduction in the number of points to be used for surface reconstruction. The algorithm of choice is an adaptive variant of the \emph{neural gas} (NG) algorithm \cite{Martinetz-etal93} and works as follows:
\begin{enumerate}
  \item initialize $\bm{W}$ with a pre-defined number $k$ of reference vectors $\bm{w}_i$ positioned at random on $\partial\excs$;
  \item generate a random point $\bm{p}$ from $\partial\excs$;
  \item find the nearest reference vector in $\bm{W}$, i.e. $\bm{w}_i  := \argmin_{\bm{w}_j\in\bm{W}} \|\bm{p} - \bm{w}_j\|$;
  \item if $\|\bm{p} - \bm{w}_i\| \le r$, where $r$ is a fixed threshold, adapt all reference vectors in $\bm{W}$ by
  $$\Delta \bm{w}_{i} = \varepsilon\cdot h_{\lambda}(k_i(\bm{p}))(\bm{p} - \bm{w}_{i})$$
  where $k_i(\bm{p}) := \#\{\bm{w}_{j} : \| \bm{p} - \bm{w}_{j} \| < \| \bm{p} - \bm{w}_{i} \|\}$ ($\#$ denotes cardinality), $\varepsilon > 0$ is a real parameter and
  $$h_{0}(n) := \delta_{0n} \quad\text{and}\quad h_{\lambda}(n) := e^{-\frac{n}{\lambda}}, \text{ for } \lambda > 0;$$
  \item otherwise, if $\bm{p}$ is farther away from $\bm{w}_i$, add a new reference vector $\bm{p}$ to $\bm{W}$;
  \item unless a maximum number of iterations $T$ has been reached, return to step 2.
\end{enumerate}

As evident from step 5, this algorithm is adaptive in the number of reference vectors in $\bm{W}$; in particular, this means that the level of refinement of the sampling of $\partial\excs$ provided by $\bm{W}$ can be controlled through the value of the fundamental threshold $r$.

In \cite{Martinetz-etal93} it is proven that, when the value of the constant $\varepsilon$ tends to $0$ as the iterations progress, the NG algorithm performs a \emph{stochastic gradient descent} towards a (local) minimum of an overall  cost function and that its configuration tends to obey the power law
\begin{equation*}
\rho(\bm{w}) \propto P(\bm{w})^{\gamma}\quad\text{with}\quad\gamma := \frax{d}{d+2},
\end{equation*}
where $d$ is the dimension of the input space being sampled, that is $d=2$ in this case. Here $\rho(\bm{w})$ is the density of reference vectors in $\bm{W}$ at $\bm{w}$ and $P(\bm{w})$ is the sampling probability. Since the exponent $\gamma$ is smaller than $1$, the overall configuration of $\bm{W}$ tends to be closer to uniformity than the sampling probability $P$. This effect is clearly visible in Fig.~\ref{fig:method-overview}(b).

\subsection{Surface reconstruction}

\begin{figure}[h]
\centering
  \subfigure[]{\includegraphics[width=0.16\linewidth]{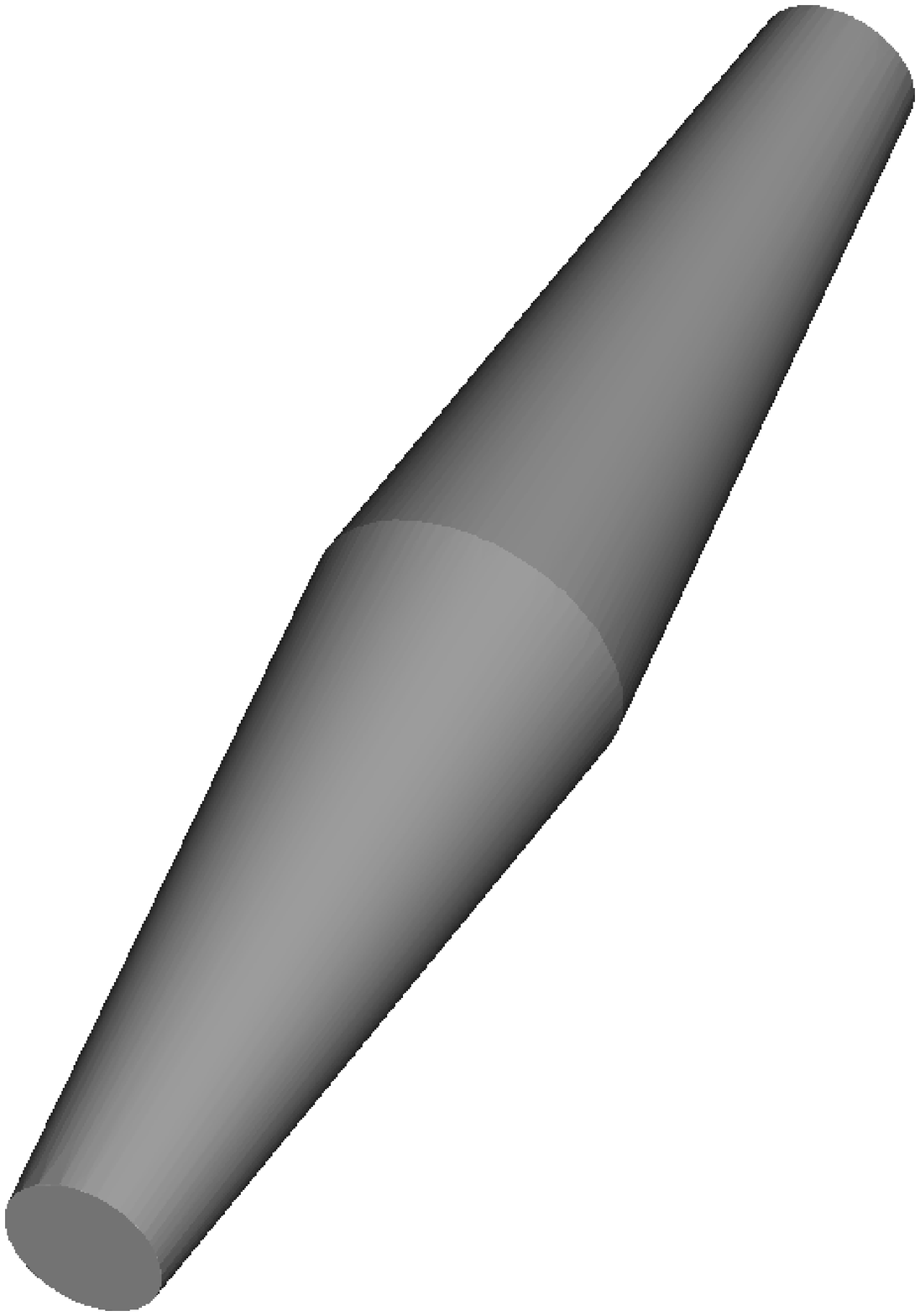}}
  \subfigure[]{\includegraphics[width=0.16\linewidth]{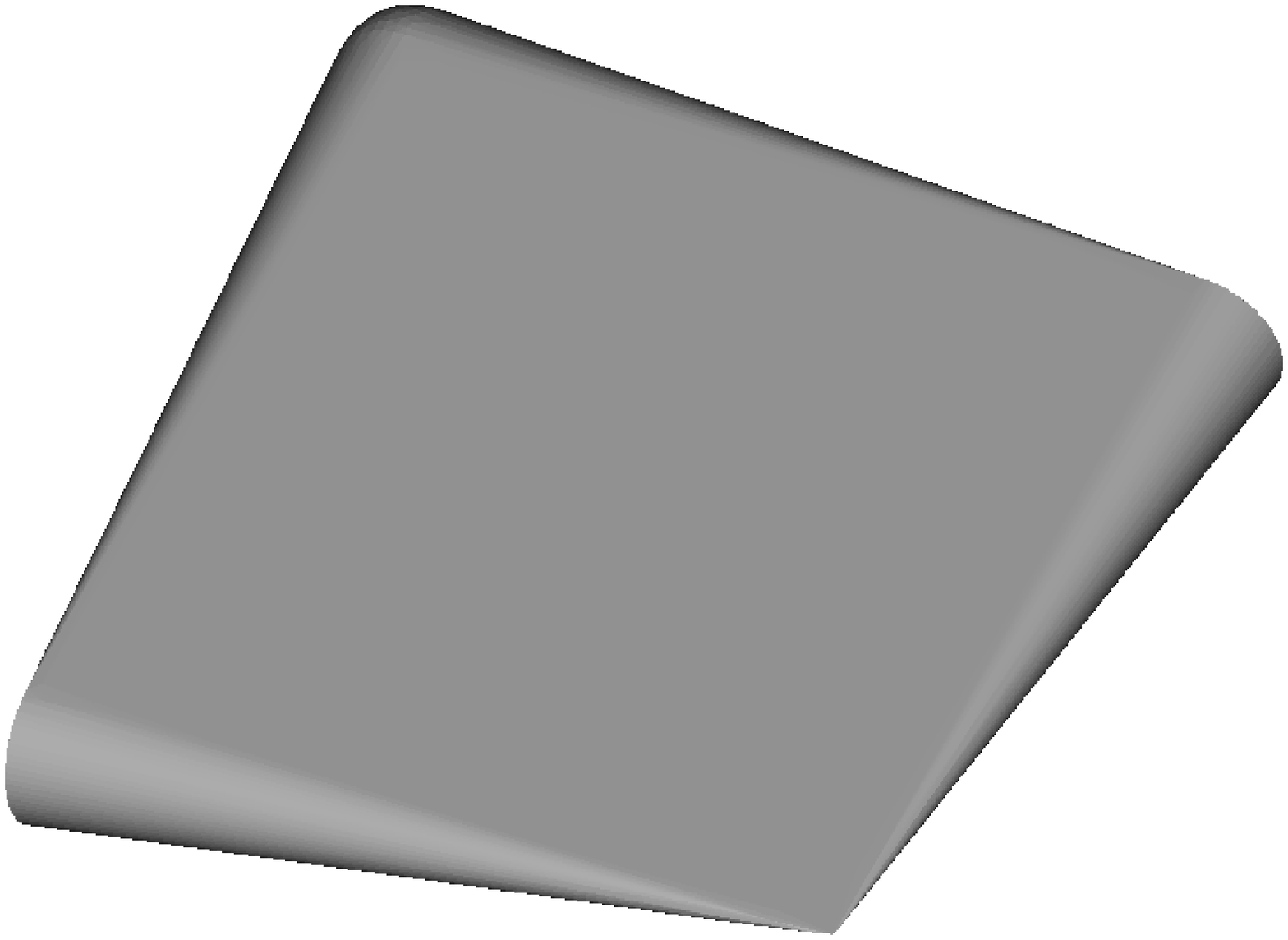}}
  \subfigure[]{\includegraphics[width=0.16\linewidth]{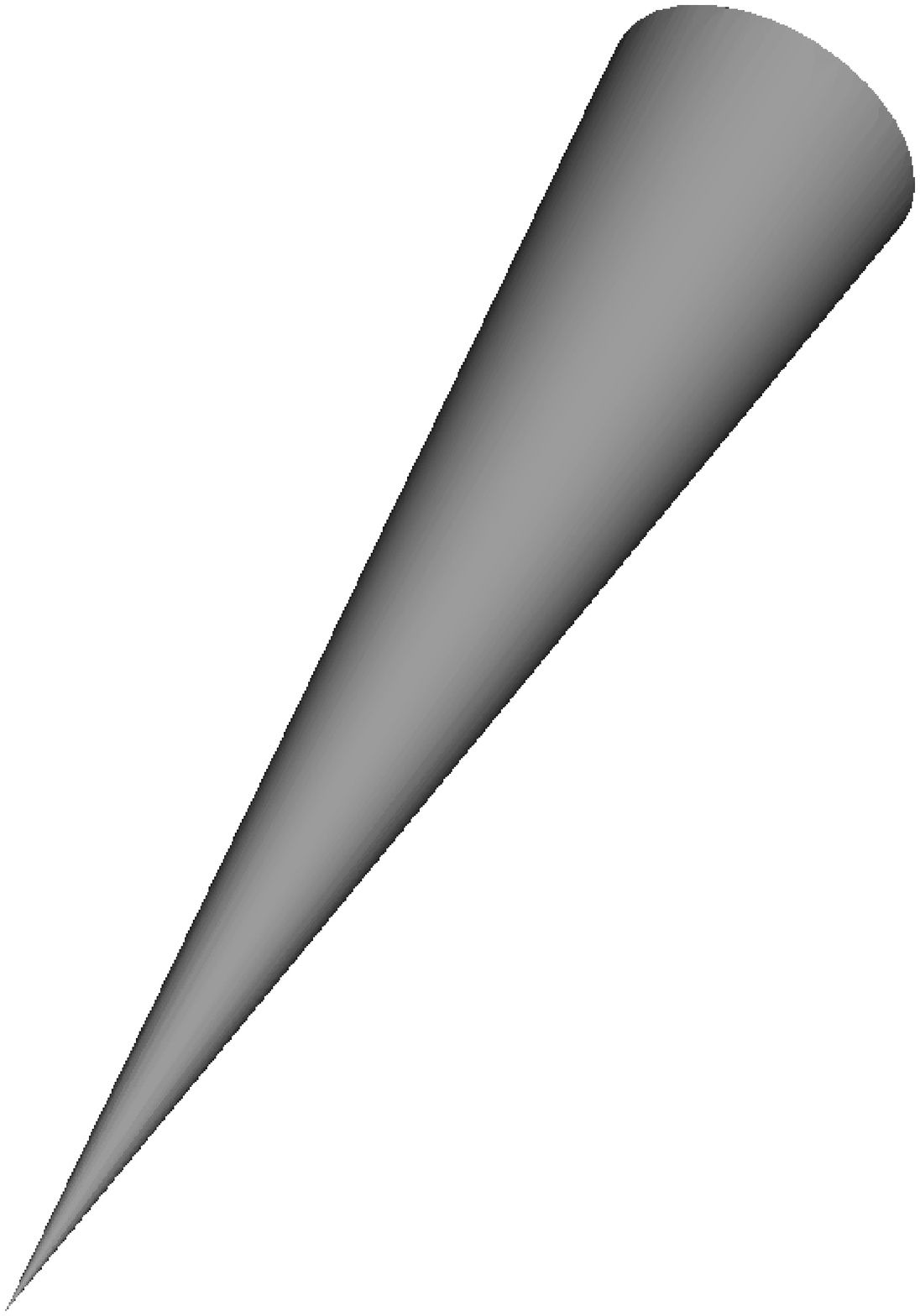}}\\
  \subfigure[]{\includegraphics[width=0.16\linewidth]{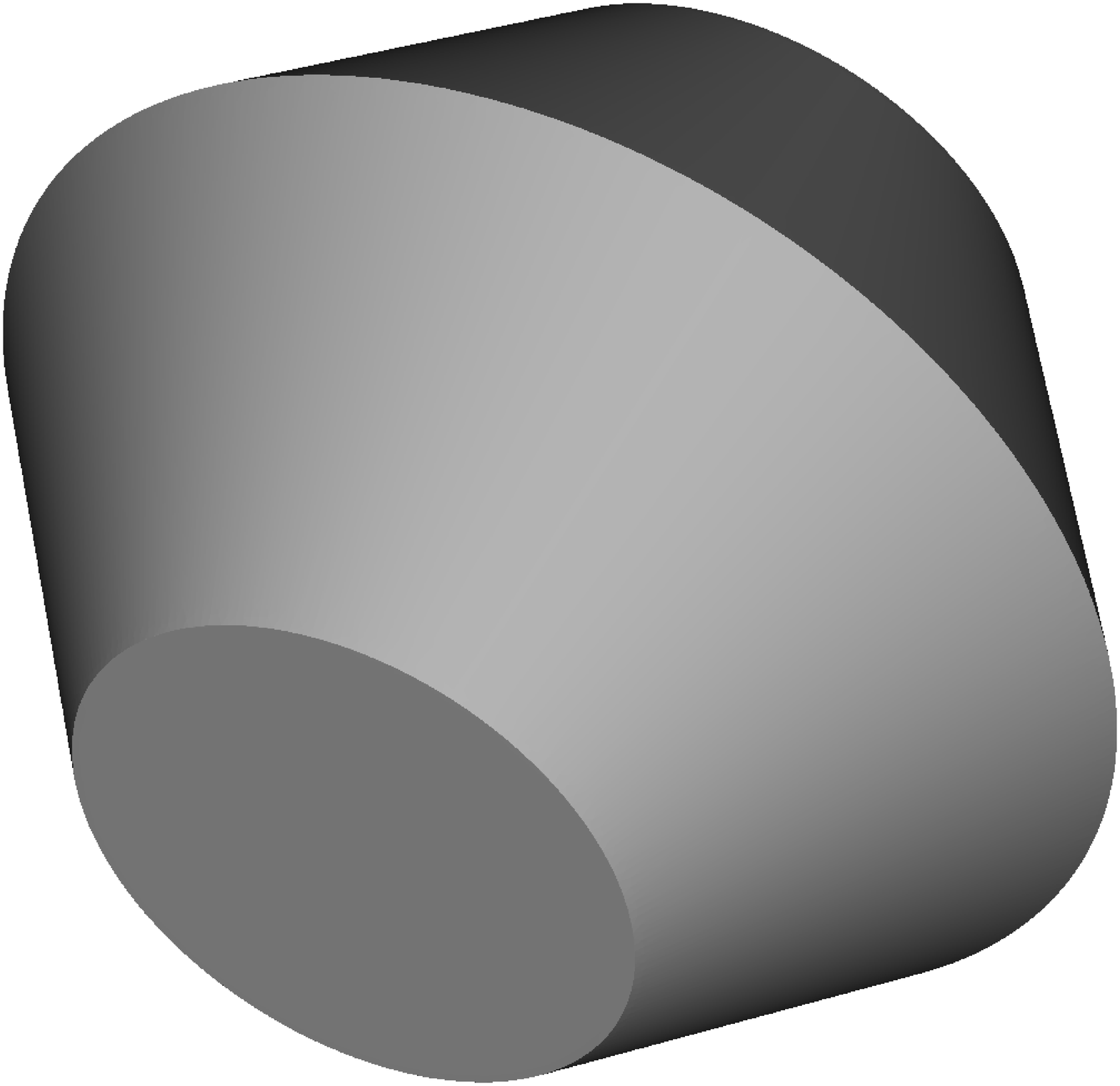}}
  \subfigure[]{\includegraphics[width=0.16\linewidth]{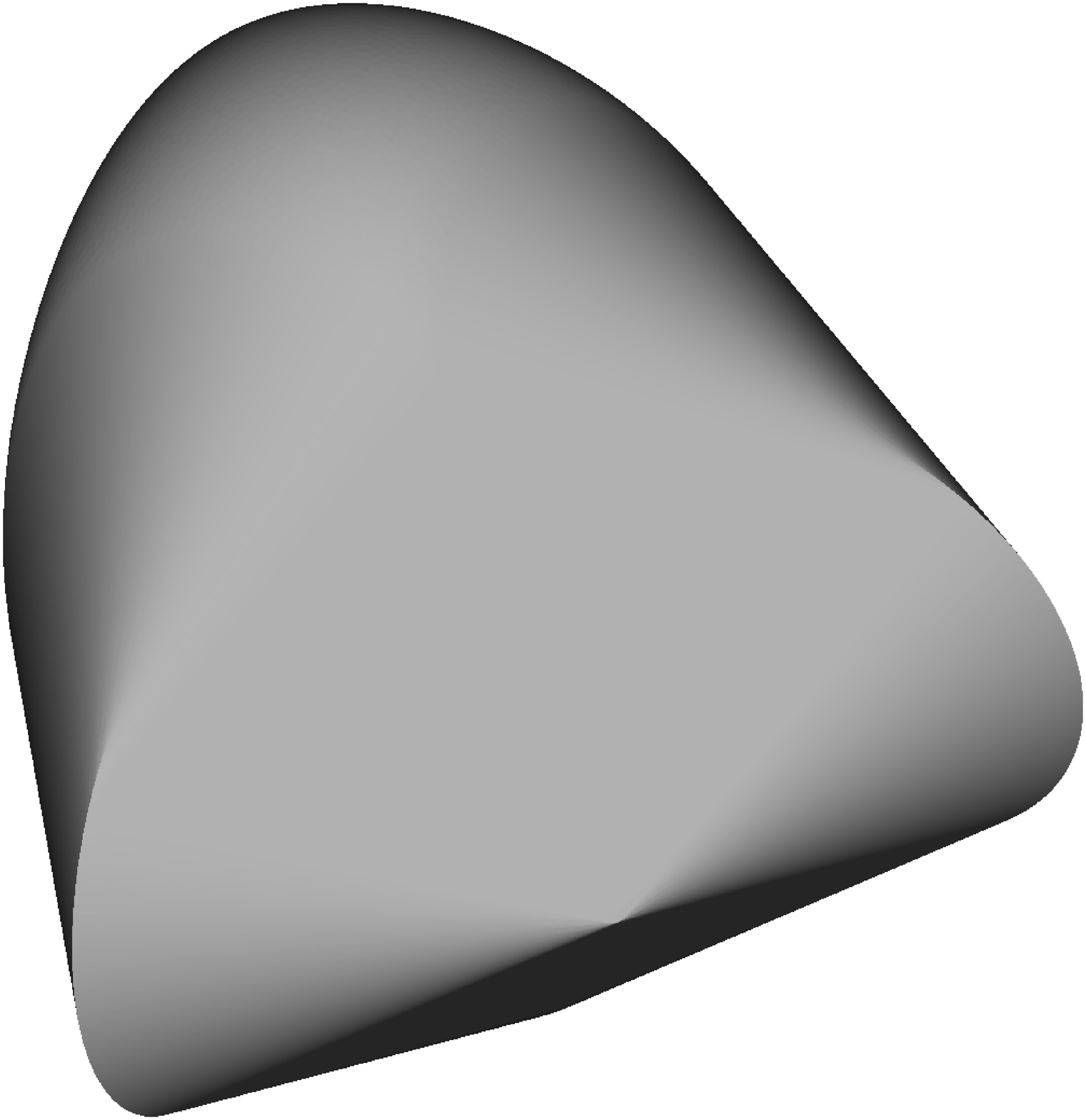}}
  \subfigure[]{\includegraphics[width=0.16\linewidth]{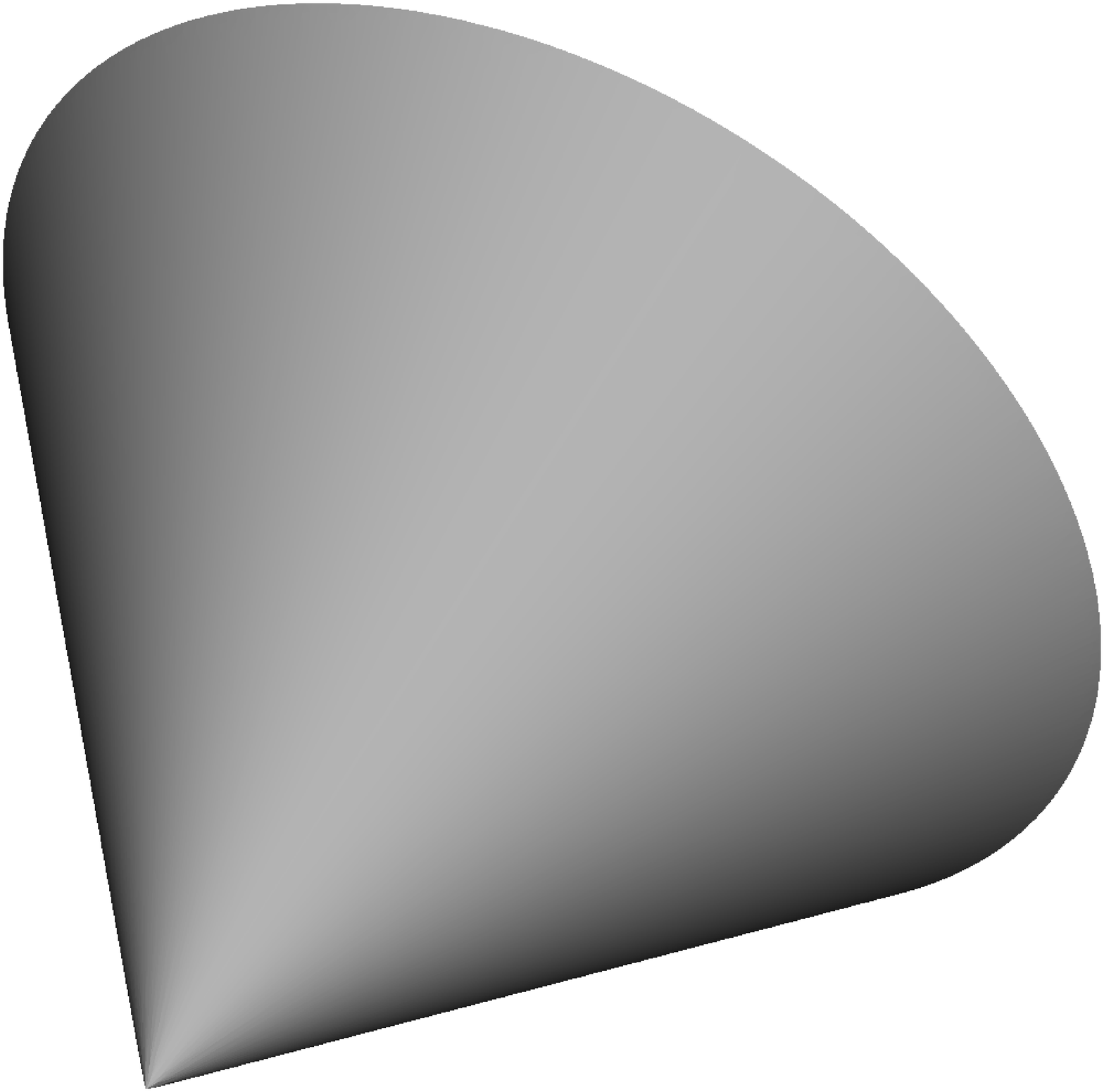}}\\
  \subfigure[]{\includegraphics[width=0.16\linewidth]{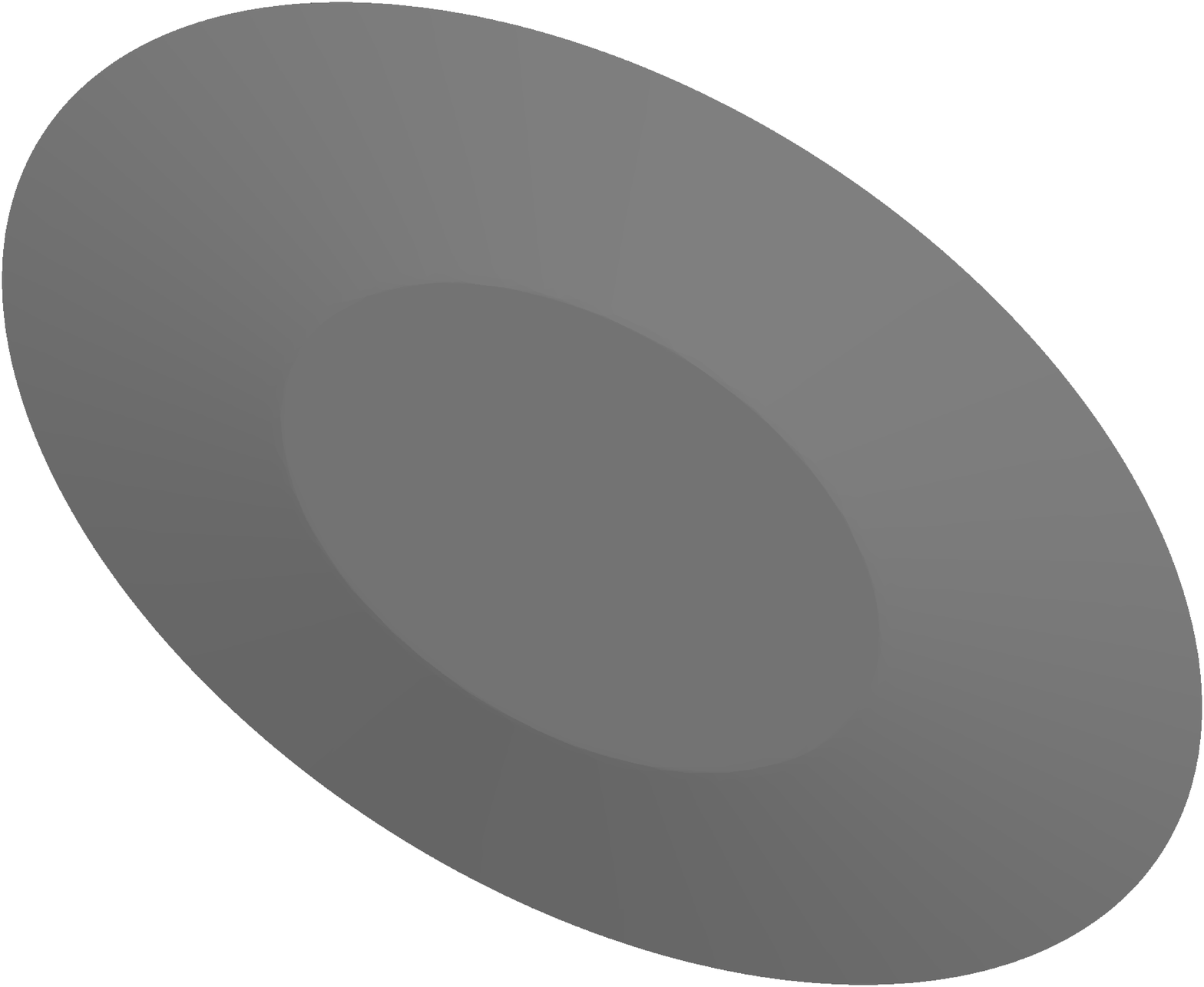}}
  \subfigure[]{\includegraphics[width=0.16\linewidth]{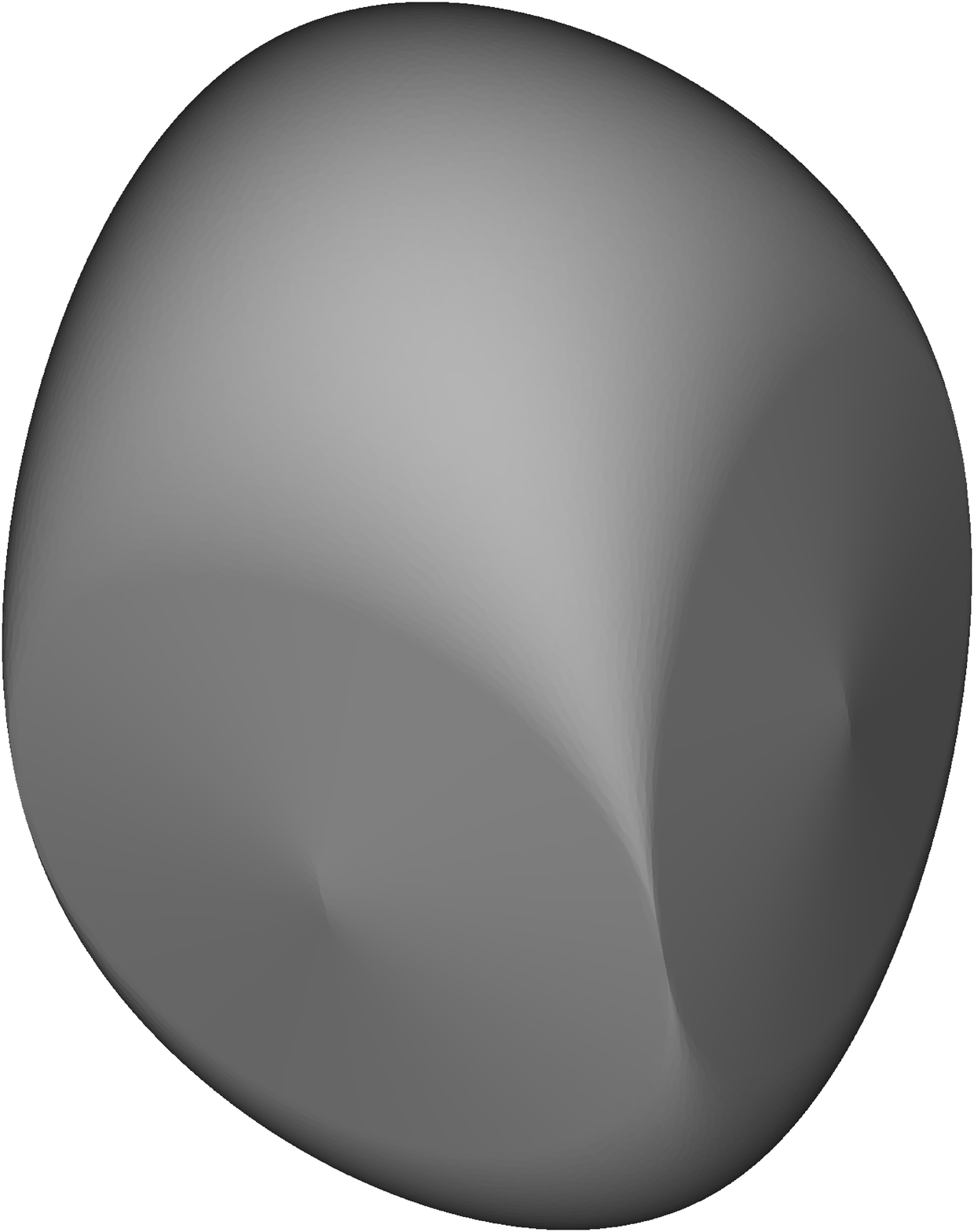}}
  \subfigure[]{\includegraphics[width=0.16\linewidth]{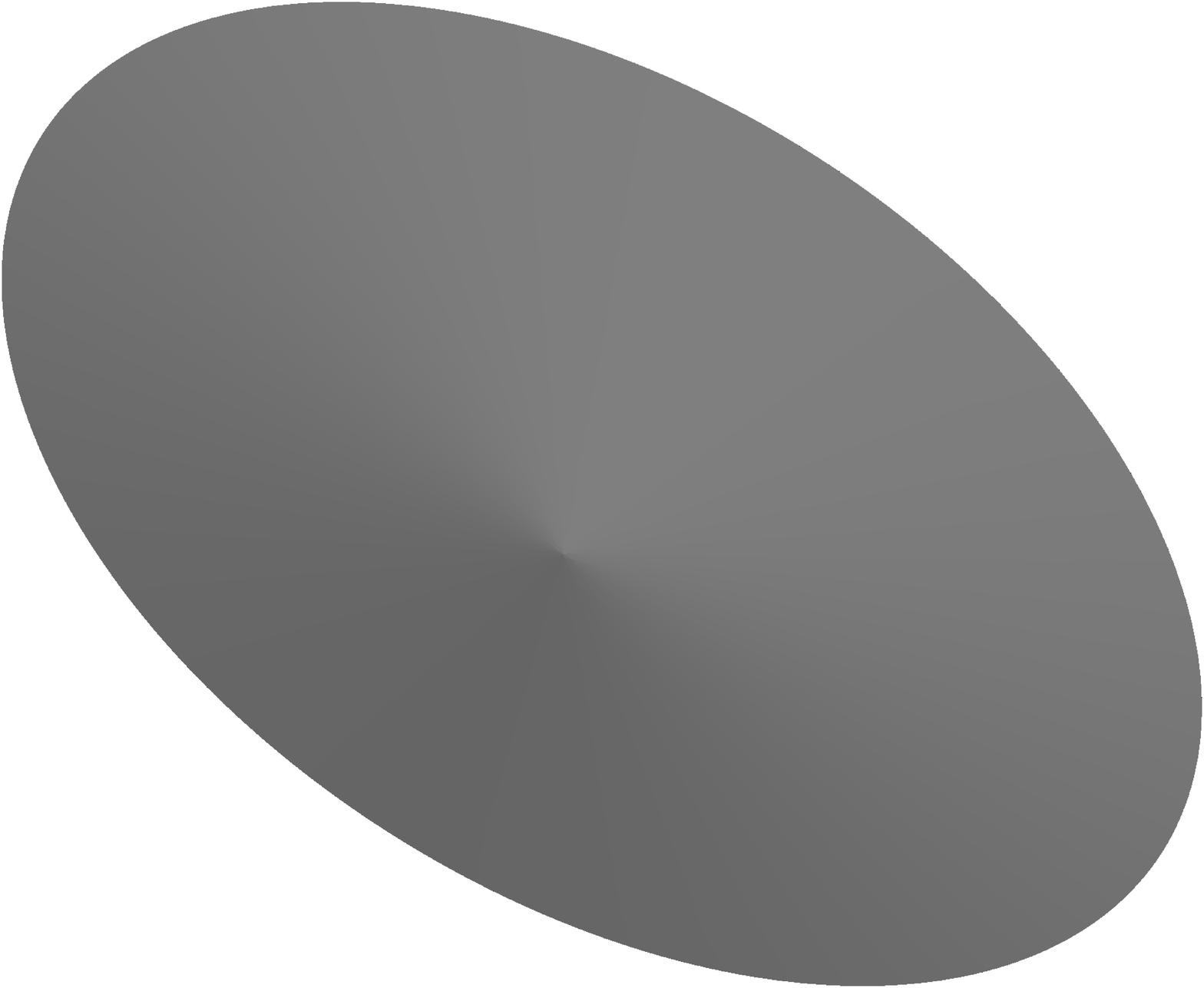}}
\caption{Gallery of reconstructed boundaries $\partial\excs$ for pairs of congruent circular cones of semi-amplitude $\alpha$. Rows correspond to values of $\alpha$ equal to $\frac{\pi}{32}$, $\frac{\pi}{6}$ and $\frac{15}{32}\pi$, respectively, while columns correspond to values of the angle $\vartheta$ between the symmetry axes $\bm{m}_1$ and $\bm{m}_2$ equal to $0$, $\frac{\pi}{2}$ and $\pi$, respectively. All figures are in the same scale and frame of reference.}
\label{fig:shape-gallery}
\end{figure}

With proper parameter settings (see below), the reconstruction of a triangular mesh from the final configuration $\bm{W}$ produced by the adaptive NG algorithm poses no particular problem and could be performed in full automation. In this work we used the ball-pivoting algorithm \cite{bernardini1999ball} which joins in a triangular face any three vectors in $\bm{W}$ whose ends are touched by a ball of a given radius $r$ that does not contain any other vector's end from the same set. One example of the results of this procedure is shown in Figs.~\ref{fig:method-overview}(c) and (d). Further examples are shown in Fig.~\ref{fig:shape-gallery}, which contains a gallery of shapes produced with the method described above.

\subsection{Implementation and benchmark}

The adaptive NG vector quantization algorithm, together with the generator of random points from $\partial\excs$, has been implemented in Java. In order to speed the execution up, the algorithm has been converted to a multi-threaded version suitable for multi-core computers, along the lines described in \cite{parigi2014multi}. For surface reconstruction, we used the implementation of the ball-pivoting algorithm included in the Meshlab open-source tool \cite{cignoni2008meshlab}.

The overall method for shape reconstruction was validated using Minkowski's formula for isotropic volume average \eqref{eq:isotropic_volume_average} together with the cone-specific functionals \eqref{eq:cone_classical_Minkowski_functionals}. For benchmarking, pairs of congruent circular cones $\coneu$ and $\conet$ having slant height $L$ and semi-amplitude $\alpha$ varying from $\frac{\pi}{32}$ to $\frac{15}{32}\pi$ with step $\frac{\pi}{32}$ were considered. For each such pair, the value of $V[\excs]$ was computed for angles $\vartheta$ between the two symmetry axes $\bm{m}_1$ and $\bm{m}_2$ varying from $0$ to $\pi$ with step $\frac{\pi}{32}$; the isotropic average of the resulting sequence of volumes was then computed and compared with the exact value of $\langle V \rangle [\excs]$. The fundamental threshold $r$, which governs the density of reference vectors in $\bm{W}$ with respect to $\partial\excs$, was determined empirically with the objective of having a difference lesser than $0.02\%$ between the exact value of each isotropic average and the corresponding value computed numerically. A value $r=\frac{1}{50}L$ was found to be adequate (see also the comparative plots in Figs.~\ref{fig:B_Zero}(a) and (b)). Also the value of $T=120\,\mathrm{M}$ maximum equivalent iterations of the NG algorithm was determined empirically. In the actual experiments, the execution was split into 4 concurrent threads, each processing in multi-signal mode (see \cite{parigi2014multi}) 250 random points per iteration.  Being dependent on the area of $\partial\excs$, the number of reference vectors in the final configurations of $\bm{W}$ varied greatly, from $3,\!592$ to $41,\!689$.

All numerical experiments were run on a workstation based on an Intel$^\circledR$ Xeon$^\circledR$ CPU E3-1240 v3, $3.4\,\mathrm{GHz}$ CPU with $8\,\mathrm{GB}$ of RAM. As for computing times, the most demanding part of the method is running the $T=120\,\mathrm{M}$ equivalent iterations of the adaptive NG algorithm. For each pair of cones and for each pose, with the precision required, this computation took on average about  $4,\!254$ seconds (i.e. about $71$ minutes) to complete.


\begin{thebibliography}{64}
\expandafter\ifx\csname natexlab\endcsname\relax\def\natexlab#1{#1}\fi
\expandafter\ifx\csname bibnamefont\endcsname\relax
  \def\bibnamefont#1{#1}\fi
\expandafter\ifx\csname bibfnamefont\endcsname\relax
  \def\bibfnamefont#1{#1}\fi
\expandafter\ifx\csname citenamefont\endcsname\relax
  \def\citenamefont#1{#1}\fi
\expandafter\ifx\csname url\endcsname\relax
  \def\url#1{\texttt{#1}}\fi
\expandafter\ifx\csname urlprefix\endcsname\relax\def\urlprefix{URL }\fi
\providecommand{\bibinfo}[2]{#2}
\providecommand{\eprint}[2][]{\url{#2}}

\bibitem[{\citenamefont{Onsager}(1949)}]{onsager:effects}
\bibinfo{author}{\bibfnamefont{L.}~\bibnamefont{Onsager}},
  \bibinfo{journal}{Ann. N.Y. Acad. Sci.} \textbf{\bibinfo{volume}{51}},
  \bibinfo{pages}{627} (\bibinfo{year}{1949}), \bibinfo{note}{reprinted in
  \cite{sluckin:crystal}, pp.~625--657.}

\bibitem[{\citenamefont{Frenkel}(2000)}]{frenkel:perspective}
\bibinfo{author}{\bibfnamefont{D.}~\bibnamefont{Frenkel}},
  \bibinfo{journal}{Theor. Chem. Acc.} \textbf{\bibinfo{volume}{103}},
  \bibinfo{pages}{212} (\bibinfo{year}{2000}).

\bibitem[{\citenamefont{Frenkel}(1999)}]{frenkel:entropy}
\bibinfo{author}{\bibfnamefont{D.}~\bibnamefont{Frenkel}},
  \bibinfo{journal}{Physica A} \textbf{\bibinfo{volume}{263}},
  \bibinfo{pages}{26} (\bibinfo{year}{1999}), \bibinfo{note}{{P}roceedings of
  the 20th IUPAP International Conference on Statistical Physics}.

\bibitem[{\citenamefont{Zheng and Palffy-Muhoray}(2007)}]{palffy:distance_2D}
\bibinfo{author}{\bibfnamefont{X.}~\bibnamefont{Zheng}} \bibnamefont{and}
  \bibinfo{author}{\bibfnamefont{P.}~\bibnamefont{Palffy-Muhoray}},
  \bibinfo{journal}{Phys. Rev. E} \textbf{\bibinfo{volume}{75}},
  \bibinfo{pages}{061709} (\bibinfo{year}{2007}).

\bibitem[{\citenamefont{Vieillard-Baron}(1972)}]{vieillard-baron:phase}
\bibinfo{author}{\bibfnamefont{J.}~\bibnamefont{Vieillard-Baron}},
  \bibinfo{journal}{J. Chem. Phys.} \textbf{\bibinfo{volume}{56}},
  \bibinfo{pages}{4729} (\bibinfo{year}{1972}).

\bibitem[{\citenamefont{Vroege and Lekkerkerker}(1992)}]{vroege:phase}
\bibinfo{author}{\bibfnamefont{G.~J.} \bibnamefont{Vroege}} \bibnamefont{and}
  \bibinfo{author}{\bibfnamefont{H.~N.~W.} \bibnamefont{Lekkerkerker}},
  \bibinfo{journal}{Rep. Prog. Phys.} \textbf{\bibinfo{volume}{55}},
  \bibinfo{pages}{1241} (\bibinfo{year}{1992}).

\bibitem[{\citenamefont{Mulder}(1986)}]{mulder:solution}
\bibinfo{author}{\bibfnamefont{B.~M.} \bibnamefont{Mulder}},
  \bibinfo{journal}{Liq. Crystals} \textbf{\bibinfo{volume}{1}},
  \bibinfo{pages}{539} (\bibinfo{year}{1986}).

\bibitem[{\citenamefont{Mulder}(2005)}]{mulder:excluded}
\bibinfo{author}{\bibfnamefont{B.~M.} \bibnamefont{Mulder}},
  \bibinfo{journal}{Mol. Phys.} \textbf{\bibinfo{volume}{103}},
  \bibinfo{pages}{1411} (\bibinfo{year}{2005}).

\bibitem[{\citenamefont{Isihara}(1951)}]{isihara:theory}
\bibinfo{author}{\bibfnamefont{A.}~\bibnamefont{Isihara}}, \bibinfo{journal}{J.
  Chem. Phys.} \textbf{\bibinfo{volume}{19}}, \bibinfo{pages}{1142}
  (\bibinfo{year}{1951}).

\bibitem[{\citenamefont{Mederos et~al.}(2014)\citenamefont{Mederos, Velasco,
  and Mart\'{\i}nez-Rat\'on}}]{mederos:hard-body}
\bibinfo{author}{\bibfnamefont{L.}~\bibnamefont{Mederos}},
  \bibinfo{author}{\bibfnamefont{E.}~\bibnamefont{Velasco}}, \bibnamefont{and}
  \bibinfo{author}{\bibfnamefont{Y.}~\bibnamefont{Mart\'{\i}nez-Rat\'on}},
  \bibinfo{journal}{J. Phys.: Condens. Matter} \textbf{\bibinfo{volume}{26}},
  \bibinfo{pages}{463101} (\bibinfo{year}{2014}).

\bibitem[{\citenamefont{Hansen and McDonald}(2013)}]{hansen:theory}
\bibinfo{author}{\bibfnamefont{J.-P.} \bibnamefont{Hansen}} \bibnamefont{and}
  \bibinfo{author}{\bibfnamefont{I.}~\bibnamefont{McDonald}},
  \emph{\bibinfo{title}{Theory of Simple Liquids}}
  (\bibinfo{publisher}{Academic Press}, \bibinfo{address}{Oxford},
  \bibinfo{year}{2013}), \bibinfo{edition}{4th} ed., \bibinfo{note}{with
  Applications to {S}oft {M}atter}.

\bibitem[{\citenamefont{Frenkel}(1987)}]{frenkel:onsager}
\bibinfo{author}{\bibfnamefont{D.}~\bibnamefont{Frenkel}}, \bibinfo{journal}{J.
  Phys. Chem.} \textbf{\bibinfo{volume}{91}}, \bibinfo{pages}{4912}
  (\bibinfo{year}{1987}).

\bibitem[{\citenamefont{Frenkel}(1988)}]{frenkel:onsager_erratum}
\bibinfo{author}{\bibfnamefont{D.}~\bibnamefont{Frenkel}}, \bibinfo{journal}{J.
  Phys. Chem.} \textbf{\bibinfo{volume}{92}}, \bibinfo{pages}{5314}
  (\bibinfo{year}{1988}).

\bibitem[{\citenamefont{Brunn}(1887)}]{brunn:thesis}
\bibinfo{author}{\bibfnamefont{H.}~\bibnamefont{Brunn}},
  \emph{\bibinfo{title}{\"{U}ber {O}vale und {E}ifl\"{a}chen}},
  \bibinfo{address}{M\"{u}nchen} (\bibinfo{year}{1887}).

\bibitem[{\citenamefont{Minkowski}(1903)}]{minkowski:volumen}
\bibinfo{author}{\bibfnamefont{H.}~\bibnamefont{Minkowski}},
  \bibinfo{journal}{Math. Ann.} \textbf{\bibinfo{volume}{57}},
  \bibinfo{pages}{447} (\bibinfo{year}{1903}).

\bibitem[{\citenamefont{Bonnesen and Fenchel}(1987)}]{bonnesen:theory}
\bibinfo{author}{\bibfnamefont{T.}~\bibnamefont{Bonnesen}} \bibnamefont{and}
  \bibinfo{author}{\bibfnamefont{W.}~\bibnamefont{Fenchel}},
  \emph{\bibinfo{title}{Theory of Convex Bodies}} (\bibinfo{publisher}{BCS
  Associates}, \bibinfo{address}{Moscow, Idaho, USA}, \bibinfo{year}{1987}),
  \bibinfo{note}{translated from the German \textit{Theorie der konvexen
  K\"{o}rper} (Springer, Berlin, 1934) and edited by L. Boron, C. Christenson,
  and B. Smith, with the collaboration of W. Fenchel.}

\bibitem[{\citenamefont{Schneider}(1993)}]{schneider:convex}
\bibinfo{author}{\bibfnamefont{R.}~\bibnamefont{Schneider}},
  \emph{\bibinfo{title}{Convex Bodies: The Brunn-Minkowski Theory}},
  vol.~\bibinfo{volume}{44} of \emph{\bibinfo{series}{Encyclopedia of
  Mathematics and its Applications}} (\bibinfo{publisher}{Cambridge University
  Press}, \bibinfo{address}{Cambridge}, \bibinfo{year}{1993}).

\bibitem[{\citenamefont{Piastra and Virga}(2013)}]{piastra:octupolar}
\bibinfo{author}{\bibfnamefont{M.}~\bibnamefont{Piastra}} \bibnamefont{and}
  \bibinfo{author}{\bibfnamefont{E.~G.} \bibnamefont{Virga}},
  \bibinfo{journal}{Phys. Rev. E} \textbf{\bibinfo{volume}{88}},
  \bibinfo{pages}{032507} (\bibinfo{year}{2013}).

\bibitem[{\citenamefont{Singh and Kumar}(2001)}]{singh:molecular}
\bibinfo{author}{\bibfnamefont{G.}~\bibnamefont{Singh}} \bibnamefont{and}
  \bibinfo{author}{\bibfnamefont{B.}~\bibnamefont{Kumar}},
  \bibinfo{journal}{Ann. Phys.} \textbf{\bibinfo{volume}{294}},
  \bibinfo{pages}{24} (\bibinfo{year}{2001}), ISSN \bibinfo{issn}{0003-4916}.

\bibitem[{\citenamefont{Gurtin et~al.}(2010)\citenamefont{Gurtin, Fried, and
  Anand}}]{gurtin:mechanics}
\bibinfo{author}{\bibfnamefont{M.~E.} \bibnamefont{Gurtin}},
  \bibinfo{author}{\bibfnamefont{E.}~\bibnamefont{Fried}}, \bibnamefont{and}
  \bibinfo{author}{\bibfnamefont{L.}~\bibnamefont{Anand}},
  \emph{\bibinfo{title}{The Mechanics and Thermodynamics of Continua}}
  (\bibinfo{publisher}{Cambridge University Press},
  \bibinfo{address}{Cambridge}, \bibinfo{year}{2010}).

\bibitem[{{\relax DLMF}()}]{NIST:DLMF}
{\relax DLMF}, \emph{\bibinfo{title}{{NIST Digital Library of Mathematical
  Functions}}}, \bibinfo{howpublished}{\url{http://dlmf.nist.gov/}, Release
  1.0.6 of 2013-05-06}, \bibinfo{note}{online companion to
  \cite{Olver:2010:NHMF}}.

\bibitem[{\citenamefont{Gradshteyn and Ryzhik}(1980)}]{gradshteyn:table}
\bibinfo{author}{\bibfnamefont{I.~S.} \bibnamefont{Gradshteyn}}
  \bibnamefont{and} \bibinfo{author}{\bibfnamefont{I.~M.}
  \bibnamefont{Ryzhik}}, \emph{\bibinfo{title}{Table of {I}ntegrals, {S}eries,
  and {P}roducts}} (\bibinfo{publisher}{Academic Press}, \bibinfo{address}{New
  York}, \bibinfo{year}{1980}), \bibinfo{note}{corrected and enlarged edition
  prepared by A. Jeffrey}.

\bibitem[{\citenamefont{Kihara}(1953{\natexlab{a}})}]{kihara:coefficients}
\bibinfo{author}{\bibfnamefont{T.}~\bibnamefont{Kihara}},
  \bibinfo{journal}{Rev. Mod. Phys.} \textbf{\bibinfo{volume}{25}},
  \bibinfo{pages}{831} (\bibinfo{year}{1953}{\natexlab{a}}).

\bibitem[{\citenamefont{Kihara}(1953{\natexlab{b}})}]{kihara:isihara}
\bibinfo{author}{\bibfnamefont{T.}~\bibnamefont{Kihara}}, \bibinfo{journal}{J.
  Phys. Soc. Japan} \textbf{\bibinfo{volume}{8}}, \bibinfo{pages}{686}
  (\bibinfo{year}{1953}{\natexlab{b}}).

\bibitem[{\citenamefont{Isihara}(1950)}]{isihara:determination}
\bibinfo{author}{\bibfnamefont{A.}~\bibnamefont{Isihara}}, \bibinfo{journal}{J.
  Chem. Phys.} \textbf{\bibinfo{volume}{18}}, \bibinfo{pages}{1446}
  (\bibinfo{year}{1950}).

\bibitem[{\citenamefont{Isihara and
  Hayashida}(1951{\natexlab{a}})}]{isihara:theory_I}
\bibinfo{author}{\bibfnamefont{A.}~\bibnamefont{Isihara}} \bibnamefont{and}
  \bibinfo{author}{\bibfnamefont{T.}~\bibnamefont{Hayashida}},
  \bibinfo{journal}{J. Phys. Soc. Japan} \textbf{\bibinfo{volume}{6}},
  \bibinfo{pages}{40} (\bibinfo{year}{1951}{\natexlab{a}}).

\bibitem[{\citenamefont{Isihara and
  Hayashida}(1951{\natexlab{b}})}]{isihara:theory_II}
\bibinfo{author}{\bibfnamefont{A.}~\bibnamefont{Isihara}} \bibnamefont{and}
  \bibinfo{author}{\bibfnamefont{T.}~\bibnamefont{Hayashida}},
  \bibinfo{journal}{J. Phys. Soc. Japan} \textbf{\bibinfo{volume}{6}},
  \bibinfo{pages}{46} (\bibinfo{year}{1951}{\natexlab{b}}).

\bibitem[{\citenamefont{Palffy-Muhoray
  et~al.}(2014)\citenamefont{Palffy-Muhoray, Virga, and
  Zheng}}]{palffy:minimum}
\bibinfo{author}{\bibfnamefont{P.}~\bibnamefont{Palffy-Muhoray}},
  \bibinfo{author}{\bibfnamefont{E.~G.} \bibnamefont{Virga}}, \bibnamefont{and}
  \bibinfo{author}{\bibfnamefont{X.}~\bibnamefont{Zheng}}, \bibinfo{journal}{J.
  Phys. A: Math. Theor.} \textbf{\bibinfo{volume}{47}}, \bibinfo{pages}{415205}
  (\bibinfo{year}{2014}).

\bibitem[{\citenamefont{Rosenfeld}(1994)}]{rosenfeld:desnity}
\bibinfo{author}{\bibfnamefont{Y.}~\bibnamefont{Rosenfeld}},
  \bibinfo{journal}{Phys. Rev. E} \textbf{\bibinfo{volume}{50}},
  \bibinfo{pages}{R3318} (\bibinfo{year}{1994}).

\bibitem[{\citenamefont{Hansen-Goos and Mecke}(2009)}]{hansen-goos:fundamental}
\bibinfo{author}{\bibfnamefont{H.}~\bibnamefont{Hansen-Goos}} \bibnamefont{and}
  \bibinfo{author}{\bibfnamefont{K.}~\bibnamefont{Mecke}},
  \bibinfo{journal}{Phys. Rev. Lett.} \textbf{\bibinfo{volume}{102}},
  \bibinfo{pages}{018302} (\bibinfo{year}{2009}).

\bibitem[{\citenamefont{Tjipto-Margo and Evans}(1991)}]{tjipto-margo:onsager}
\bibinfo{author}{\bibfnamefont{B.}~\bibnamefont{Tjipto-Margo}}
  \bibnamefont{and} \bibinfo{author}{\bibfnamefont{G.~T.} \bibnamefont{Evans}},
  \bibinfo{journal}{J. Chem. Phys.} \textbf{\bibinfo{volume}{94}},
  \bibinfo{pages}{4546} (\bibinfo{year}{1991}).

\bibitem[{\citenamefont{Ho{\l}yst and Poniewierski}(1990)}]{holyst:study}
\bibinfo{author}{\bibfnamefont{R.}~\bibnamefont{Ho{\l}yst}} \bibnamefont{and}
  \bibinfo{author}{\bibfnamefont{A.}~\bibnamefont{Poniewierski}},
  \bibinfo{journal}{Mol. Phys.} \textbf{\bibinfo{volume}{69}},
  \bibinfo{pages}{193} (\bibinfo{year}{1990}).

\bibitem[{\citenamefont{Mulder}(1989)}]{mulder:isotropic}
\bibinfo{author}{\bibfnamefont{B.}~\bibnamefont{Mulder}},
  \bibinfo{journal}{Phys. Rev. A} \textbf{\bibinfo{volume}{39}},
  \bibinfo{pages}{360} (\bibinfo{year}{1989}).

\bibitem[{\citenamefont{Rigby}(1989)}]{rigby:hard}
\bibinfo{author}{\bibfnamefont{M.}~\bibnamefont{Rigby}}, \bibinfo{journal}{Mol.
  Phys.} \textbf{\bibinfo{volume}{66}}, \bibinfo{pages}{1261}
  (\bibinfo{year}{1989}).

\bibitem[{\citenamefont{Ogston and Winzor}(1975)}]{ogston:treatment}
\bibinfo{author}{\bibfnamefont{A.~G.} \bibnamefont{Ogston}} \bibnamefont{and}
  \bibinfo{author}{\bibfnamefont{D.~J.} \bibnamefont{Winzor}},
  \bibinfo{journal}{J. Phys. Chem.} \textbf{\bibinfo{volume}{79}},
  \bibinfo{pages}{2496} (\bibinfo{year}{1975}).

\bibitem[{\citenamefont{Ambrosetti et~al.}(2008)\citenamefont{Ambrosetti,
  Johner, Grimaldi, Danani, and Ryser}}]{ambrosetti:percolative}
\bibinfo{author}{\bibfnamefont{G.}~\bibnamefont{Ambrosetti}},
  \bibinfo{author}{\bibfnamefont{N.}~\bibnamefont{Johner}},
  \bibinfo{author}{\bibfnamefont{C.}~\bibnamefont{Grimaldi}},
  \bibinfo{author}{\bibfnamefont{A.}~\bibnamefont{Danani}}, \bibnamefont{and}
  \bibinfo{author}{\bibfnamefont{P.}~\bibnamefont{Ryser}},
  \bibinfo{journal}{Phys. Rev. E} \textbf{\bibinfo{volume}{78}},
  \bibinfo{pages}{061126} (\bibinfo{year}{2008}).

\bibitem[{\citenamefont{Masters}(2008)}]{masters:virial}
\bibinfo{author}{\bibfnamefont{A.~J.} \bibnamefont{Masters}},
  \bibinfo{journal}{J. Phys.: Condens. Matter} \textbf{\bibinfo{volume}{20}},
  \bibinfo{pages}{283102} (\bibinfo{year}{2008}).

\bibitem[{\citenamefont{Wertheim}(2001)}]{wertheim:third}
\bibinfo{author}{\bibfnamefont{M.~S.} \bibnamefont{Wertheim}},
  \bibinfo{journal}{Mol. Phys.} \textbf{\bibinfo{volume}{99}},
  \bibinfo{pages}{187} (\bibinfo{year}{2001}).

\bibitem[{\citenamefont{Wertheim}(1996)}]{wertheim:fluids_3}
\bibinfo{author}{\bibfnamefont{M.}~\bibnamefont{Wertheim}},
  \bibinfo{journal}{Mol. Phys.} \textbf{\bibinfo{volume}{89}},
  \bibinfo{pages}{1005} (\bibinfo{year}{1996}).

\bibitem[{\citenamefont{Singh and Kumar}(1996)}]{singh:geometry}
\bibinfo{author}{\bibfnamefont{G.~S.} \bibnamefont{Singh}} \bibnamefont{and}
  \bibinfo{author}{\bibfnamefont{B.}~\bibnamefont{Kumar}}, \bibinfo{journal}{J.
  Chem. Phys.} \textbf{\bibinfo{volume}{105}}, \bibinfo{pages}{2429}
  (\bibinfo{year}{1996}).

\bibitem[{\citenamefont{Rigby}(1993)}]{rigby:virial}
\bibinfo{author}{\bibfnamefont{M.}~\bibnamefont{Rigby}}, \bibinfo{journal}{Mol.
  Phys.} \textbf{\bibinfo{volume}{78}}, \bibinfo{pages}{21}
  (\bibinfo{year}{1993}).

\bibitem[{\citenamefont{Baus et~al.}(1987)\citenamefont{Baus, Colot, Wu, and
  Xu}}]{baus:finite}
\bibinfo{author}{\bibfnamefont{M.}~\bibnamefont{Baus}},
  \bibinfo{author}{\bibfnamefont{J.-L.} \bibnamefont{Colot}},
  \bibinfo{author}{\bibfnamefont{X.-G.} \bibnamefont{Wu}}, \bibnamefont{and}
  \bibinfo{author}{\bibfnamefont{H.}~\bibnamefont{Xu}}, \bibinfo{journal}{Phys.
  Rev. Lett.} \textbf{\bibinfo{volume}{59}}, \bibinfo{pages}{2184}
  (\bibinfo{year}{1987}).

\bibitem[{\citenamefont{Colot et~al.}(1988)\citenamefont{Colot, Wu, Xu, and
  Baus}}]{colot:desnity}
\bibinfo{author}{\bibfnamefont{J.-L.} \bibnamefont{Colot}},
  \bibinfo{author}{\bibfnamefont{X.-G.} \bibnamefont{Wu}},
  \bibinfo{author}{\bibfnamefont{H.}~\bibnamefont{Xu}}, \bibnamefont{and}
  \bibinfo{author}{\bibfnamefont{M.}~\bibnamefont{Baus}},
  \bibinfo{journal}{Phys. Rev. A} \textbf{\bibinfo{volume}{38}},
  \bibinfo{pages}{2022} (\bibinfo{year}{1988}).

\bibitem[{\citenamefont{Perram and Wertheim}(1985)}]{perram:statistical}
\bibinfo{author}{\bibfnamefont{J.~W.} \bibnamefont{Perram}} \bibnamefont{and}
  \bibinfo{author}{\bibfnamefont{M.}~\bibnamefont{Wertheim}},
  \bibinfo{journal}{J. Comp. Phys.} \textbf{\bibinfo{volume}{58}},
  \bibinfo{pages}{409} (\bibinfo{year}{1985}).

\bibitem[{\citenamefont{Berne and Pechukas}(1972)}]{berne:gaussian}
\bibinfo{author}{\bibfnamefont{B.~J.} \bibnamefont{Berne}} \bibnamefont{and}
  \bibinfo{author}{\bibfnamefont{P.}~\bibnamefont{Pechukas}},
  \bibinfo{journal}{The Journal of Chemical Physics}
  \textbf{\bibinfo{volume}{56}}, \bibinfo{pages}{4213} (\bibinfo{year}{1972}).

\bibitem[{\citenamefont{Lee}(1988)}]{lee:onsager}
\bibinfo{author}{\bibfnamefont{S.-D.} \bibnamefont{Lee}}, \bibinfo{journal}{J.
  Chem. Phys.} \textbf{\bibinfo{volume}{89}} (\bibinfo{year}{1988}).

\bibitem[{\citenamefont{Bhethanabotla and
  Steele}(1987)}]{bhethanabotla:comparison}
\bibinfo{author}{\bibfnamefont{V.~R.} \bibnamefont{Bhethanabotla}}
  \bibnamefont{and} \bibinfo{author}{\bibfnamefont{W.}~\bibnamefont{Steele}},
  \bibinfo{journal}{Mol. Phys.} \textbf{\bibinfo{volume}{60}},
  \bibinfo{pages}{249} (\bibinfo{year}{1987}).

\bibitem[{\citenamefont{Singh}(2005)}]{singh:structure}
\bibinfo{author}{\bibfnamefont{R.~C.} \bibnamefont{Singh}},
  \bibinfo{journal}{J. Mol. Liquids} \textbf{\bibinfo{volume}{122}},
  \bibinfo{pages}{1} (\bibinfo{year}{2005}).

\bibitem[{\citenamefont{Zheng et~al.}(2008)\citenamefont{Zheng, Iglesias, and
  Palffy-Muhoray}}]{palffy:distance_3D}
\bibinfo{author}{\bibfnamefont{X.}~\bibnamefont{Zheng}},
  \bibinfo{author}{\bibfnamefont{W.}~\bibnamefont{Iglesias}}, \bibnamefont{and}
  \bibinfo{author}{\bibfnamefont{P.}~\bibnamefont{Palffy-Muhoray}},
  \bibinfo{journal}{electronic-Liquid Crystals Communications}
  (\bibinfo{year}{2008}),
  \bibinfo{note}{\url{http://www.e-lc.org/docs/2008_10_12_23_11_56}}.

\bibitem[{\citenamefont{Bawden et~al.}(1936)\citenamefont{Bawden, Pirie,
  Bernal, and Fankuchen}}]{bawden:liquid}
\bibinfo{author}{\bibfnamefont{F.~C.} \bibnamefont{Bawden}},
  \bibinfo{author}{\bibfnamefont{N.~W.} \bibnamefont{Pirie}},
  \bibinfo{author}{\bibfnamefont{J.~D.} \bibnamefont{Bernal}},
  \bibnamefont{and}
  \bibinfo{author}{\bibfnamefont{I.}~\bibnamefont{Fankuchen}},
  \bibinfo{journal}{Nature} \textbf{\bibinfo{volume}{138}},
  \bibinfo{pages}{1051} (\bibinfo{year}{1936}).

\bibitem[{\citenamefont{Dogic and Fraden}(2006)}]{dogic:ordered}
\bibinfo{author}{\bibfnamefont{Z.}~\bibnamefont{Dogic}} \bibnamefont{and}
  \bibinfo{author}{\bibfnamefont{S.}~\bibnamefont{Fraden}},
  \bibinfo{journal}{Curr. Opin. Colloid Interface Sci.}
  \textbf{\bibinfo{volume}{11}}, \bibinfo{pages}{47} (\bibinfo{year}{2006}).

\bibitem[{\citenamefont{Sonnet and Virga}(2012)}]{sonnet:dissipative}
\bibinfo{author}{\bibfnamefont{A.~M.} \bibnamefont{Sonnet}} \bibnamefont{and}
  \bibinfo{author}{\bibfnamefont{E.~G.} \bibnamefont{Virga}},
  \emph{\bibinfo{title}{Dissipative Ordered Fluids. {T}heories for Liquid
  Crystals}} (\bibinfo{publisher}{Springer}, \bibinfo{address}{New York},
  \bibinfo{year}{2012}).

\bibitem[{\citenamefont{Wertheim}(1994)}]{wertheim:fluids_1}
\bibinfo{author}{\bibfnamefont{M.}~\bibnamefont{Wertheim}},
  \bibinfo{journal}{Mol. Phys.} \textbf{\bibinfo{volume}{83}},
  \bibinfo{pages}{519} (\bibinfo{year}{1994}).

\bibitem[{\citenamefont{Piastra}(2013)}]{piastra:soam}
\bibinfo{author}{\bibfnamefont{M.}~\bibnamefont{Piastra}},
  \bibinfo{journal}{Neural Networks} \textbf{\bibinfo{volume}{41}},
  \bibinfo{pages}{96} (\bibinfo{year}{2013}).

\bibitem[{\citenamefont{de~Graaf et~al.}(2011)\citenamefont{de~Graaf, van Roij,
  and Dijkstra}}]{deGraaf2011prl}
\bibinfo{author}{\bibfnamefont{J.}~\bibnamefont{de~Graaf}},
  \bibinfo{author}{\bibfnamefont{R.}~\bibnamefont{van Roij}}, \bibnamefont{and}
  \bibinfo{author}{\bibfnamefont{M.}~\bibnamefont{Dijkstra}},
  \bibinfo{journal}{Phys. Rev. Lett.} \textbf{\bibinfo{volume}{107}},
  \bibinfo{pages}{155501} (\bibinfo{year}{2011}),
  \urlprefix\url{http://link.aps.org/doi/10.1103/PhysRevLett.107.155501}.

\bibitem[{\citenamefont{de~Graaf et~al.}(2012)\citenamefont{de~Graaf, Filion,
  Marechal, van Roij, and Dijkstra}}]{deGraaf2012method}
\bibinfo{author}{\bibfnamefont{J.}~\bibnamefont{de~Graaf}},
  \bibinfo{author}{\bibfnamefont{L.}~\bibnamefont{Filion}},
  \bibinfo{author}{\bibfnamefont{M.}~\bibnamefont{Marechal}},
  \bibinfo{author}{\bibfnamefont{R.}~\bibnamefont{van Roij}}, \bibnamefont{and}
  \bibinfo{author}{\bibfnamefont{M.}~\bibnamefont{Dijkstra}},
  \bibinfo{journal}{The Journal of Chemical Physics}
  \textbf{\bibinfo{volume}{137}}, \bibinfo{eid}{214101} (\bibinfo{year}{2012}),
  \urlprefix\url{http://scitation.aip.org/content/aip/journal/jcp/137/21/10.10%
63/1.4767529}.

\bibitem[{\citenamefont{Arndt}(2008)}]{Arndt2008509}
\bibinfo{author}{\bibfnamefont{M.~F.} \bibnamefont{Arndt}},
  \bibinfo{journal}{Nuclear Instruments and Methods in Physics Research Section
  A: Accelerators, Spectrometers, Detectors and Associated Equipment}
  \textbf{\bibinfo{volume}{588}}, \bibinfo{pages}{509 } (\bibinfo{year}{2008}),
  ISSN \bibinfo{issn}{0168-9002},
  \urlprefix\url{http://www.sciencedirect.com/science/article/pii/S01689002080%
01939}.

\bibitem[{\citenamefont{Dey}(2006)}]{dey:CurveandSurfaceReconstruction}
\bibinfo{author}{\bibfnamefont{T.~K.} \bibnamefont{Dey}},
  \emph{\bibinfo{title}{Curve and Surface Reconstruction}}
  (\bibinfo{publisher}{Cambridge University Press}, \bibinfo{year}{2006}), ISBN
  \bibinfo{isbn}{9780511546860},
  \urlprefix\url{http://dx.doi.org/10.1017/CBO9780511546860}.

\bibitem[{\citenamefont{Martinetz et~al.}(1993)\citenamefont{Martinetz,
  Berkovich, and Schulten}}]{Martinetz-etal93}
\bibinfo{author}{\bibfnamefont{T.}~\bibnamefont{Martinetz}},
  \bibinfo{author}{\bibfnamefont{S.}~\bibnamefont{Berkovich}},
  \bibnamefont{and} \bibinfo{author}{\bibfnamefont{K.}~\bibnamefont{Schulten}},
  \bibinfo{journal}{IEEE Trans. Neural Networks} \textbf{\bibinfo{volume}{4}},
  \bibinfo{pages}{558} (\bibinfo{year}{1993}), ISSN \bibinfo{issn}{1045-9227}.

\bibitem[{\citenamefont{Bernardini et~al.}(1999)\citenamefont{Bernardini,
  Mittleman, Rushmeier, Silva, and Taubin}}]{bernardini1999ball}
\bibinfo{author}{\bibfnamefont{F.}~\bibnamefont{Bernardini}},
  \bibinfo{author}{\bibfnamefont{J.}~\bibnamefont{Mittleman}},
  \bibinfo{author}{\bibfnamefont{H.}~\bibnamefont{Rushmeier}},
  \bibinfo{author}{\bibfnamefont{C.}~\bibnamefont{Silva}}, \bibnamefont{and}
  \bibinfo{author}{\bibfnamefont{G.}~\bibnamefont{Taubin}},
  \bibinfo{journal}{Visualization and Computer Graphics, IEEE Transactions on}
  \textbf{\bibinfo{volume}{5}}, \bibinfo{pages}{349} (\bibinfo{year}{1999}).

\bibitem[{\citenamefont{Parigi et~al.}(2014)\citenamefont{Parigi, Stramieri,
  Pau, and Piastra}}]{parigi2014multi}
\bibinfo{author}{\bibfnamefont{G.}~\bibnamefont{Parigi}},
  \bibinfo{author}{\bibfnamefont{A.}~\bibnamefont{Stramieri}},
  \bibinfo{author}{\bibfnamefont{D.}~\bibnamefont{Pau}}, \bibnamefont{and}
  \bibinfo{author}{\bibfnamefont{M.}~\bibnamefont{Piastra}}, in
  \emph{\bibinfo{booktitle}{Informatics in Control, Automation and Robotics}}
  (\bibinfo{publisher}{Springer}, \bibinfo{year}{2014}), pp.
  \bibinfo{pages}{83--100}.

\bibitem[{\citenamefont{Cignoni et~al.}(2008)\citenamefont{Cignoni, Corsini,
  and Ranzuglia}}]{cignoni2008meshlab}
\bibinfo{author}{\bibfnamefont{P.}~\bibnamefont{Cignoni}},
  \bibinfo{author}{\bibfnamefont{M.}~\bibnamefont{Corsini}}, \bibnamefont{and}
  \bibinfo{author}{\bibfnamefont{G.}~\bibnamefont{Ranzuglia}},
  \bibinfo{journal}{Ercim news} \textbf{\bibinfo{volume}{73}},
  \bibinfo{pages}{45} (\bibinfo{year}{2008}).

\bibitem[{\citenamefont{Sluckin et~al.}(2004)\citenamefont{Sluckin, Dunmur, and
  Stegemeyer}}]{sluckin:crystal}
\bibinfo{author}{\bibfnamefont{T.~J.} \bibnamefont{Sluckin}},
  \bibinfo{author}{\bibfnamefont{D.~A.} \bibnamefont{Dunmur}},
  \bibnamefont{and}
  \bibinfo{author}{\bibfnamefont{H.}~\bibnamefont{Stegemeyer}},
  \emph{\bibinfo{title}{Crystals that Flow}} (\bibinfo{publisher}{Taylor \&
  Francis}, \bibinfo{address}{London, New York}, \bibinfo{year}{2004}).

\bibitem[{\citenamefont{Olver et~al.}(2010)\citenamefont{Olver, Lozier,
  Boisvert, and Clark}}]{Olver:2010:NHMF}
\bibinfo{editor}{\bibfnamefont{F.~W.~J.} \bibnamefont{Olver}},
  \bibinfo{editor}{\bibfnamefont{D.~W.} \bibnamefont{Lozier}},
  \bibinfo{editor}{\bibfnamefont{R.~F.} \bibnamefont{Boisvert}},
  \bibnamefont{and} \bibinfo{editor}{\bibfnamefont{C.~W.} \bibnamefont{Clark}},
  eds., \emph{\bibinfo{title}{{NIST Handbook of Mathematical Functions}}}
  (\bibinfo{publisher}{Cambridge University Press}, \bibinfo{address}{New York,
  NY}, \bibinfo{year}{2010}), \bibinfo{note}{print companion to
  \cite{NIST:DLMF}}.

\end{thebibliography}

\end{document}